\documentclass[12pt]{article}
\usepackage{amsmath,amssymb,epsfig}
\makeatletter \@addtoreset{equation}{section} \makeatother
\renewcommand{\theequation}{\thesection.\arabic{equation}}
\newcommand{\ba}{\begin{array}}
\newcommand{\ea}{\end{array}}
\newcommand{\beq}{\begin{equation}}
\newcommand{\eeq}{\end{equation}}
\newcommand{\bea}{\begin{eqnarray}}
\newcommand{\eea}{\end{eqnarray}}
\def\bce{\begin{center}}
\def\ece{\end{center}}

\def\nonu{\nonumber}

\def\be{\beta}

\newcommand{\tr}{\mbox{Tr}}

\def\eps6{{\displaystyle \mathop{\epsilon}^{6}}{}}

\def\nab6{{\displaystyle \mathop{\nabla}^{6}}{}}
\def\0{{\sst{(0)}}}
\def\1{{\sst{(1)}}}
\def\2{{\sst{(2)}}}
\def\3{{\sst{(3)}}}
\def\4{{\sst{(4)}}}
\def\5{{\sst{(5)}}}
\def\6{{\sst{(6)}}}
\def\7{{\sst{(7)}}}
\def\8{{\sst{(8)}}}

\def\ba{\begin{array}}
\def\ea{\end{array}}
\def\beq{\begin{equation}}
\def\eeq{\end{equation}}
\def\be{\begin{equation}}
\def\ee{\end{equation}}

\def\tr{\mathop{\rm tr}}

\def\eps{\epsilon}

\def\ba{\begin{array}}
\def\ea{\end{array}}
\def\beq{\begin{equation}}
\def\eeq{\end{equation}}
\def\be{\begin{equation}}
\def\ee{\end{equation}}

\def\tr{\mathop{\rm tr}}

\def\eps{\epsilon}

\newcommand{\bean}{\begin{eqnarray*}}
\newcommand{\eean}{\end{eqnarray*}}

\begin{document}
\thispagestyle{empty} \addtocounter{page}{-1}
   \begin{flushright}
%KIAS-P08019 \\
%CALT-68-nnnn \\
%{\tt hep-th/yymmnnn}\\
\end{flushright}

\vspace*{1.3cm}

\centerline{ \Large \bf  More on Meta-Stable Brane Configurations  }
\vspace{.3cm} 
\centerline{ \Large \bf  by Dualizing the Multiple Gauge Groups  } 
\vspace*{1.5cm}
\centerline{{\bf Changhyun Ahn} 
%and {\bf Yutaka Ookouchi $^{2}$}
} 
\vspace*{1.0cm} 
\centerline{\it 
Department of Physics, Kyungpook National University, Taegu
702-701, Korea} 
%\centerline{\it $^{2}$ California Institute of 
%Technology, Pasadena, CA91125, USA }
\vspace*{0.8cm} 
\centerline{\tt ahn@knu.ac.kr} 
%\qquad
%yutaka@caltech.edu} 
\vskip2cm

\centerline{\bf Abstract}
\vspace*{0.5cm}

We reexamine
the ${\cal N}=1$ supersymmetric gauge theories with product gauge groups
by adding the mass terms and the quartic terms
for the flavors: two gauge group theory 
with fundamentals, bifundamentals and adjoints, 
three gauge group theory with fundamentals and bifundamentals,
and their orientifold 4-plane generalizations. 
By moving the branes appropriately,    
we obtain the corresponding dual gauge theories. 
By analyzing the dual superpotentials, we present the type IIA
nonsupersymmetric meta-stable brane configurations.
 
\baselineskip=18pt
\newpage
\renewcommand{\theequation}
{\arabic{section}\mbox{.}\arabic{equation}}

%%%%%%%%%%%%%%%%%%%%%%%%%%%%%%%%%%%%%%%%%%%%%%%%%%%%%%%%%%%%%%%%%%%%%
%%%%%%%%%%%%%%%%%%%%%%%%%%%%%%%%%%%%%%%%%%%%%%%%%%%%%%%%%%%%%%%%%%%%%%
\section{Introduction}
%%%%%%%%%%%%%%%%%%%%%%%%%%%%%%%%%%%%%%%%%%%%%%%%%%%%%%%%%%%%%%%%%%%%%%
%%%%%%%%%%%%%%%%%%%%%%%%%%%%%%%%%%%%%%%%%%%%%%%%%%%%%%%%%%%%%%%%%%%%%

The dynamical supersymmetry breaking in meta-stable vacua \cite{ISS,IS} 
occurs
in the ${\cal N}=1$ supersymmetric gauge theory with massive fundamental 
quarks.  
Further quartic deformation for the quarks in the 
superpotential \cite{GK0710,GK0710-1} also has lead to  
the supersymmetry breaking meta-stable ground states 
when 
the gravitational attraction of NS5-brane \cite{GK} is considered.
It is known, in the construction of 
supersymmetric ground states or its type IIA brane
configurations \cite{GK98},  
that  a number of gauge theory duals(magnetic theory) 
involving
product gauge groups can be interpreted in terms of branes of type IIA
string theory. 
The question for the dynamical supersymmetry
breaking in meta-stable vacua in the simplest 
${\cal N}=1$ supersymmetric
product gauge theories
which have both mass terms \cite{ISS,OO1,FGU,BGHSS} and quartic terms
\cite{GK0710,GK0710-1} is answered recently in \cite{Ahn08-4} 
when one dualizes the whole two gauge
groups.   

Although a single gauge group theory with two adjoint fields, where the
superpotential has order $(k+1)$ and $(k'+1)$ for these adjoint fields, does
not have dual theory correctly so far \cite{EGKRS},  
the two gauge group theory with two adjoint fields as well as fundamentals
and bifundamentals does have its dual description \cite{Brodie}.
For $k=2$ and $k'=1$ case(i.e., two NS5-branes and one NS5'-brane), 
the meta-stable brane configuration for a single gauge group theory 
was found in \cite{Ahn06} and 
for the $k=1$ and $k'=2$ case(i.e., one NS5-brane and two
NS5'-branes), 
the meta-stable brane configuration for a single gauge group theory
was studied in \cite{Ahn08-3}.
The type IIA brane configuration corresponding to two gauge
group theory with fundamentals, bifundamentals and two adjoints 
was found in \cite{BH} some time ago and 
the brane configuration on triple product gauge group theory 
with fundamentals and bifundamentals
was also studied
by using the brane motion and linking number counting: the peculiar
thing was the presence of full D4-branes ranging from 
$x^6=-\infty$ to $x^6=\infty$ in the brane 
configuration without changing the linking numbers. 

In this paper, one reexamines these supersymmetric brane
configurations \cite{BH}
as well as their orientifold 4-plane generalizations
and extracts the possible brane motions, during the dual process, for 
new meta-stable brane configurations, along the lines of 
\cite{Ahn07-11,Ahn08-1,Ahn08-1two,Ahn08-2,Ahn08-3,Ahn08-4}. 
The common feature with the work of \cite{Ahn08-4} and different
feature with the previous works of 
\cite{Ahn07-11,Ahn08-1,Ahn08-1two,Ahn08-2,Ahn08-3} is that one
dualizes the {\it whole product gauge groups}, not a single gauge group. 
The order of NS-branes is reversed completely after duality.
The geometrical positions of the
branes and the creation of D4-branes, during the dual process, 
play the important role for
removing the unwanted gauge singlets \cite{Ahn07,Ahn07-1} and selecting 
the wanted gauge singlets.  The magnetic theories we obtained are
different from the one \cite{BH} in the sense that the dual color
numbers are different and the dual superpotentials have different
form: some of the gauge singlets do not appear and we add the mass
terms and quartic terms for the quarks. These deformation terms are
realized geometrically in type IIA string theory. 

In section 2, we review the type IIA brane configuration corresponding
to the ${\cal N}=1$ $SU(N_c) \times SU(N_c')$ gauge theory 
with fundamentals, bifundamentals and {\it adjoints} and deform this theory 
by adding both the mass terms
and the quartic terms for the fundamentals. 
The extra adjoint fields are realized by the multiple NS-branes in
brane configuration.
Then we describe the dual 
${\cal N}=1$ $SU(\widetilde{N}_c) \times SU(\widetilde{N}_c')$ gauge 
theory with corresponding dual
matter as well as the gauge singlets. 
We discuss the nonsupersymmetric meta-stable
minimum  by looking at the reduced dual superpotential and present 
the corresponding 
intersecting brane configuration of type IIA string
theory.

In section 3,
we review the type IIA brane configuration corresponding
to the ${\cal N}=1$ $SO(2N_c) \times Sp(N_c')$ gauge theory 
with vectors, fundamentals, bifundamentals and {\it adjoints} 
and deform this theory 
by adding both the mass terms
and the quartic terms for the vectors and fundamentals. 
Then we describe the dual 
${\cal N}=1$ $SO(2\widetilde{N}_c) \times Sp(\widetilde{N}_c')$ gauge 
theory with corresponding dual
matter as well as the gauge singlets. 
We describe the nonsupersymmetric meta-stable
minimum  and 
the corresponding 
intersecting brane configuration of type IIA string
theory which is nothing but the brane configuration of section 3 with
the addition of O4-plane.

In section 4,
 we review the type IIA brane configuration corresponding
to the ${\cal N}=1$ $SU(N_c) \times SU(N_c') \times SU(N_c'')$ gauge theory 
with fundamentals, bifundamentals and deform this theory 
by adding both the mass terms
and the quartic terms for the fundamentals. 
Then we describe the dual 
${\cal N}=1$ $SU(\widetilde{N}_c) \times SU(\widetilde{N}_c') 
\times SU(\widetilde{N}_c'')$ gauge 
theory with corresponding dual
matter as well as the gauge singlets. 
We discuss the nonsupersymmetric meta-stable
minimum  from the reduced dual superpotential and present 
the corresponding 
intersecting brane configuration of type IIA string
theory.

In section 5,
we review the type IIA brane configuration corresponding
to the ${\cal N}=1$ $Sp(N_c) \times SO(2N_c') \times Sp(N_c'')$ gauge theory 
with fundamentals, vectors, bifundamentals and deform this theory 
by adding both the mass terms
and the quartic terms for the fundamentals. 
Then we describe the dual 
${\cal N}=1$ $Sp(\widetilde{N}_c) \times SO(2\widetilde{N}_c') 
\times Sp(\widetilde{N}_c'')$ gauge 
theory with corresponding dual
matter as well as the gauge singlets. 
We discuss the nonsupersymmetric meta-stable
minimum  and present 
the corresponding 
intersecting brane configuration of type IIA string
theory which is nothing but the brane configuration of section 4 with
the addition of O4-plane.

In section 6, we summarize the results of this paper and 
comment on the future directions.

%%%%%%%%%%%%%%%%%%%%%%%%%%%%%%%%%%%%%%%%%%%%%%%%%%%%%%%%%%%%%%%%
%%%%%%%%%%%%%%%%%%%%%%%%%%%%%%%%%%%%%%%%%%%%%%%%%%%%%%%%%%%%%%%%
\section{$SU(N_c) \times SU(N_c')$ with $N_f$-, $N_f'$-fund., two
adjoints, and bifund.}
%%%%%%%%%%%%%%%%%%%%%%%%%%%%%%%%%%%%%%%%%%%%%%%%%%%%%%%%%%%%%%%%
%%%%%%%%%%%%%%%%%%%%%%%%%%%%%%%%%%%%%%%%%%%%%%%%%%%%%%%%%%%%%%%%

%%%%%%%%%%%%%%%%%%%%%%%%%%%%%%%%%%%%%%%%%%%%%%%%%%%%%%%%%%%%%%%%%%%%%%%%%%%%
\subsection{Electric theory}
%%%%%%%%%%%%%%%%%%%%%%%%%%%%%%%%%%%%%%%%%%%%%%%%%%%%%%%%%%%%%%%%%%%%%%%%%%%%

The type IIA supersymmetric electric
brane configuration \cite{BH,AT97,BHKL} corresponding to 
${\cal N}=1$ $SU(N_c) \times SU(N_c')$ gauge theory  \cite{Brodie} with  
$N_f$-fundamental flavors $Q, \widetilde{Q}$,
$N_f'$-fundamental flavors $Q', \widetilde{Q}'$,
bifundamentals $F, \widetilde{F}$ and two adjoint fields $X_1, X_2$ 
can be described as two middle NS5-branes, two
left NS5'-branes, two right NS5'-branes, 
$N_c$- and $N_c'$-D4-branes, and $N_f$- and 
$N_f'$-D6-branes for the cubic superpotential of the adjoints.
The $X_1$ is in the representation $\bf{(N_c^2-1,1)}$
while the $X_2$ is in the representation $\bf{(1, {N_c'}^2-1)}$, under
the gauge group. 
The $F$ is in the representation $\bf{(N_c,N_c')}$ while 
the $\widetilde{F}$ is in the representation $\bf{(\overline{N_c},
\overline{N_c'})}$,
under the gauge group. 
The quarks $Q$ and $\widetilde{Q}$ are in the representation 
$\bf{(N_c, 1)}$ and $\bf{(\overline{N_c}, 1)}$ 
respectively
and
the quarks $Q'$ and $\widetilde{Q}'$ are in the representation 
$\bf{(1, N_c')}$ and $\bf{(1, \overline{N_c'})}$ 
respectively, under the gauge group.

The mass terms for each fundamental quarks 
can be added in the superpotential by displacing each D6-branes along 
\bea
v \equiv x^4 + i x^5
\nonu
\eea
direction leading to their coordinates 
$v = + 
v_{D6_{-\theta}}(+v_{D6_{-\theta'}})$ respectively  
while the quartic terms for each fundamental quarks 
can be added also by rotating each D6-branes \cite{GK0710-1}
by an angle 
$-\theta(-\theta')$ in $(w,v)$-plane respectively.
Here we define the complex coordinate $w$ as 
\bea
w \equiv x^8 + i x^9.
\nonu 
\eea 
Then, the general 
superpotential from the one \cite{Brodie,BH,BHKL} 
by adding the above deformations is
given by
\bea
W_{elec} & = &
 \left[    \frac{s_1}{3} \tr X_1^3 + \frac{s_2}{3} \tr X_2^3 + 
\tr X_1  \widetilde{F} F
  + \tr X_2  F \widetilde{F}  + \lambda_1 Q X_1 \widetilde{Q} +
  \lambda_2 Q' X_2
  \widetilde{Q}'  \right] \nonu \\
 &+& \frac{\alpha}{2} \tr (Q \widetilde{Q})^2 - m \tr Q
\widetilde{Q} 
 + \frac{\alpha'}{2} \tr (Q' \widetilde{Q}')^2 - m' \tr Q'
\widetilde{Q}'  
\label{elecsuper}
\eea 
where the parameters are described as the following 
geometric quantities
\bea
\alpha \equiv \frac{\tan \theta}{\Lambda}, \qquad
 \alpha' \equiv \frac{\tan \theta'}{\Lambda'}, \qquad
m \equiv \frac{v_{D6_{-\theta}}}{2\pi \ell_s^2}, \qquad
m' \equiv \frac{v_{D6_{-\theta'}}}{2\pi \ell_s^2}, \qquad
\lambda_1 \equiv \sin \theta, \qquad
\lambda_2 \equiv \sin \theta'.
\nonu
\eea
The first two terms of (\ref{elecsuper}), in general, 
are due to the rotation angles $\omega_L$ and
$\omega_R$  
of two left and right NS5'-branes 
with respect to the middle NS5-branes:
$s_1 \equiv \tan \omega_L$ and $ s_2 \equiv \tan \omega_R$. 
We consider the case where $\omega_L=\omega_R=\frac{\pi}{2}$.
Although 
the relative displacement of two color D4-branes, where the mass for the
bifundamentals $m_F
\equiv \frac{v_{NS5'}}{2\pi \ell_s^2}$
is the distance of D4-branes 
along the $v$-direction, can be added in the superpotential,
we focus on the particular limit $m_F =0$.
Note that we also put the perturbations by $Q X_1 \widetilde{Q}$ and 
$Q' X_2 \widetilde{Q}'$ in the superpotential \cite{BHKL} which will
arise as the mesons in the magnetic theory.
The lower order terms for the adjoints can occur when we displace the
coincident 
NS5-branes and NS5'-branes in $w$ direction and in $v$ direction respectively.

Then the ${\cal N}=1$ supersymmetric electric brane
configuration for the superpotential 
(\ref{elecsuper}) 
in type IIA string theory is given as follows and let us draw this
brane structure in
Figure 1 explicitly:

$\bullet$
Two middle NS5-branes in $(012345)$ directions 

$\bullet$ 
Two left NS5'-branes in  $(012389)$ directions 

$\bullet$ 
Two right NS5'-branes in  $(012389)$ directions 

$\bullet$
$N_f$ $D6_{-\theta}$-branes in (01237)
directions and
two other directions in $(v,w)$-plane

$\bullet$
$N_f'$ 
$D6_{-\theta'}$-branes in (01237)
directions and
two other directions in $(v,w)$-plane

$\bullet$
$N_c$- and $N_c'$-color D4-branes in $(01236)$ directions   

%%%%%%%%%%%%%%%%%
%Figure 1
%%%%%%%%%%%%%%%%
%%%%%%%%%%%%%%%%%%%%%%%%%%%%%%%%%%%%%%%%%%%%%%%%%%%%%%%%%%%%%%%%%%%%%%
%%%%%%%%%%%%%%%%%%%%%%%%%%%%%%%%%%%%%%%%%%%%%%%%%%%%%%%%%%%%%%%%%%%%%%
\begin{figure}[ht]
   \epsfxsize=4.5in 
\centerline{\epsffile{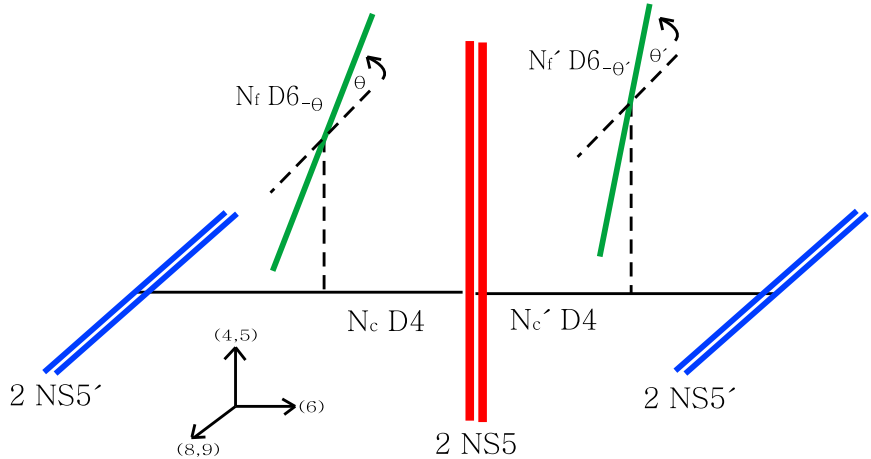}}
   \caption[FIG. \arabic{figure}.]{ 
The  ${\cal N}=1$ supersymmetric 
electric brane configuration for the gauge group $SU(N_c) \times SU(N_c')$ 
with bifundamentals $F, \widetilde{F}$, 
fundamentals $Q, \widetilde{Q}, Q', \widetilde{Q}'$, and adjoint fields
$X_1, X_2$. 
The powers 
of $X_1$ and $X_2$ in the superpotential (\ref{elecsuper}) are related
to the number of NS-branes. A 
rotation of coincident $N_f(N_f')$ D6-branes in $(w,v)$-plane
corresponds to 
a quartic term for the fundamentals $Q, \widetilde{Q}(Q', \widetilde{Q}')$ while 
a displacement of $N_f(N_f')$ D6-branes in $+v$ direction corresponds to a
mass term for the fundamentals $Q, \widetilde{Q}(Q', \widetilde{Q}')$.
 Let us assume that $v_{D6_{-\theta}}
< v_{D6_{-\theta'}}$ and 
$\theta < \theta'$.}
\end{figure}
%%%%%%%%%%%%%%%%%%%%%%%%%%%%%%%%%%%%%%%%%%%%%%%%%%%%%%%%%%%%%%%%%%%%%
%%%%%%%%%%%%%%%%%%%%%%%%%%%%%%%%%%%%%%%%%%%%%%%%%%%%%%%%%%%%%%%%%%%%%

%%%%%%%%%%%%%%%%%%%%%%%%%%%%%%%%%%%%%%%%%%%%%%%%%%%%%%%%%%%%%%%%%%%%%%%%%%%
\subsection{Magnetic theory }
%%%%%%%%%%%%%%%%%%%%%%%%%%%%%%%%%%%%%%%%%%%%%%%%%%%%%%%%%%%%%%%%%%%%%%%%

For the meta-stable brane configurations, we should 
go to the magnetic theory corresponding to 
the previous electric theory.
The left NS5'-branes 
start out with linking number $l_e=-\frac{2N_f'}{2} + N_c$
from Figure 1 and after duality 
these left NS5'-branes end up with linking number 
$l_m = \frac{2N_f'}{2} -\widetilde{N}_c'+2N_f$ from Figure 2.
In arriving at this, we consider the coincident NS-branes for the time
being
and assume that there is no splitting and reconnection 
between D4-branes in Figure 2.
We consider only the particular brane motion where
$N_f$ $D6_{-\theta}$-branes meet 
the middle NS5-branes with {\it no angles}. That is, 
the  $D6_{-\theta}$-branes become $D6_{-\frac{\pi}{2}}$-branes 
when they meet with the middle NS5-branes instantaneously and then
after that
they come back to the original $D6_{-\theta}$-branes \cite{Ahn08-4}.
Therefore, in this dual process, there is no creation of D4-branes.
That is the reason for the $2N_f$ factor in the $l_m$, not $4N_f$.
Then the dual color number $\widetilde{N}_c'$
is given by $\widetilde{N}_c' = 2N_f+2N_f'-N_c$. 

The right NS5'-branes 
start out with linking number $l_e=\frac{2N_f}{2} - N_c'$
and after duality 
these right NS5'-branes end up with linking number 
$l_m = -\frac{2N_f}{2} +\widetilde{N}_c-2N_f'$.
We consider only the particular brane motion where
the  $D6_{-\theta'}$-branes become $D6_{-\frac{\pi}{2}}$-branes 
when they meet with the middle NS5-branes instantaneously and after that 
they come back to the original  $D6_{-\theta'}$-branes.
Therefore, in this dual process, there is {\it no} creation of D4-branes.
That is the reason for the $2N_f'$ factor in the $l_m$, not $4N_f'$.
Then it turns out that the dual color number $\widetilde{N}_c$
is given by $\widetilde{N}_c = 2N_f'+2N_f-N_c'$. 
Finally, one has the following dual color numbers   
\bea
\widetilde{N}_c = 2N_f'+2N_f-N_c', 
\qquad \widetilde{N}_c' = 2N_f+2N_f'-N_c.
\nonu
\eea

The low energy theory \cite{Brodie,BH} on the two color D4-branes 
has $SU(\widetilde{N}_c) \times SU(\widetilde{N}_c')$ gauge group and  
$N_f$-fundamental dual quarks $q', \widetilde{q}'$, 
$N_f'$-fundamental dual quarks $q, \widetilde{q}$,
bifundamentals $f, \widetilde{f}$, two adjoints $x_1, x_2$  
and various gauge singlets.
The $f$ is in the representation $\bf{(\widetilde{N}_c, \widetilde{N}_c')}$ while 
the $\widetilde{f}$ is in the representation $\bf{(\overline{\widetilde{N}_c},
\overline{\widetilde{N}_c'})}$,
under the dual gauge group. 
The $N_f'$ flavors $q$ and $\widetilde{q}$ are in the representation 
$\bf{(\widetilde{N}_c, 1)}$ and $\bf{(\overline{\widetilde{N}_c}, 1)}$ 
respectively under the gauge group and 
 in the representation 
$\bf{(\overline{N_f'}, 1)}$ and $\bf{(1, \overline{N_f'})}$ 
respectively under the flavor group $SU(N_f')_L \times SU(N_f')_R$.
Similarly, 
the $N_f$ flavors $q'$ and $\widetilde{q}'$ are in the representation 
$\bf{(1, \widetilde{N}_c')}$ and $\bf{(1, \overline{\widetilde{N}_c'})}$ 
respectively under the gauge group
and 
in the representation 
$\bf{(\overline{N_f}, 1)}$ and $\bf{(1, \overline{N_f})}$ 
respectively under the flavor group $SU(N_f)_L \times SU(N_f)_R$.
In particular, a magnetic meson field 
$
M_0 \equiv Q \widetilde{Q}
$
is $N_f \times N_f$ matrix and comes from 
4-4 strings of $N_f$ flavor D4-branes, created from the intersection
of $D6_{-\theta}$-branes and one of the right NS5'-branes, while
a magnetic meson field 
$
M'_0 \equiv Q' \widetilde{Q}'
$
is $N_f' \times N_f'$ matrix and comes from 
4-4 strings of $N_f'$ flavor D4-branes, created from 
 the intersection
of $D6_{-\theta'}$-branes and one of the left NS5'-branes.
The adjoint fields $x_1, x_2$ correspond to the motion of two
left and right NS5'-branes and two NS5-branes in $(v,w)$-plane.

Then the most general magnetic superpotential, when we consider  
the case where $N_f(N_f')$ $D6_{-\theta}$-branes($D6_{-\theta'}$-branes) meet 
the middle NS5-branes {\it with angles}, compared to the previous paragraph,
is given by  
\bea
W_{dual} & = & 
\left[ \frac{s_1}{3} x_1^3 + \frac{s_2}{3} x_2^3 + x_1  \widetilde{f} f
+
x_2  f \widetilde{f}  + \lambda_1
M_1 + \lambda_2 M_1'  \right]
\label{dualdual} \\
& + & 
\left( \frac{\alpha}{2} \tr M_0^2 - m M_0  \right) + \left(
\frac{\alpha'}{2} 
\tr {M_0'}^2 - m' M_0' \right) \nonu \\
& +& \left[ M_0 q' x_2 f \widetilde{f} \widetilde{q}' +
  M_0' q x_1  \widetilde{f} f   \widetilde{q}  + M_1 q'  
f \widetilde{f} \widetilde{q}' +  M_1' q   \widetilde{f} f  
\widetilde{q} \right] \nonu \\ 
& + & \left[ M_2 \widetilde{q}' x_2 q' + M_3 \widetilde{q}' q' 
 +   
M_2' \widetilde{q} x_1 q + M_3' \widetilde{q} q
+ P_1 q x_1 \widetilde{f} \widetilde{q}' + 
\widetilde{P}_1  q' x_2 f \widetilde{q} +P_2  q 
\widetilde{f} \widetilde{q}'  + 
\widetilde{P}_2  q'  f \widetilde{q} \right]
\nonu
\eea
where the mesons are given by \cite{Brodie,BH}
\bea
M_0 & \equiv &  Q  \widetilde{Q},\quad  
M_0'  \equiv Q' \widetilde{Q}', \quad 
M_1 \equiv Q X_1 \widetilde{Q}, \quad
M_1'   \equiv  Q'  X_2 \widetilde{Q}',
\nonu \\
M_2  & \equiv & Q  \widetilde{F} F \widetilde{Q}, \quad 
M_3   \equiv Q \widetilde{F} F X_1 \widetilde{Q},
  \quad
M_2'  \equiv Q'  F \widetilde{F}  \widetilde{Q}', \quad
M_3'   \equiv Q'  F \widetilde{F}  X_2 \widetilde{Q}', 
\nonu \\
\quad 
P_1   & \equiv &   Q \widetilde{F} \widetilde{Q}', 
\quad
\widetilde{P}_1   \equiv    Q' F \widetilde{Q}, \quad
P_2 \equiv Q X_1 \widetilde{F}  \widetilde{Q}', \quad
\widetilde{P}_2  \equiv  Q' X_2 F \widetilde{Q}.
\nonu
\eea
The first two lines of (\ref{dualdual}) are dual expressions for the electric
superpotential (\ref{elecsuper}) and the corresponding meson fields
$M_0, M_0', M_1$ and $M_1'$ are replaced and the third and fourth
lines of (\ref{dualdual}) are the analogs of the cubic term
superpotential between
the meson and dual quarks in Seiberg duality. 
Compared with the theory \cite{ILS,Ahn08-4} without two adjoints fields, 
there exist the extra meson fields coming from the adjoint fields 
$X_1$ and $X_2$:$M_1, M_1',M_3,M_3',P_2$ and $\widetilde{P}_2$.

Now we want to find out the reduced magnetic superpotential which is
relevant to the meta-stable brane configuration we are interested in.

As observed in \cite{Ahn08-4}, 
when the $N_f$ $D6_{-\theta}$-branes meet the middle NS5-branes,
{\it no} creation of D4-branes  
implies that there is no $M_2$- or $M_3$-term 
in the above superpotential
(\ref{dualdual}).
The mesons  
$M_2$ and $M_3$ originate from  $SU(N_c)$ chiral mesons
$Q\widetilde{Q}$  when one
dualizes the $SU(N_c)$ gauge group first by moving the middle NS5-branes
to the left of the left NS5'-branes 
\cite{Ahn07-3,Ahn07-4,Ahn07-8,Ahn07-9,BIWW}. That is, the fluctuations of
strings stretching between the $2N_f$ ``flavor'' D4-branes correspond
to these meson fields. 
After two additional dual procedures, 
$SU(N_c')$ and $SU(\widetilde{n}_c)$, the cubic terms in the
superpotential arise as 
$M_2$-dependent  and $M_3$-dependent terms 
where $M_2$ has 
extra $\widetilde{F} F$ fields and 
$M_3$ has 
extra $\widetilde{F} F X_1 $ fields, besides $Q \widetilde{Q}$, 
due to the further $SU(N_c')$-dualization. The $M_2$-term in the
superpotential has an extra 
$x_2$ factor besides $\widetilde{q}' q'$.

Similarly, 
when 
the $N_f'$ $D6_{-\theta'}$-branes 
meet the middle NS5-branes with {\it no angles}, 
there is no $M_2'$- or $M_3'$-term in the above superpotential
(\ref{dualdual}).
These meson fields 
$M_2'$ and $M_3'$ originate from  $SU(N_c')$ chiral mesons
$Q'\widetilde{Q}'$  when one
dualizes the $SU(N_c')$ gauge group first by moving the middle NS5-branes
to the right of the right NS5'-branes. The
strings stretching between the $2N_f'$ ``flavor'' D4-branes provide
these mesons. 
After two additional dual procedures, 
$SU(N_c)$ and $SU(\widetilde{n}_c')$, 
the cubic terms in the superpotential arise as 
$M_2'$-term and $M_3'$-term  where $M_2'$ has 
extra $F \widetilde{F} $ fields 
and $M_3'$ has 
extra $F \widetilde{F} X_2$ fields, besides $Q' \widetilde{Q}'$, 
due to the further $SU(N_c)$-dualization.
The $M_2'$-term in the superpotential has an extra 
$x_1$ factor  besides $\widetilde{q} q$.

Furthermore, 
when the $N_f$ $D6_{-\theta}$-branes, the 
$N_f'$ $D6_{-\theta'}$-branes and the middle NS5-branes
meet each other with no angles,
{\it no} $P_1$- and $P_2$- or 
$\widetilde{P}_1$- and $\widetilde{P}_2$-dependent terms arise in the
superpotential (\ref{dualdual}).
These mesons 
originate from  $SU(N_c')$ chiral mesons
$\widetilde{F} \widetilde{Q}'$ and $F Q'$ when one
dualizes the $SU(N_c')$  first by moving the middle NS5-branes
to the right of the right NS5'-branes. 
The
strings stretching between the $2N_f'$ flavor D4-branes and $N_c$
color D4-branes give rise to these 
$2N_f'$ $SU(N_c)$ fundamentals and 
$2N_f'$ $SU(N_c)$ antifundamentals. 
After two additional dual procedures, 
$SU(N_c)$ and $SU(\widetilde{n}_c')$, these cubic terms arise as 
these meson terms  where there exist  extra 
$q x_1, \widetilde{q} x_2, q$ 
and $\widetilde{q}$ in the interactions of $P_1, \widetilde{P}_1, P_2$ and 
$\widetilde{P}_2$ in the superpotential respectively and these mesons have 
extra $Q, \widetilde{Q}, Q X_1, \widetilde{Q} X_2$ fields respectively, 
due to the further $SU(N_c)$-dualization. 

Then the reduced magnetic superpotential in our case  
by taking the first three lines of (\ref{dualdual}) 
is given by 
\bea
W_{dual} & = & 
\left[ \frac{s_1}{3} x_1^3 + \frac{s_2}{3} x_2^3 + x_1  \widetilde{f} f
+
  \widetilde{f} x_2 f  + M_1 ( q'  
f \widetilde{f} \widetilde{q}' +\lambda_1) +  
M_1' ( q   \widetilde{f} f \widetilde{q} + \lambda_2) \right]
\nonu \\ 
& + & 
\left[  M_0 q' x_2 f \widetilde{f} \widetilde{q}' +
\frac{\alpha}{2} \tr M_0^2 - m M_0  \right] + \left[
M_0' q x_1  \widetilde{f} f  \widetilde{q} + \frac{\alpha'}{2} 
\tr {M_0'}^2 - m' M_0' \right].
\label{dualW}
\eea
Let us describe the meta-stable brane configuration with this magnetic 
superpotential.
For the supersymmetric vacua, one can compute the F-term equations for
this superpotential (\ref{dualW}) 
and the F-terms for $M_0, q', \widetilde{q}', M_0', q,
\widetilde{q}, f, \widetilde{f}, M_1, M_1', x_1$ 
and $x_2$ are given by
\bea
&& q' x_2 f \widetilde{f} \widetilde{q}'-m + \alpha M_0 =0, \qquad 
 x_2 f \widetilde{f} \widetilde{q}' M_0 +   
f \widetilde{f} \widetilde{q}' M_1 =0, \qquad
(M_0 q' x_2 + M_1 q') f \widetilde{f}  =0,
\nonu \\
&& q x_1  \widetilde{f} f   \widetilde{q} -m' + \alpha' M_0'=0, \qquad 
x_1 \widetilde{f} f \widetilde{q} M_0'  +   
\widetilde{f} f \widetilde{q} M_1' =0,
\qquad 
(M_0' q x_1 +  M_1' q ) \widetilde{f} f =0,
\nonu \\
&& (x_1  \widetilde{f} + \widetilde{f} x_2) + \widetilde{f}
\widetilde{q}' 
( M_1 q'+
 M_0 q' x_2) + \widetilde{q}  (M_1' q  +
 M_0' q x_1) \widetilde{f} =0, 
\nonu \\
&& (f x_1 +  x_2 f) +  \widetilde{q}' (M_1 q'  
 +    M_0 q' x_2) f + f \widetilde{q} (M_1' q   +  M_0' 
q x_1)  =0,
\qquad  q'  
f \widetilde{f} \widetilde{q}' +\lambda_1 =0, \nonu \\  
&& q  
\widetilde{f}  f \widetilde{q} +
\lambda_2 =0,
\qquad 
s_1 x_1^2 +  \widetilde{f} f +   \widetilde{f} f   \widetilde{q} M_0' q
=0, \qquad
s_2 x_2^2 + f \widetilde{f}  + f \widetilde{f} \widetilde{q}'  M_0 q'=0.
\label{fterm}
\eea

%%%%%%%%%%%%%%%%%
%Figure 2
%%%%%%%%%%%%%%%%%
%%%%%%%%%%%%%%%%%%%%%%%%%%%%%%%%%%%%%%%%%%%%%%%%%%%%%%%%%%%%%%%%%%%%%%
%%%%%%%%%%%%%%%%%%%%%%%%%%%%%%%%%%%%%%%%%%%%%%%%%%%%%%%%%%%%%%%%%%%%%%
\begin{figure}[ht]
   \epsfxsize=4.5in 
\centerline{\epsffile{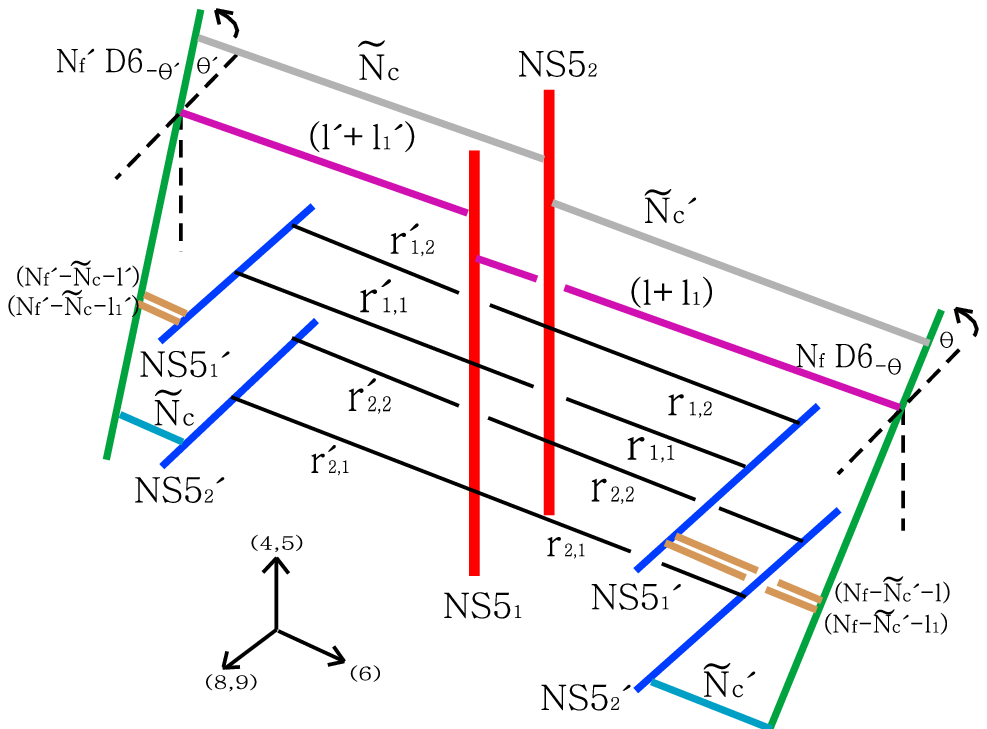}}
   \caption[FIG. \arabic{figure}.]{ 
The  ${\cal N}=1$ supersymmetric
magnetic brane configuration corresponding to Figure 1 with 
a splitting and a reconnection 
between D4-branes when the gravitational potential of
the NS5-branes is ignored.
The $NS5_1$-brane is located at $w=0$.  
The $N_f$ flavor D4-branes connecting between
$D6_{-\theta}$-branes and $NS5_1'$-brane are splitting into 
$(N_f-\widetilde{N}_c'-l_1)$-, $l_1$-, and 
$\widetilde{N}_c'$-D4-branes while
the other $N_f$ flavor D4-branes connecting between
them are splitting into 
$\widetilde{N}_c'$-, $(N_f-\widetilde{N}_c'-l)$-, and
$l$- D4-branes. Also other $2N_f'$ flavor D4-branes can split
similarly. 
The $v$ and $w$ coordinates of these D4-branes are increasing and
decreasing, in that order.  
Note that the numbers of dual colors connecting 
$i$-th $NS5_i'$-brane and $j$-th $NS5_j$-brane $r_{i,j}$ and
$r'_{i,j}$ have the following relations  
$\sum_{i,j=1}^2 r_{i,j} = \widetilde{N}_c'-l-l_1$ and 
$\sum_{i,j=1}^2 r'_{i,j} = \widetilde{N}_c-l'-l_1'$.}
\end{figure}
%%%%%%%%%%%%%%%%%%%%%%%%%%%%%%%%%%%%%%%%%%%%%%%%%%%%%%%%%%%%%%%%%%%%%
%%%%%%%%%%%%%%%%%%%%%%%%%%%%%%%%%%%%%%%%%%%%%%%%%%%%%%%%%%%%%%%%%%%%%

The third, sixth, seventh and eighth equations of (\ref{fterm}) 
are satified if  
the following equations hold 
\bea
f x_1= -x_2 f, \qquad x_1 \widetilde{f} = - \widetilde{f} x_2,
\qquad M_1 q' = -M_0 q' x_2, \qquad M_1' q = - M_0' q x_1.
\label{fterm1}
\eea
By multiplying $f$ to the second equation of 
(\ref{fterm1}) from the left, one obtains
$
f x_1 \widetilde{f} = - f \widetilde{f} x_2$.
Using the first equation of (\ref{fterm1}) one gets
$
(-x_2 f) \widetilde{f} = -  f \widetilde{f} x_2$ and this leads to 
$
x_2 f \widetilde{f} =   f \widetilde{f} x_2$.
Then one simplifies the second equation of (\ref{fterm}) as
\bea 
f \widetilde{f} (x_2 \widetilde{q}' M_0 +   
 \widetilde{q}' M_1) =0 \rightarrow  
 \widetilde{q}' M_1 = -x_2 \widetilde{q}' M_0
\nonu 
\eea 
by moving $x_2$ to the right.
Similarly, 
by multiplying $\widetilde{f}$ to the first equation of 
(\ref{fterm1}) from the left, one obtains
$
\widetilde{f} f x_1  = - \widetilde{f} x_2 f$.
Using the second equation of (\ref{fterm1}) one gets
$
\widetilde{f} f x_1  =   (x_1 \widetilde{f}) f$ and this leads to 
$
x_1 \widetilde{f} f =   \widetilde{f} f x_1$.
Then one simplifies the fifth equation of (\ref{fterm}) as
\bea
\widetilde{f} f (x_1 \widetilde{q} M_0'  +   
\widetilde{q} M_1') =0 \rightarrow      
\widetilde{q} M_1' = -x_1 \widetilde{q} M_0'
\nonu
\eea
by moving $x_1$ to the right.

Then the remaining F-term equations can be summarized as
\bea
&& q'  f \widetilde{f} x_2 \widetilde{q}'-m + \alpha M_0  =  0, \qquad
q   \widetilde{f} f x_1   \widetilde{q} -m' + \alpha' M_0'=0, 
\qquad
q'  
f \widetilde{f} \widetilde{q}' +\lambda_1  =  0, \nonu \\
&& q  
\widetilde{f}  f \widetilde{q} +
\lambda_2 =0,
\qquad
s_1 x_1^2 +  \widetilde{f} f (1 +   \widetilde{q} M_0' q)
 =  0, \qquad
s_2 x_2^2 + f \widetilde{f}  (1+  \widetilde{q}'  M_0 q')=0
\label{reducedfterm}
\eea
where we used the identities for $x_1$ and $x_2$ with $f,
\widetilde{f}$ 
we have discussed.

The theory has many nonsupersymmetric meta-stable ground states and 
when we rescale the meson fields as
$
M_0 = h \Lambda \Phi_0$ 
and 
$M'_0 = h' \Lambda' \Phi'_0$,
then the Kahler potential for $\Phi_0$ and $\Phi'_0$ 
is canonical and the magnetic
quarks are canonical near the origin of field space \cite{ISS}.
Then the magnetic superpotential (\ref{dualW}) can be rewritten as
\bea
W_{mag} & = & \left[ h \Phi_0  q' x_2   f \widetilde{f}  \widetilde{q}' 
 +  
\frac{h^2 \mu_{\phi}}{2} \tr \Phi_0^2- h \mu^2 \tr \Phi_0 \right]
+
\left[ h' \Phi_0'  q x_1 \widetilde{f} f \widetilde{q} 
 +  
\frac{{h'}^2 \mu_{\phi}'}{2} \tr {\Phi_0'}^2- 
h' {\mu'}^2 \tr \Phi_0'\right]
\nonu \\
& + & 
\left[ \frac{s_1}{3} x_1^3 + \frac{s_2}{3} x_2^3 + x_1  \widetilde{f} f
+
  \widetilde{f} x_2 f  + h \Phi_1 ( q'  
f \widetilde{f} \widetilde{q}' +\lambda_1) +  
h' \Phi_1' ( q   \widetilde{f} f \widetilde{q} + \lambda_2) \right]
\label{finalW}
\eea
where
$
\mu^2 = m \Lambda, {\mu'}^2 =m' \Lambda'$ and  
$\mu_{\phi} = \alpha \Lambda^2, \mu_{\phi}' = \alpha' {\Lambda'}^2$.

Now one splits 
the $(N_f-\widetilde{N}_c'-l) \times (N_f-\widetilde{N}_c'-l)$
block  at the lower right corner of $h\Phi_0$ and $q' f \widetilde{f} 
x_2 \widetilde{q}'$ of supersymmetric solutions
into blocks of 
size $n$ and $(N_f-\widetilde{N}_c'-l-n)$ and 
 one decomposes 
the $(N_f'-\widetilde{N}_c-l') \times (N_f'-\widetilde{N}_c-l')$
block  at the lower right corner of $h' \Phi'_0$ and $q x_1  
\widetilde{f} f  \widetilde{q}$ of supersymmetric solutions
into blocks of 
size $n'$ and $(N_f'-\widetilde{N}_c-l'-n')$
as follows \cite{GK0710}:
\bea
h\Phi_0 & = &  \left(
\begin{array}{cccc}
\frac{\lambda_1}{\alpha} X_{\widetilde{N}_c'}  & 0 & 0 & 0  \\
0 & 0_l & 0 & 0 \\
0 & 0 & h \Phi_n & 0 \\
0 & 0 & 0 & \frac{\mu^2}{\mu_{\phi}} {\bf 1}_{N_f-\widetilde{N}_c'-l-n}
\end{array}
\right), \nonu \\
h' \Phi_0'   & = &    \left(
\begin{array}{cccc}
\frac{\lambda_2}{\alpha'} Y_{\widetilde{N}_c}  & 0 & 0 & 0  \\
0 & 0_{l'}  & 0 & 0  \\
0 & 0 & h' \Phi_{n'}' & 0 \\
0 & 0 & 0 & \frac{{\mu'}^2}{\mu_{\phi}'} {\bf 1}_{N_f'-\widetilde{N}_c-l'-n'}
\end{array}
\right), \nonu \\
q'  f \widetilde{f} x_2 \widetilde{q}'  & = & \left(
\begin{array}{cccc}
\mu^2 {\bf 1}_{\widetilde{N}_c'} -
\lambda_1 X_{\widetilde{N}_c'}  & 0 & 0 & 0  \\
0 & \mu^2 {\bf 1}_l & 0 & 0  \\
0 & 0 & { \varphi}' g \widetilde{g} y_2  \widetilde{\varphi}'  &  0 \\
0 & 0 & 0 &  0_{N_f-\widetilde{N}_c'-l-n}
\end{array}
\right), \nonu \\
q x_1 \widetilde{f} f \widetilde{q}  & = & \left(
\begin{array}{cccc}
{\mu'}^2 {\bf 1}_{\widetilde{N}_c} -
\lambda_2 Y_{\widetilde{N}_c}  & 0 & 0 & 0  \\
0 & {\mu'}^2 {\bf 1}_{l'} & 0 & 0   \\
0 & 0 & \varphi y_1 \widetilde{g} g   \widetilde{\varphi}  &  0 \\
0 & 0 & 0 & 0_{N_f'-\widetilde{N}_c-l'-n'}
\end{array}
\right),
\label{vac}
\eea
with the expectation values of two adjoint fields 
$x_2$ and $x_1$ by 
$X_{\widetilde{N}_c'} = \mbox{diag}(a_1, a_2, \cdots, 
a_{\widetilde{N}_c'})$ and  
$Y_{\widetilde{N}_c} = \mbox{diag}(b_1, b_2, \cdots, 
b_{\widetilde{N}_c})$ which are traceless.
We used the first four equations of (\ref{reducedfterm}) in order to
obtain 
these expectation values. 
Here $\varphi'$ and $\widetilde{\varphi}'$ 
are $n \times (\widetilde{N}_c'-l)$
dimensional matrices  and 
$\varphi$ and $\widetilde{\varphi}$ are $n' \times (\widetilde{N}_c-l')$
dimensional matrices. 
In the brane configuration shown in Figure 3, 
$\varphi'$ and $\widetilde{\varphi}'$ correspond to 
fundamental strings connecting the $n$ flavor D4-branes and
$(\widetilde{N}_c'-l)$
color D4-branes and 
$\varphi$ and $\widetilde{\varphi}$ correspond to 
fundamental strings connecting the $n'$ flavor D4-branes and
$(\widetilde{N}_c-l')$
color D4-branes. 
The $\Phi_n$ and ${ \varphi}' g \widetilde{g}  
y_2 \widetilde{\varphi}'$
are $n \times n$ matrices while 
$\Phi'_{n'}$ and $\varphi y_1 \widetilde{g} g \widetilde{\varphi}$
are $n' \times n'$ matrices.

The supersymmetric ground state, in Figure 2,  corresponds to
$
h\Phi_n= \frac{\mu^2}{\mu_{\phi}} {\bf 1}_{n}, 
\varphi'  g y_2  =0=y_2 \widetilde{g} \widetilde{\varphi}'$ 
and 
$h'\Phi'_n= \frac{{\mu'}^2}{\mu_{\phi}'} {\bf 1}_{n'}, 
\varphi \widetilde{g} y_1 =0=y_1 g \widetilde{\varphi}$.
Let us make the two NS5'-branes and two NS5-branes be separated along 
$v$- and $w$-directions respectively. Let us put the $NS5_1'$-brane at
$v=0$
and the $NS5_1$-brane at $w=0$. Then the dual color numbers for each factor 
are distributed as $r_{i,j}$ and $r_{i,j}'$ 
connecting between the $NS5_i'$-brane and the
$NS5_j$-brane. Then we have  $\sum_{i,j=1}^2 r_{i,j} = \widetilde{N}_c'$ and 
$\sum_{i,j=1}^2 r'_{i,j} = \widetilde{N}_c$.

The $l$ of the
$N_f$-flavor D4-branes are reconnected with $l$-color
D4-branes
and the resulting $l$ D4-branes 
stretch from the $D6_{-\theta}$-branes to
the $NS5_1$-brane directly 
and the intersection point between the 
$l$ D4-branes and the $D6_{-\theta}$-branes is given by 
$(v, w)=(+v_{D6_{-\theta}}, 0)$.
This
corresponds to  exactly the $l$'s eigenvalues from 
zeros of 
$h\Phi_0$ in (\ref{vac}).
Now the remaining $(N_f-\widetilde{N}_c'-l)$-flavor D4-branes between 
the $D6_{-\theta}$-branes and 
the $NS5_1'$-brane correspond to the eigenvalues 
of $h\Phi_0$ in (\ref{vac}), i.e.,   
$\frac{\mu^2}{\mu_{\phi}} {\bf 1}_{N_f-\widetilde{N}_c'-l}$.
The intersection point between the 
$(N_f-\widetilde{N}_c'-l)$ D4-branes and the $NS5_1'$-branes is given 
by $(v, w)=(0, +v_{D6_{-\theta}} \cot \theta)$ from trigonometric 
geometry.
Finally, the remnant $\widetilde{N}_c'$-flavor D4-branes
between 
the $D6_{-\theta}$-branes and 
the $NS5'_2$-brane
correspond to the eigenvalues 
$\frac{\lambda_1}{\alpha} X_{\widetilde{N}_c'}
$ in (\ref{vac}) providing the positive $w$ coordinates that are 
$w=+(v_{D6_{-\theta}} - v_{NS5_2'}) \cot \theta$ for
these flavor D4-branes.  

Similarly, 
the $l'$ of the
$N_f'$-flavor D4-branes are reconnected with $l'$-color
D4-branes
and the resulting $l'$ D4-branes 
stretch from the $D6_{-\theta'}$-branes to
the $NS5_1$-brane directly 
and the intersection point between the 
$l'$ D4-branes and the $D6_{-\theta'}$-branes is given by 
$(v, w)=(+v_{D6_{-\theta'}}, 0)$.
This
corresponds to  exactly the $l'$'s eigenvalues from 
zeros of 
$h'\Phi'_0$ in (\ref{vac}).
Now the remaining $(N_f'-\widetilde{N}_c-l')$-flavor D4-branes between 
the $D6_{-\theta'}$-branes and 
the $NS5_1'$-brane correspond to the eigenvalues 
of $h'\Phi'_0$ in (\ref{vac}), i.e.,   
$\frac{{\mu'}^2}{\mu_{\phi}'} {\bf 1}_{N_f'-\widetilde{N}_c-l'}$.
The intersection point between the 
$(N_f'-\widetilde{N}_c-l')$ D4-branes and the $NS5_1'$-branes is given 
by $(v, w)=(0, +v_{D6_{-\theta'}} \cot \theta')$ from trigonometric 
geometry.
Finally, the remnant $\widetilde{N}_c$-flavor D4-branes
between 
the $D6_{-\theta'}$-branes and 
the $NS5'_2$-brane
correspond to the eigenvalues 
$
\frac{\lambda_2}{\alpha'} Y_{\widetilde{N}_c}
$ in (\ref{vac}) providing the positive $w$ coordinates 
 that are 
$w=+(v_{D6_{-\theta'}} - v_{NS5'_2}) \cot \theta$
for
these flavor D4-branes. 

Furthermore, one gets the following expectation values by using 
(\ref{fterm1}) 
\bea
h\Phi_1   =    \left(
\begin{array}{ccc}
-\frac{\lambda_1}{\alpha} X^2_{\widetilde{N}_c'} &0   & 0   \\
0 & 0_{l_1} & 0  \\
0 & 0   & \frac{\mu^2}{\mu_{\phi}} {\bf 1}_{N_f-\widetilde{N}_c'-l_1}
\end{array}
\right), \qquad
h' \Phi_1'    =     \left(
\begin{array}{ccc}
-\frac{\lambda_2}{\alpha'} Y^2_{\widetilde{N}_c}  & 0 & 0   \\
0 & 0_{l_1'}  & 0    \\
0 & 0   & \frac{{\mu'}^2}{\mu_{\phi}'} {\bf 1}_{N_f'-\widetilde{N}_c-l_1'}
\end{array}
\right).
\label{othermeson}
\eea
Finally the other two last equations of (\ref{reducedfterm}) and other
equations of (\ref{fterm1}) provide
the expectation values for $f$ and $\widetilde{f}$. 
Note that the superpotential (\ref{finalW}) has only the linear term
in $\Phi_1$ and $\Phi_1'$, contrary to the $M_0$ and $M_0'$.
In Figure 2 or 3, 
the $l_1$ of the
$N_f$-flavor D4-branes are reconnected with $l_1$-color
D4-branes
and the resulting $l_1$ D4-branes 
stretch from the $D6_{-\theta}$-branes to
the $NS5_1$-brane directly 
and the intersection point between the 
$l_1$ D4-branes and the $D6_{-\theta}$-branes is given by 
$(v, w)=(+v_{D6_{-\theta}}, 0)$.
This
corresponds to  exactly the $l_1$'s eigenvalues from 
zeros of 
$h\Phi_1$ in (\ref{othermeson}).
Now the remaining $(N_f-\widetilde{N}_c'-l_1)$-flavor D4-branes between 
the $D6_{-\theta}$-branes and 
the $NS5_1'$-brane correspond to the eigenvalues 
of $h\Phi_1$ in (\ref{othermeson}), i.e.,   
$\frac{\mu^2}{\mu_{\phi}} {\bf 1}_{N_f-\widetilde{N}_c'-l_1}$.
The intersection point between the 
$(N_f-\widetilde{N}_c'-l_1)$ D4-branes and the $NS5_1'$-branes is given 
by $(v, w)=(0, +v_{D6_{-\theta}} \cot \theta)$ from trigonometric 
geometry \cite{GK0710-1}.
Finally, the remnant $\widetilde{N}_c'$-flavor D4-branes
between 
the $D6_{-\theta}$-branes and 
the $NS5_2$-brane
correspond to the eigenvalues 
$-\frac{\lambda_1}{\alpha} X^2_{\widetilde{N}_c'}
$ in (\ref{othermeson}) providing the negative $w$ coordinates for
these flavor D4-branes, that is, the distance between the $NS5_1$-brane
and the $NS5_2$-brane.  

Similarly, 
the $l'_1$ of the
$N_f'$-flavor D4-branes are reconnected with $l'_1$-color
D4-branes
and the resulting $l'_1$ D4-branes 
stretch from the $D6_{-\theta'}$-branes to
the $NS5_1$-brane directly 
and the intersection point between the 
$l'_1$ D4-branes and the $D6_{-\theta'}$-branes is given by 
$(v, w)=(+v_{D6_{-\theta'}}, 0)$.
This
corresponds to  exactly the $l'_1$'s eigenvalues from 
zeros of 
$h'\Phi'_1$ in (\ref{othermeson}).
Now the remaining $(N_f'-\widetilde{N}_c-l'_1)$-flavor D4-branes between 
the $D6_{-\theta'}$-branes and 
the $NS5_1'$-brane correspond to the eigenvalues 
of $h'\Phi'_1$ in (\ref{othermeson}), i.e.,   
$\frac{{\mu'}^2}{\mu_{\phi}'} {\bf 1}_{N_f'-\widetilde{N}_c-l'_1}$.
The intersection point between the 
$(N_f'-\widetilde{N}_c-l_1')$ D4-branes and the $NS5_1'$-branes is given 
by $(v, w)=(0, +v_{D6_{-\theta'}} \cot \theta')$ from trigonometric 
geometry.
Finally, the remnant $\widetilde{N}_c$-flavor D4-branes
between 
the $D6_{-\theta'}$-branes and 
the $NS5_2$-brane
correspond to the eigenvalues 
$
-\frac{\lambda_2}{\alpha'} Y^2_{\widetilde{N}_c}
$ in (\ref{othermeson}) providing the negative $w$ coordinates for
these flavor D4-branes, that is, 
the distance between the $NS5_1$-brane
and the $NS5_2$-brane.  

Now the full one loop potential containing $\Phi_n, \Phi'_{n'},
\Phi_n^{\dagger}$ or  ${\Phi'}^{\dagger}_{n'}$ with $\mu_{\phi} << \mu
<< \Lambda_m$ and $\mu_{\phi}' << \mu' << \Lambda_m'$, 
by combining the superpotential 
and the vacuum expectation values for the fields, 
takes the form
\bea
V  & = &  
|h \Phi_n \varphi' y_2 g \widetilde{g}|^2   
+  |h  g \widetilde{g} y_2 \widetilde{\varphi}' \Phi_n |^2
  +  
| h \varphi'  g \widetilde{g} y_2 \widetilde{\varphi}'-h \mu^2 {\bf 1}_{n} + 
h^2 \mu_{\phi} \Phi_n|^2 
 +  b |h^2 \mu|^2 \tr \Phi_n^{\dagger} \Phi_n 
 \nonu \\
& + &  |h' \Phi_{n'}'  \varphi y_1 \widetilde{g} g  |^2   
+  |h'  \widetilde{g} g y_1 \widetilde{\varphi} \Phi_{n'}'|^2
 +   
|   h' \varphi y_1 \widetilde{g} g \widetilde{\varphi} -h' {\mu'}^2 {\bf 1}_{n'} + 
{h'}^2 \mu_{\phi}' \Phi_{n'}'|^2 
 +  b' |{h'}^2 \mu'|^2 
\tr {\Phi'}^{\dagger}_{n'} 
\Phi_{n'}'  \nonu \\
&+& 
|y_1  \widetilde{g} + \widetilde{g} y_2 + h \widetilde{g}
\widetilde{\varphi}' 
 \Phi_n \varphi' y_2 
+ h' \widetilde{\varphi}
 \Phi_n' \varphi y_1 \widetilde{g}|^2 
+ 
|g y_1 +  y_2 g 
+  h \widetilde{\varphi}' \Phi_n \varphi' y_2 g  
+ h' g \widetilde{\varphi} \Phi_n' 
\varphi y_1|^2
\nonu \\
& + & |s_2 y_2^2 + g \widetilde{g} + h g \widetilde{g}
\widetilde{\varphi}' \Phi_n \varphi'|^2 + 
|s_1 y_1^2 + \widetilde{g} g +  h' \widetilde{g} g
\widetilde{\varphi} \Phi'_n \varphi|^2 + \cdots
\nonu
\eea
where $b = \frac{(\ln 4-1)}{8\pi^2} \widetilde{N}_c$ and 
$b' = \frac{(\ln 4-1)}{8\pi^2} \widetilde{N}_c'$ \cite{ISS} and we did
not write down the irrelevant terms which do not depend on 
$\Phi_n$ and $\Phi'_{n'}$ and their conjugate fields explicitly.
Differentiating this potential with respect to 
$\Phi_n^{\dagger}$ and ${\Phi'}^{\dagger}_{n'}$ and putting 
$\varphi' g  =0 =   \widetilde{g} \widetilde{\varphi}'$ and 
$\varphi \widetilde{g}  =0 =   g
\widetilde{\varphi}$ 
\cite{GK0710}, 
one obtains
\bea
h \Phi_n 
& \simeq &  \frac{ \mu_{\phi}}{b }
{\bf 1}_n   \qquad  \mbox{or} \qquad
M_n \simeq \frac{\alpha \Lambda^3}{\widetilde{N}_c} {\bf 1}_{n}, 
\nonu \\
h' \Phi_{n'}' 
& \simeq &  \frac{ \mu_{\phi}'}{b' }
{\bf 1}_{n'}  \qquad  \mbox{or} \qquad
M'_{n'} \simeq \frac{\alpha' {\Lambda'}^3}{\widetilde{N}_c'} {\bf
  1}_{n'} 
\nonu
\eea
corresponding to the $w$ coordinates of $n$ curved flavor D4-branes between 
the $D6_{-\theta}$-branes and the $NS5_1'$-brane and  
the $w$ coordinates of $n'$ curved flavor D4-branes between 
the $D6_{-\theta'}$-branes and the other $NS5_1'$-brane respectively, in Figure 3.
Since $ \frac{ \mu_{\phi}}{b } << \frac{\mu^2}{\mu_{\phi}}$ and 
$ \frac{ \mu_{\phi}'}{b' } << \frac{{\mu'}^2}{\mu_{\phi}'}$, the $n$-
and $n'$-
curved D4-branes are nearer to $w=0$ at which the $NS5_1$-brane is located.

One can also consider the fluctuations on the other meson fields 
$\Phi_1$ and $\Phi_1'$ (\ref{othermeson}) in addition to the above
fluctuations.
However, the superpotential (\ref{finalW}) does not contain the quadratic
terms for these meson fields, the stable points arise as
zero values of $w$ corresponding to supersymmetric vacua.    
That is, as $n_1$ or $n_1'$ D4-branes approach to the $NS5_1$-brane,
they reconnect with the same number of color D4-branes respectively 
and those each combined
D4-branes will be connected between
$D6_{-\theta}$-branes($D6_{-\theta'}$-branes) and the $NS5_1$-brane directly. 

Therefore, the meta-stable states, for cubic superpotential of 
the adjoint fields with rotation angle $\theta(\theta')$
of $D6_{-\theta}(D6_{-\theta'})$-branes, 
are classified by the number of various D4-branes $
(r_{i,j}, r'_{i,j}, l, l', l_1, l_1', n, n')$, the position of
$D6_{-\theta}$-branes $v_{D6_{-\theta}}$, 
the position of $D6_{-\theta'}$-branes $v_{D6_{-\theta'}}$, 
the positions of $NS5_j'$-branes $ v_{NS5_{j}'}$  
and the positions of $NS5_j$-branes $w_{NS5_{j}}$.

%%%%%%%%%%%%%
%Figure 3
%%%%%%%%%%%%%
%%%%%%%%%%%%%%%%%%%%%%%%%%%%%%%%%%%%%%%%%%%%%%%%%%%%%%%%%%%%%%%%%%%%%%
%%%%%%%%%%%%%%%%%%%%%%%%%%%%%%%%%%%%%%%%%%%%%%%%%%%%%%%%%%%%%%%%%%%%%%
\begin{figure}[ht]
   \epsfxsize=4.5in 
\centerline{\epsffile{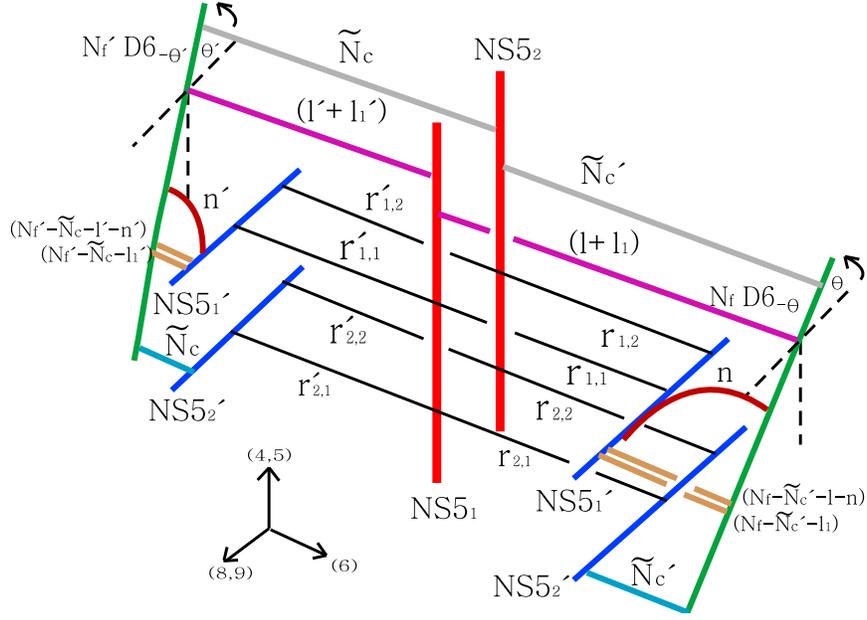}}
   \caption[FIG. \arabic{figure}.]{ 
The  nonsupersymmetric meta-stable
magnetic brane configuration corresponding to Figure 1 with 
a misalignment between D4-branes when the gravitational potential of
the NS5-brane is considered. 
The $(N_f-\widetilde{N}_c'-l)$ flavor D4-branes in Figure 2 connecting between
$D6_{-\theta}$-branes and $NS5_1'$-brane are 
further splitting into $(N_f-\widetilde{N}_c'-l-n)$- and
$n$-curved D4-branes while
the $(N_f'-\widetilde{N}_c-l')$ flavor D4-branes in Figure 2 connecting between
$D6_{-\theta'}$-branes and $NS5_1'$-brane are further 
splitting into $(N_f'-\widetilde{N}_c-l'-n')$- and
$n'$-curved D4-branes.   }
\end{figure}
%%%%%%%%%%%%%%%%%%%%%%%%%%%%%%%%%%%%%%%%%%%%%%%%%%%%%%%%%%%%%%%%%%%%%
%%%%%%%%%%%%%%%%%%%%%%%%%%%%%%%%%%%%%%%%%%%%%%%%%%%%%%%%%%%%%%%%%%%%%

%%%%%%%%%%%%%%%%%%%%%%%%%%%%%%%%%%%%%%%%%%%%%%%%%%%%%%%%%%%%%%%%%%%%%%%%%%%
\subsection{Higher order superpotential for adjoints}
%%%%%%%%%%%%%%%%%%%%%%%%%%%%%%%%%%%%%%%%%%%%%%%%%%%%%%%%%%%%%%%%%%%%%%%%%%%

When we consider higher order superpotential whose degree is greater than 
three(i.e., the number of NS5-branes or NS5'-branes $k > 2$), 
there exist more meson fields.
In general, it is not known how  
the deformations for the meson fields $Q X_1^j \widetilde{Q}$ or $Q'
X_2^j \widetilde{Q}'$ where $j \geq 2$ arise geometrically 
in the type IIA brane 
configuration.
Therefore, it is not clear how to add these deformations in an
electric theory or a magnetic theory.
Recall that the perturbations by $Q X_1 \widetilde{Q}$ and 
$Q' X_2 \widetilde{Q}'$ when $j=1$ in the superpotential were realized by the
relative rotations of $D6_{-\theta}(D6_{-\theta'})$-branes in $(w,v)$-plane.

%%%%%%%%%%%%%%%%%%%%%%%%%%%%%%%%%%%%%%%%%%%%%%%%%%%%%%%%%%%%%%%%
%%%%%%%%%%%%%%%%%%%%%%%%%%%%%%%%%%%%%%%%%%%%%%%%%%%%%%%%%%%%%%%%
\subsection{
$SU(N_c)$ with $N_f$-fund., symmetric and conjugate symmetric tensors, 
and adjoint.}
%%%%%%%%%%%%%%%%%%%%%%%%%%%%%%%%%%%%%%%%%%%%%%%%%%%%%%%%%%%%%%%%
%%%%%%%%%%%%%%%%%%%%%%%%%%%%%%%%%%%%%%%%%%%%%%%%%%%%%%%%%%%%%%%%

When we add an orientifold 6-plane to the above brane configuration, 
then we obtain this theory. Due to the orientifold 6-plane, the number
of gauge group, fundamentals, and adjoint fields are reduced by half.   
Moreover, the bifundamentals become symmetric and conjugate symmetric 
tensors. 
The type IIA brane configuration for this theory is
already known in \cite{BHKL}. The gauge theory analysis was found in
\cite{BS}
where the power of adjoint field in the superpotential 
appears in $k+1$ and they found that $k$ should be odd. 
For the simplest nontrivial case, one sets $k=3$.
The number of NS5-branes or NS5'-branes is equal to three also.
Then there exist three meson fields:
$
M_0  \equiv   Q  \widetilde{Q}, 
M_1  \equiv Q X \widetilde{Q}$, and 
$M_2 \equiv Q X X \widetilde{Q}$.
Since we do not know how  
the deformation for $M_2$ arises geometrically 
in the type IIA brane 
configuration,
it is not clear how to add the deformation corresponding to $M_2$ in an
electric theory or a magnetic theory.

%%%%%%%%%%%%%%%%%%%%%%%%%%%%%%%%%%%%%%%%%%%%%%%%%%%%%%%%%%%%%%%%
%%%%%%%%%%%%%%%%%%%%%%%%%%%%%%%%%%%%%%%%%%%%%%%%%%%%%%%%%%%%%%%%
\section{$SO(2N_c) \times Sp(N_c')$ with $2N_f$-vectors, $2N_f'$-fund., two
adjoints, and bifund.}
%%%%%%%%%%%%%%%%%%%%%%%%%%%%%%%%%%%%%%%%%%%%%%%%%%%%%%%%%%%%%%%%
%%%%%%%%%%%%%%%%%%%%%%%%%%%%%%%%%%%%%%%%%%%%%%%%%%%%%%%%%%%%%%%%

Let us add the O4-plane to the brane configuration of previous section. 

%%%%%%%%%%%%%%%%%%%%%%%%%%%%%%%%%%%%%%%%%%%%%%%%%%%%%%%%%%%%%%%%%%%%%%%%%%%%
\subsection{Electric theory}
%%%%%%%%%%%%%%%%%%%%%%%%%%%%%%%%%%%%%%%%%%%%%%%%%%%%%%%%%%%%%%%%%%%%%%%%%%%%

The type IIA supersymmetric electric
brane configuration  corresponding to 
${\cal N}=1$ $SO(2N_c) \times Sp(N_c')$ gauge theory  \cite{Ahn97} with  
$2N_f$-vector $Q$,
$2N_f'$-fundamental fields $Q'$,
bifundamentals $F$ and two adjoint fields $X_1, X_2$ 
can be described as two middle NS5-branes, two
left NS5'-branes, two right NS5'-branes, 
$2N_c$- and $2N_c'$-D4-branes, $2N_f$- and 
$2N_f'$-D6-branes and an O4-plane for the cubic superpotential of the adjoints.
The $X_1$ is in the representation $\bf{(N_c(2N_c-1),1)}$
while the $X_2$ is in the representation $\bf{(1, N_c'(2N_c'+1))}$, under
the gauge group. 
The $F$ is in the representation $\bf{(2N_c,2N_c')}$ 
under the gauge group. 
The quarks $Q$ is in the representation 
$\bf{(2N_c, 1)}$ and
the quarks $Q'$  is in the representation 
$\bf{(1, 2N_c')}$, under the gauge group.

The mass terms for each quarks 
can be added by displacing each D6-branes along 
$v$
direction leading to their coordinates 
$v = \pm
v_{D6_{-\theta}}( \pm v_{D6_{-\theta'}})$ respectively  
while the quartic terms for each quarks 
can be added also by rotating each D6-branes 
by an angle 
$-\theta(-\theta')$ in $(w,v)$-plane respectively. 
Then, the general 
superpotential by adding the above deformations is
given by
\bea
W_{elec} & = & 
\left[    \frac{s_1}{3} \tr X_1^3 + \frac{s_2}{3} \tr X_2^3 + 
\tr X_1  F^2 
  + \tr X_2  F^2   + \lambda_1 Q X_1 Q +
  \lambda_2 Q' X_2 Q'  \right]
\nonu \\
& + & \frac{\alpha}{2} \tr (Q Q)^2 - m \tr Q
Q 
 + \frac{\alpha'}{2} \tr (Q' Q')^2 - m' \tr Q' Q'.  
\label{elecsuper1}
\eea 
The parameters are the same as before and 
the rotation angles $\omega_L$ and
$\omega_R$  
of two left and right NS5'-branes 
with respect to the middle NS5-branes are given by 
$\omega_L=\omega_R=\frac{\pi}{2}$.
Although 
the relative displacement of two color D4-branes
can be added in the superpotential,
we focus on the particular zero limit of bifundamental mass.
There are the perturbations by $Q X_1 Q$ and 
$Q' X_2 Q'$ in the superpotential which will
arise as the mesons in the magnetic theory.
The mass matrix $m$ is symmetric and the mass matrix $m'$ is 
antisymmetric.

Then the ${\cal N}=1$ supersymmetric electric brane
configuration for the superpotential 
(\ref{elecsuper1}) 
in type IIA string theory is given as follows:

$\bullet$
Two middle NS5-branes in $(012345)$ directions 

$\bullet$ 
Two left NS5'-branes in  $(012389)$ directions 

$\bullet$ 
Two right NS5'-branes in  $(012389)$ directions 

$\bullet$
$2N_f$ $D6_{-\theta}$-branes in (01237)
directions and
two other directions in $(v,w)$-plane

$\bullet$
$2N_f'$ 
$D6_{-\theta'}$-branes in (01237)
directions and
two other directions in $(v,w)$-plane

$\bullet$
$2N_c$- and $2N_c'$-color D4-branes in $(01236)$ directions 

$\bullet$ $O4^{\pm}$-planes in (01236) directions

The corresponding brane configuration can be obtained from the
previous section by considering the correct mirrors based on the
O4-plane action.

%%%%%%%%%%%%%%%%%%%%%%%%%%%%%%%%%%%%%%%%%%%%%%%%%%%%%%%%%%%%%%%%%%%%%%%%%%%
\subsection{Magnetic theory }
%%%%%%%%%%%%%%%%%%%%%%%%%%%%%%%%%%%%%%%%%%%%%%%%%%%%%%%%%%%%%%%%%%%%%%%%

It is straightforward to compute the 
dual color numbers by considering the D4-brane charge of an
orientifold 4-plane and realizing that the number of D6-branes are 
doubled.

The left NS5'-branes 
start out with linking number $l_e=-\frac{4N_f'}{2} + 2N_c-2$
and after duality 
these left NS5'-branes end up with linking number 
$l_m = \frac{4N_f'}{2} -2\widetilde{N}_c'+4N_f-2$.
We consider only the particular brane motion where
$N_f$ $D6_{-\theta}$-branes meet 
the middle NS5-branes with no angles(and their mirrors). That is, 
the  $D6_{-\theta}$-branes become $D6_{-\frac{\pi}{2}}$-branes 
when they meet with the middle NS5-branes instantaneously and then
after that
they come back to the original $D6_{-\theta}$-branes.
Therefore, in this dual process, there is no creation of D4-branes.
That is the reason for the $4N_f$ factor in the $l_m$, not $8N_f$.
Then the dual color number $2\widetilde{N}_c'$
is given by $2\widetilde{N}_c' = 4N_f+4N_f'-2N_c$. 

The right NS5'-branes 
start out with linking number $l_e=\frac{4N_f}{2} - 2N_c'-2$
and after duality 
these right NS5'-branes end up with linking number 
$l_m = -\frac{4N_f}{2} +2\widetilde{N}_c-4N_f'-2$.
We consider only the particular brane motion where
the  $D6_{-\theta'}$-branes become $D6_{-\frac{\pi}{2}}$-branes 
when they meet with the middle NS5-branes instantaneously and after that 
they come back to the original  $D6_{-\theta'}$-branes(and their mirrors).
Therefore, in this dual process, there is no creation of D4-branes.
That is the reason for the $4N_f'$ factor in the $l_m$, not $8N_f'$.
Then it turns out that the dual color number $2\widetilde{N}_c$
is given by $2\widetilde{N}_c = 4N_f'+4N_f-2N_c'$. 
Finally, one has the following dual color numbers   
\bea
2\widetilde{N}_c = 4N_f'+4N_f-2N_c', \qquad 
2\widetilde{N}_c'= 4N_f+ 4N_f'-2N_c.
\nonu
\eea

The low energy theory on the two color D4-branes 
has $SO(2\widetilde{N}_c) \times Sp(\widetilde{N}_c')$ gauge group and  
$2N_f$-vector dual quarks $q'$, 
$2N_f'$-fundamental dual quarks $q$,
bifundamental $f$  
and various gauge singlets.
The $f$ is in the representation 
$\bf{(2\widetilde{N}_c, 2\widetilde{N}_c')}$ 
under the dual gauge group. 
The $2N_f'$ fields $q$  are in the representation 
$\bf{(2\widetilde{N}_c, 1)}$
and similarly, 
the $2N_f$ fields $q'$ are in the representation 
$\bf{(1, 2\widetilde{N}_c')}$, under the gauge group.
In particular, a magnetic meson field 
$
M_0 \equiv Q Q
$
is $2N_f \times 2N_f$ matrix and comes from 
4-4 strings of $2N_f$ flavor D4-branes created when $2N_f$
$D6_{-\theta}$-branes meet the one of the right NS5'-branes  while
a magnetic meson field 
$
M'_0 \equiv Q' Q'
$
is $2N_f' \times 2N_f'$ matrix and comes from 
4-4 strings of $2N_f'$ flavor D4-branes created
when $2N_f'$
$D6_{-\theta'}$-branes meet the one of the left NS5'-branes.
The adjoint fields $x_1, x_2$ correspond to the motion of the 
NS5'-branes and the NS5-branes in $(v,w)$-plane.

Then the most general magnetic superpotential, when we consider  
the case where $N_f(N_f')$ $D6_{-\theta}$-branes($D6_{-\theta'}$-branes) meet 
the middle NS5-branes {\it with angles}(and their mirrors),
is given by  
\bea
W_{dual} & = & 
\left[ \frac{s_1}{3} x_1^3 + \frac{s_2}{3} x_2^3 + x_1 f^2
+
x_2  f^2   + 
 \lambda_1
M_1 + \lambda_2 M_1'  
 \right]
\nonu \\ 
& + & 
\left( \frac{\alpha}{2} \tr M_0^2 - m M_0  \right) + \left(
\frac{\alpha'}{2} 
\tr {M_0'}^2 - m' M_0' \right) \nonu \\
& + &  \left[ M_0 q' x_2 f^2 q' +
  M_0' q x_1  f^2  q + M_1 q'  
f^2 q' +  M_1' q  f^2  
q \right] \nonu \\
& + & \left[ M_2 q' x_2 q' + M_3 q' q' 
 +   
M_2' q x_1 q + M_3' q q
+ P_1 q x_1 f q'  +P_2  q 
f q'  + \widetilde{P}_2  q'  f q \right]
\label{dualdualone}
\eea
where the mesons are given by 
\bea
M_0 & \equiv &  Q  Q,\quad  
M_0'  \equiv Q' Q', \quad 
M_1 \equiv Q X_1 Q, \quad
M_1'   \equiv  Q'  X_2 Q',
\nonu \\
M_2  & \equiv & Q F^2 Q, \quad 
M_3   \equiv Q F^2 X_1 Q,
  \quad
M_2'  \equiv Q'  F^2 Q', \quad
M_3'   \equiv Q'  F^2  X_2 Q', 
\nonu \\
\quad 
P_1   & \equiv &   Q F Q', 
\quad
P_2 \equiv Q X_1 F Q', \quad
\widetilde{P}_2  \equiv  Q' X_2 F Q.
\nonu
\eea
The first two lines of (\ref{dualdualone}) are dual expressions for the electric
superpotential (\ref{elecsuper1}) and the corresponding meson fields
$M_0, M_0', M_1$ and $M_1'$ are replaced and the third and fourth
lines of (\ref{dualdualone}) are the analogs of the cubic term
superpotential between
the meson and dual quarks in Seiberg duality. 
Compared with the theory \cite{ILS,Ahn08-4} without two adjoints fields, 
there exist the extra meson fields coming from the adjoint fields 
$X_1$ and $X_2$:$M_1, M_1',M_3,M_3',P_2$ and $\widetilde{P}_2$.

Let us find out the relevant magnetic superpotential for the
meta-stable brane configuration.
 
When the $N_f$ $D6_{-\theta}$-branes 
meet the middle NS5-branes(and their mirrors),
no creation of D4-branes  
implies that there is no $M_2$- or $M_3$-term 
in the above superpotential
(\ref{dualdualone}).
The mesons  
$M_2$ and $M_3$ originate from  $SO(2N_c)$ chiral mesons
$Q Q$  when one
dualizes the $SO(2N_c)$ gauge group first by moving the middle NS5-branes
to the left of the left NS5'-branes \cite{Ahn07-2}. 
That is, the fluctuations of
strings stretching between the $2N_f$ ``flavor'' D4-branes correspond
to these meson fields(and their mirrors). 
After the additional dual procedures, the cubic terms in the
superpotential arise as 
$M_2$-dependent  and $M_3$-dependent terms 
where $M_2$ has 
extra $F^2$ fields and 
$M_3$ has 
extra $F^2 X_1 $ fields, besides $Q Q$,  
due to the further dualization. The $M_2$-term in the 
superpotential has an extra 
$x_2$ factor  besides $q' q'$.

Similarly, 
when 
the $N_f'$ $D6_{-\theta'}$-branes 
meet the middle NS5-branes with no angles(and their mirrors), 
there is no $M_2'$- or $M_3'$-term in the above superpotential
(\ref{dualdualone}).
These meson fields 
$M_2'$ and $M_3'$ originate from  $Sp(N_c')$ chiral mesons
$Q' Q'$  when one
dualizes the $Sp(N_c')$ gauge group first by moving the middle NS5-branes
to the right of the right NS5'-branes. The
strings stretching between the $2N_f'$ ``flavor'' D4-branes provide
these mesons(and their mirrors). 
After the additional dual procedures, 
the cubic terms in the superpotential arise as 
$M_2'$-term and $M_3'$-term  where $M_2'$ has 
extra $F^2 $ fields
and $M_3'$ has 
extra $F^2 X_2$ fields, besides $Q' Q'$, 
due to the further dualization.
The $M_2'$-term in the superpotential has an extra 
$x_1$ factor  besides $q q$.

Furthermore, 
when the $N_f$ $D6_{-\theta}$-branes, the 
$N_f'$ $D6_{-\theta'}$-branes and the middle NS5-branes
meet each other with no angles(and their mirrors),
no $P_1$- and $P_2$- or $\widetilde{P}_2$-dependent terms occur in the
superpotential (\ref{dualdualone}).
These mesons 
originate from  $Sp(N_c')$ chiral mesons
$F Q'$ when one
dualizes the $Sp(N_c')$  first by moving the middle NS5-branes
to the right of the right NS5'-branes. 
The
strings stretching between the $2N_f'$ flavor D4-branes and $2N_c$
color D4-branes give rise to these 
$2N_f'$ $SO(2N_c)$ vectors(and their mirrors). 
After the additional dual procedures, these cubic terms arise as 
these meson terms  where there exist  extra 
$q x_1, q$ 
and $q$ in the interactions of $P_1, P_2$ and 
$\widetilde{P}_2$ in the superpotential and the mesons have 
extra $Q, Q X_1, Q X_2$ fields respectively, 
due to the further dualization. 

Then the reduced magnetic superpotential in our case  
by taking the first three lines of (\ref{dualdualone}) 
is given by 
\bea
W_{dual} & = & 
\left[ \frac{s_1}{3} x_1^3 + \frac{s_2}{3} x_2^3 + x_1 f^2
+
  f x_2 f  + M_1 ( q'  
f^2 q' +\lambda_1) +  
M_1' ( q  f^2 q + \lambda_2) \right]
\nonu \\ 
& + & 
\left[  M_0 q' x_2 f^2 q' +
\frac{\alpha}{2} \tr M_0^2 - m M_0  \right] + \left[
M_0' q x_1  f^2 q + \frac{\alpha'}{2} 
\tr {M_0'}^2 - m' M_0' \right].
\label{Ddual}
\eea

Let us describe the meta-stable brane configuration with this magnetic 
superpotential.

For the supersymmetric vacua, one can compute the F-term equations for
this superpotential (\ref{Ddual}) 
and the F-terms for $M_0, q', M_0', q, f, M_1, M_1', x_1$ 
and $x_2$ are given by
\bea
&& q' x_2 f^2 q'-m + \alpha M_0 =0, \qquad 
 x_2 f^2 q' M_0 +   
f^2 q' M_1 + 
(M_0 q' x_2 + M_1 q') f^2  =0,
\nonu \\
&& q x_1  f^2 q -m' + \alpha' M_0'=0, \qquad 
x_1 f^2 q M_0'  +   
f^2 q M_1' +
(M_0' q x_1 +  M_1' q ) f^2 =0,
\nonu \\
&& (x_1  f + f x_2) + f q' 
( M_1 q'+
 M_0 q' x_2) + q  (M_1' q  +
 M_0' q x_1) f  
\nonu \\
&& + (f x_1 +  x_2 f) +  q' (M_1 q'  
 +    M_0 q' x_2) f + f q (M_1' q   +  M_0' 
q x_1)  =0,
\nonu \\
&&  q'  
f^2 q' +\lambda_1 =0, \qquad  q  
f^2 q +
\lambda_2 =0,
\nonu \\
&& s_1 x_1^2 +  f^2 +   f^2  q M_0' q
=0, \qquad
s_2 x_2^2 + f^2  + f^2 q'  M_0 q'=0.
\label{ftermterm}
\eea

The fifth equation of (\ref{ftermterm}) 
is satified if  
the following equations hold 
\bea
f x_1= -x_2 f, \qquad x_1 f = - f x_2,
\qquad M_1 q' = -M_0 q' x_2, \qquad M_1' q = - M_0' q x_1.
\label{fterm1term}
\eea
By multiplying $f$ to the second equation of 
(\ref{fterm1term}), one obtains
$
f x_1 f = - f^2 x_2$.
Using the first equation of (\ref{fterm1term}) one gets
$
(-x_2 f) f = -  f^2 x_2$ and this leads to 
$
x_2 f^2  =   f^2 x_2$.
Then one simplifies the second equation of (\ref{ftermterm}) 
with (\ref{fterm1term}) as
\bea 
f^2 (x_2 q' M_0 +   
 q' M_1) =0 \rightarrow  
 q' M_1 = -x_2 q' M_0
\nonu 
\eea 
by moving $x_2$ to the right.
Similarly, 
by multiplying $f$ to the first equation of 
(\ref{fterm1term}), one obtains
$
f^2 x_1  = - f x_2 f$.
Using the second equation of (\ref{fterm1term}) one gets
$
f^2 x_1  =   (x_1 f) f$ and this leads to 
$
x_1 f^2 =   f^2 x_1$.
Then one simplifies the fourth equation of (\ref{ftermterm}) with 
(\ref{fterm1term}) as
\bea
f^2 (x_1 q M_0'  +   
q M_1') =0 \rightarrow      
q M_1' = -x_1 q M_0'
\nonu
\eea
by moving $x_1$ to the right.

Then the remaining F-term equations can be summarized as
\bea
&& q'  f^2 x_2 q'-m + \alpha M_0  =  0, \qquad
q   f^2 x_1  q -m' + \alpha' M_0'=0, 
\qquad
q'  
f^2 q' +\lambda_1  =  0, \nonu \\
&&  q  
f^2 q +
\lambda_2 =0,
\qquad
s_1 x_1^2 +  f^2 (1 + q M_0' q)
 =  0, \qquad
s_2 x_2^2 + f^2  (1+ q'  M_0 q')=0
\label{reducedfterm1}
\eea
where we used the identities for $x_1$ and $x_2$ with $f$ 
we discussed.

The theory has many nonsupersymmetric meta-stable ground states and 
when we rescale the meson fields as
$
M_0 = h \Lambda \Phi_0$ 
and 
$M'_0 = h' \Lambda' \Phi'_0$,
then the Kahler potential for $\Phi_0$ and $\Phi'_0$ 
is canonical and the magnetic
quarks are canonical near the origin of field space.
Then the magnetic superpotential (\ref{Ddual}) can be rewritten as
\bea
W_{mag} & = & \left[ h \Phi_0  q' x_2   f^2  q' 
 +  
\frac{h^2 \mu_{\phi}}{2} \tr \Phi_0^2- h \mu^2 \tr \Phi_0 \right]
 +  
\left[ h' \Phi_0'  q x_1 f^2 q 
 +  
\frac{{h'}^2 \mu_{\phi}'}{2} \tr {\Phi_0'}^2- 
h' {\mu'}^2 \tr \Phi_0'\right]
\nonu \\
& + & 
\left[ \frac{s_1}{3} x_1^3 + \frac{s_2}{3} x_2^3 + x_1 f^2
+
f x_2 f  + h \Phi_1 ( q'  
f^2 q' +\lambda_1) +  
h' \Phi_1' ( q  f^2 q + \lambda_2) \right]
\label{finalW2}
\eea
where
$
\mu^2 = m \Lambda, {\mu'}^2 =m' \Lambda'$ and  
$\mu_{\phi} = \alpha \Lambda^2, \mu_{\phi}' = \alpha' {\Lambda'}^2$.

Now one splits 
the $2(N_f-\widetilde{N}_c'-l) \times 2(N_f-\widetilde{N}_c'-l)$
block  at the lower right corner of $h\Phi_0$ and $q' f^2 
x_2 q'$ 
into blocks of 
size $2n$ and $2(N_f-\widetilde{N}_c'-l-n)$ and 
 one decomposes 
the $2(N_f'-\widetilde{N}_c-l') \times 2(N_f'-\widetilde{N}_c-l')$
block  at the lower right corner of $h' \Phi'_0$ and $q x_1  
f^2 q$ 
into blocks of 
size $2n'$ and $2(N_f'-\widetilde{N}_c-l'-n')$
as follows:
\bea
h\Phi_0 & = &  \left(
\begin{array}{cccc}
\frac{\lambda_1}{\alpha} X_{2\widetilde{N}_c'}  & 0 & 0 & 0  \\
0 & 0_{2l} & 0 & 0 \\
0 & 0 & h \Phi_{2n} & 0 \\
0 & 0 & 0 & \frac{\mu^2}{\mu_{\phi}} {\bf
  1}_{N_f-\widetilde{N}_c'-l-n} \otimes \sigma_3
\end{array}
\right), \nonu \\
h' \Phi_0'  & = &   \left(
\begin{array}{cccc}
\frac{\lambda_2}{\alpha'} Y_{2\widetilde{N}_c}  & 0 & 0 & 0  \\
0 & 0_{2l'}  & 0 & 0  \\
0 & 0 & h' \Phi_{2n'}' & 0 \\
0 & 0 & 0 & \frac{{\mu'}^2}{\mu_{\phi}'} 
{\bf 1}_{N_f'-\widetilde{N}_c-l'-n'} \otimes i \sigma_2
\end{array}
\right), \nonu \\
q'  f^2 x_2 q'  & = & \left(
\begin{array}{cccc}
\mu^2 {\bf 1}_{2\widetilde{N}_c'} -
\lambda_1 X_{2\widetilde{N}_c'}  & 0 & 0 & 0  \\
0 & \mu^2 {\bf 1}_{2l} & 0 & 0  \\
0 & 0 & { \varphi}' g^2 y_2 \varphi'  &  0 \\
0 & 0 & 0 &  0_{2(N_f-\widetilde{N}_c'-l-n)}
\end{array}
\right), \nonu \\
q x_1 f^2 q  & = & \left(
\begin{array}{cccc}
{\mu'}^2 {\bf 1}_{2\widetilde{N}_c} -
\lambda_2 Y_{2\widetilde{N}_c}  & 0 & 0 & 0  \\
0 & {\mu'}^2 {\bf 1}_{2l'} & 0 & 0   \\
0 & 0 & \varphi y_1 g^2 \varphi  &  0 \\
0 & 0 & 0 & 0_{2(N_f'-\widetilde{N}_c-l'-n')}
\end{array}
\right),
\label{vac1}
\eea
with $X_{2\widetilde{N}_c'} = \mbox{diag}(a_{1}, a_{2}, \cdots, 
a_{\widetilde{N}_c'}) \otimes \sigma_3$ and  
$Y_{2\widetilde{N}_c} = \mbox{diag}(b_{1}, b_{2}, \cdots, 
b_{\widetilde{N}_c}) \otimes i \sigma_2$.
We used the first four equations of (\ref{reducedfterm1}) in order to
obtain 
these expectation values.
Here $\varphi'$  
is $2n \times 2(\widetilde{N}_c'-l)$
dimensional matrices  and 
$\varphi$ is  $2n' \times 2(\widetilde{N}_c-l')$
dimensional matrices. 
The $\varphi'$ corresponds to 
fundamental strings connecting the $2n$ flavor D4-branes and
$2(\widetilde{N}_c'-l)$
color D4-branes and 
$\varphi$ corresponds to 
fundamental strings connecting the $2n'$ flavor D4-branes and
$2(\widetilde{N}_c-l')$
color D4-branes.
The $\Phi_{2n}$ and $\varphi' g^2  
y_2 \varphi'$
are $2n \times 2n$ matrices while 
$\Phi'_{2n'}$ and $\varphi y_1 g^2 \varphi$
are $2n' \times 2n'$ matrices.

The supersymmetric ground state corresponds to
$
h\Phi_{2n}= \frac{\mu^2}{\mu_{\phi}} {\bf 1}_{n} \otimes \sigma_3, 
\varphi'  g y_2  =0=y_2 g \varphi'$ 
and 
$h'\Phi'_{2n'}= \frac{{\mu'}^2}{\mu_{\phi}'} {\bf 1}_{n'} \otimes i \sigma_2, 
\varphi g y_1 =0=y_1 g \varphi$.
The $l$ of the upper
$N_f$-flavor D4-branes are reconnected with $l$-color
D4-branes
and the resulting $l$ D4-branes 
stretch from the upper $D6_{-\theta}$-branes to
the $NS5_1$-brane directly 
and the intersection point between the 
$l$ D4-branes and the upper $D6_{-\theta}$-branes is given by 
$(v, w)=(+v_{D6_{-\theta}}, 0)$.
The mirrors are located at 
$(v, w)=(-v_{D6_{-\theta}}, 0)$.
This
corresponds to  exactly the $2l$'s eigenvalues from 
zeros of 
$h\Phi_0$ in (\ref{vac1}).
Now the remaining upper 
$(N_f-\widetilde{N}_c'-l)$-flavor D4-branes between 
the upper $D6_{-\theta}$-branes and 
the $NS5_1'$-brane correspond to the positive eigenvalues 
of $h\Phi_0$ in (\ref{vac1}), i.e.,   
$\frac{\mu^2}{\mu_{\phi}} {\bf 1}_{N_f-\widetilde{N}_c'-l}$.
The intersection point between the upper  
$(N_f-\widetilde{N}_c'-l)$ D4-branes and the $NS5_1'$-branes is given 
by $(v, w)=(0, +v_{D6_{-\theta}} \cot \theta)$ from trigonometric 
geometry.
The mirrors are located at 
$(v, w)=(0, -v_{D6_{-\theta}} \cot \theta)$ corresponding to
 the negative eigenvalues 
of $h\Phi_0$ in (\ref{vac1}), i.e.,   
$-\frac{\mu^2}{\mu_{\phi}} {\bf 1}_{N_f-\widetilde{N}_c'-l}$.
Finally, the remnant $2\widetilde{N}_c'$-flavor D4-branes
between 
the $D6_{-\theta}$-branes and 
the $NS5_2'$-brane
correspond to the eigenvalues 
$\frac{\lambda_1}{\alpha} X_{2\widetilde{N}_c'}
$ in (\ref{vac1}) providing the $w$ coordinates 
 that are 
$w=+(v_{D6_{-\theta}} - v_{NS5'_2}) \cot \theta$
for
these flavor D4-branes.  

Similarly, 
the $l'$ of the upper 
$N_f'$-flavor D4-branes are reconnected with $l'$-color
D4-branes
and the resulting $l'$ D4-branes 
stretch from the upper $D6_{-\theta'}$-branes to
the $NS5_1$-brane directly 
and the intersection point between the 
$l'$ D4-branes and the upper $D6_{-\theta'}$-branes is given by 
$(v, w)=(+v_{D6_{-\theta'}}, 0)$.
The mirrors are located at 
$(v, w)=(-v_{D6_{-\theta'}}, 0)$.
This
corresponds to  exactly the $2l'$'s eigenvalues from 
zeros of 
$h'\Phi'_0$ in (\ref{vac1}).
Now the remaining upper 
$(N_f'-\widetilde{N}_c-l')$-flavor D4-branes between 
the upper $D6_{-\theta'}$-branes and 
the $NS5_1'$-brane correspond to the positive eigenvalues 
of $h'\Phi'_0$ in (\ref{vac1}), i.e.,   
$\frac{{\mu'}^2}{\mu_{\phi}'} {\bf 1}_{N_f'-\widetilde{N}_c-l'}$.
The intersection point between the 
$(N_f'-\widetilde{N}_c-l')$ D4-branes and the $NS5_1'$-branes is given 
by $(v, w)=(0, +v_{D6_{-\theta'}} \cot \theta')$ from trigonometric 
geometry. The mirrors are located at 
$(v, w)=(0, -v_{D6_{-\theta'}} \cot \theta')$
corresponding to  the negative eigenvalues 
of $h'\Phi'_0$ in (\ref{vac1}), i.e.,   
$-\frac{{\mu'}^2}{\mu_{\phi}'} {\bf 1}_{N_f'-\widetilde{N}_c-l'}$.
Finally, the remnant $2\widetilde{N}_c$-flavor D4-branes
between 
the $D6_{-\theta'}$-branes and 
the $NS5_2'$-brane
correspond to the eigenvalues 
$
\frac{\lambda_2}{\alpha'} Y_{2\widetilde{N}_c}
$ in (\ref{vac1}) providing the $w$ coordinates 
 that are 
$w=+(v_{D6_{-\theta'}} - v_{NS5'_2}) \cot \theta'$
for
these flavor D4-branes. 

Furthermore, one gets the following expectation values by using 
(\ref{fterm1term}) 
\bea
h\Phi_1  & = &   \left(
\begin{array}{ccc}
-\frac{\lambda_1}{\alpha} X^2_{2\widetilde{N}_c'}   & 0 & 0  \\
0 & 0_{2l_1} & 0  \\
0 & 0  & \frac{\mu^2}{\mu_{\phi}} {\bf 1}_{N_f-\widetilde{N}_c'-l_1}
\otimes \sigma_3
\end{array}
\right), \nonu \\
h' \Phi_1'  & = &   \left(
\begin{array}{ccc}
-\frac{\lambda_2}{\alpha'} Y^2_{2\widetilde{N}_c}  & 0 & 0  \\
0 & 0_{2l_1'}  & 0   \\
0 & 0  & \frac{{\mu'}^2}{\mu_{\phi}'} {\bf
  1}_{N_f'-\widetilde{N}_c-l_1'} \otimes i \sigma_2
\end{array}
\right).
\label{othermeson1}
\eea
Finally the other two last equations of (\ref{reducedfterm1}) and other
equation of (\ref{fterm1term}) provide
the expectation value for $f$. 
Note that the superpotential (\ref{finalW2}) has only the linear term
in $\Phi_1$ and $\Phi_1'$, contrary to the $M_0$ and $M_0'$.
The $l_1$ of the upper 
$N_f$-flavor D4-branes are reconnected with $l_1$-color
D4-branes
and the resulting $l_1$ D4-branes 
stretch from the upper $D6_{-\theta}$-branes to
the $NS5_1$-brane directly 
and the intersection point between the 
$l_1$ D4-branes and the upper $D6_{-\theta}$-branes is given by 
$(v, w)=(+v_{D6_{-\theta}}, 0)$.
The mirrors are located at 
$(v, w)=(-v_{D6_{-\theta}}, 0)$.
This
corresponds to  exactly the $2l_1$'s eigenvalues from 
zeros of 
$h\Phi_1$ in (\ref{othermeson1}).
Now the remaining upper $(N_f-\widetilde{N}_c'-l_1)$-flavor D4-branes between 
the upper $D6_{-\theta}$-branes and 
the $NS5_1'$-brane correspond to the eigenvalues 
of $h\Phi_1$ in (\ref{othermeson1}), i.e.,   
$\frac{\mu^2}{\mu_{\phi}} {\bf 1}_{N_f-\widetilde{N}_c'-l_1}$.
The intersection point between the upper  
$(N_f-\widetilde{N}_c'-l_1)$ D4-branes and the $NS5_1'$-branes is given 
by $(v, w)=(0, +v_{D6_{-\theta}} \cot \theta)$ from trigonometric 
geometry.
The mirrors are located at 
 $(v, w)=(0, -v_{D6_{-\theta}} \cot \theta)$.
Finally, the remnant $2\widetilde{N}_c'$-flavor D4-branes
between 
the $D6_{-\theta}$-branes and 
the $NS5_2$-brane
correspond to the eigenvalues 
$-\frac{\lambda_1}{\alpha} X^2_{2\widetilde{N}_c'}
$ in (\ref{othermeson1}) providing the negative $w$ coordinates for
these flavor D4-branes.  

Similarly, 
the $l'_1$ of the upper
$N_f'$-flavor D4-branes are reconnected with $l'_1$-color
D4-branes
and the resulting $l'_1$ D4-branes 
stretch from the upper $D6_{-\theta'}$-branes to
the $NS5_1$-brane directly 
and the intersection point between the 
$l'_1$ D4-branes and the $D6_{-\theta'}$-branes is given by 
$(v, w)=(+v_{D6_{-\theta'}}, 0)$.
The mirrors are located at 
$(v, w)=(-v_{D6_{-\theta'}}, 0)$.
This
corresponds to  exactly the $2l'_1$'s eigenvalues from 
zeros of 
$h'\Phi'_1$ in (\ref{othermeson1}).
Now the remaining upper 
$(N_f'-\widetilde{N}_c-l'_1)$-flavor D4-branes between 
the upper $D6_{-\theta'}$-branes and 
the $NS5_1'$-brane correspond to the eigenvalues 
of $h'\Phi'_1$ in (\ref{othermeson1}), i.e.,   
$\frac{{\mu'}^2}{\mu_{\phi}'} {\bf 1}_{N_f'-\widetilde{N}_c-l'_1}$.
The intersection point between the 
$(N_f'-\widetilde{N}_c-l_1')$ D4-branes and the $NS5_1'$-branes is given 
by $(v, w)=(0, +v_{D6_{-\theta'}} \cot \theta')$ from trigonometric 
geometry and 
the mirrors are located at 
 $(v, w)=(0, -v_{D6_{-\theta'}} \cot \theta')$.
Finally, the remnant $2\widetilde{N}_c$-flavor D4-branes
between 
the $D6_{-\theta'}$-branes and 
the $NS5_2$-brane
correspond to the eigenvalues 
$
-\frac{\lambda_2}{\alpha'} Y^2_{2\widetilde{N}_c}
$ in (\ref{othermeson1}) providing the negative $w$ coordinates for
these flavor D4-branes.  

Now the full one loop potential containing $\Phi_{2n}, \Phi'_{2n'}$, 
by combining the superpotential 
and the vacuum expectation values for the fields, 
takes the form
\bea
V  & = &  
|h \Phi_{2n} \varphi' y_2 g^2    
+  h  g^2 y_2 \varphi' \Phi_{2n}|^2
  +  
| h \varphi'  g^2 y_2 \varphi'-h \mu^2 {\bf 1}_{2n} + 
h^2 \mu_{\phi} \Phi_{2n}|^2 + b |h^2 \mu|^2 \tr \Phi_{2n} \Phi_{2n}
\nonu \\ 
& + & |h' \Phi_{2n'}'  \varphi y_1 g^2    
+  h'  g^2 y_1 \varphi \Phi_{2n'}' |^2
  +  
|   h' \varphi y_1 g^2 \varphi -h' {\mu'}^2 {\bf 1}_{2n'} + 
{h'}^2 \mu_{\phi}' \Phi_{2n'}'|^2 + b' |{h'}^2 \mu'|^2 
\tr {\Phi'}_{2n'} 
\Phi_{2n'}'
\nonu \\
&+& 
|y_1  g + g y_2 + h g \varphi' 
 \Phi_{2n} \varphi' y_2 + h' \varphi
 \Phi_{2n'} \varphi y_1 g +
g y_1 +  y_2 g +  h \varphi' \Phi_{2n} \varphi' y_2 g 
+ h' g \varphi \Phi_{2n'} 
\varphi y_1|^2
\nonu \\
& + & |s_2 y_2^2 + g^2 + h g^2 \varphi' \Phi_{2n} \varphi'|^2 + 
|s_1 y_1^2 + g^2 +  h' g^2
\varphi \Phi'_{2n} \varphi|^2 + \cdots
\nonu
\eea
where $b = \frac{(\ln 4-1)}{8\pi^2} \widetilde{N}_c$ and 
$b' = \frac{(\ln 4-1)}{8\pi^2} \widetilde{N}_c'$ \cite{ISS,Ahn06-1} 
and we did
not write down the irrelevant terms which do not depend on 
$\Phi_{2n}$ and $\Phi'_{2n'}$ explicitly.
Differentiating this potential with respect to 
$\Phi_{2n}$ and ${\Phi'}_{2n'}$ and putting 
$\varphi' g =0 =   g \varphi'$ and 
$\varphi g  =0 =   g \varphi$, 
one obtains
\bea
h \Phi_{2n} 
& \simeq & \frac{ \mu_{\phi}}{b }
{\bf 1}_n \otimes  \sigma_3 \qquad \mbox{or} \qquad
M_{2n} \simeq \frac{\alpha \Lambda^3}{\widetilde{N}_c} {\bf 1}_{n}
\otimes  \sigma_3, 
\nonu \\
h' \Phi'_{2n'} 
& \simeq & \frac{ \mu_{\phi}'}{b' }
{\bf 1}_{n'} \otimes i \sigma_2 \qquad \mbox{or} \qquad
M'_{2n'} \simeq \frac{\alpha' {\Lambda'}^3}{\widetilde{N}_c'} {\bf 1}_{n'}
\otimes i \sigma_2 
\nonu
\eea
corresponding to the $w$ coordinates of $2n$ curved flavor D4-branes between 
the $D6_{-\theta}$-branes and the $NS5_1'$-brane and  
the $w$ coordinates of $2n'$ curved flavor D4-branes between 
the $D6_{-\theta'}$-branes and the $NS5_1'$-brane respectively.

One can also consider the fluctuations on the other meson fields 
$\Phi_1$ and $\Phi_1'$ (\ref{othermeson1}) in addition to the above
fluctuations.
Since the superpotential (\ref{finalW2}) does not contain the quadratic
terms for these meson fields, the stable points arise as
zero values of $w$ corresponding to supersymmetric vacua.    
That is, as $2n_1$ or $2n_1'$ D4-branes approach to the $NS5_1$-brane,
they reconnect with the same number of color D4-branes respectively 
and those each combined
D4-branes will be connected between
$D6_{-\theta}$-branes($D6_{-\theta'}$-branes) and the $NS5_1$-brane directly.

%%%%%%%%%%%%%%%%%%%%%%%%%%%%%%%%%%%%%%%%%%%%%%%%%%%%%%%%%%%%%%%%%%%%%%%%%%%
\subsection{Higher order superpotential for adjoints}
%%%%%%%%%%%%%%%%%%%%%%%%%%%%%%%%%%%%%%%%%%%%%%%%%%%%%%%%%%%%%%%%%%%%%%%%%%%

When we consider higher order superpotential whose degree is greater than 
three(i.e., the number of NS5-branes or NS5'-branes $k > 2$), 
there exist more meson fields.
In general, it is not known how  
the deformations for the meson fields $Q X_1^j Q$ or $Q'
X_2^j Q'$ where $j \geq 2$ arise geometrically 
in the type IIA brane 
configuration, as in previous section.
Therefore, it is not clear how to add these deformations in an
electric theory or a magnetic theory.
Recall that the perturbations by $Q X_1 Q$ and 
$Q' X_2 Q'$ when $j=1$ in the superpotential were realized by the
relative rotations of 
$D6_{-\theta}(D6_{-\theta'})$-branes in $(w,v)$-plane.

%%%%%%%%%%%%%%%%%%%%%%%%%%%%%%%%%%%%%%%%%%%%%%%%%%%%%%%%%%%%%%%%
%%%%%%%%%%%%%%%%%%%%%%%%%%%%%%%%%%%%%%%%%%%%%%%%%%%%%%%%%%%%%%%%
\section{$SU(N_c) \times SU(N_c') \times SU(N_c'')$ 
with $N_f$-, $N_f'$-, and $N_f''$-fund., and bifund.}
%%%%%%%%%%%%%%%%%%%%%%%%%%%%%%%%%%%%%%%%%%%%%%%%%%%%%%%%%%%%%%%%
%%%%%%%%%%%%%%%%%%%%%%%%%%%%%%%%%%%%%%%%%%%%%%%%%%%%%%%%%%%%%%%%

%%%%%%%%%%%%%%%%%%%%%%%%%%%%%%%%%%%%%%%%%%%%%%%%%%%%%%%%%%%%%%%%%%%%%%%%%%%%
\subsection{Electric theory}
%%%%%%%%%%%%%%%%%%%%%%%%%%%%%%%%%%%%%%%%%%%%%%%%%%%%%%%%%%%%%%%%%%%%%%%%%%%%

The type IIA supersymmetric electric
brane configuration \cite{BH,AT97} corresponding to 
${\cal N}=1$ $SU(N_c) \times SU(N_c') \times SU(N_c'')$ 
gauge theory  with  
$N_f$-fundamental flavors $Q, \widetilde{Q}$,
$N_f'$-fundamental flavors $Q', \widetilde{Q}'$,
$N_f''$-fundamental flavors $Q'', \widetilde{Q}''$,
bifundamentals $F, \widetilde{F}$ and $G, \widetilde{G}$  
can be described as two NS5-branes, two
NS5'-branes, 
$N_c$-, $N_c'$- and $N_c''$-D4-branes, and $N_f$-, $N_f'$- and 
$N_f''$-D6-branes. 
The $F$ is in the representation $\bf{(N_c,\overline{N_c'}, 1)}$ and 
the $\widetilde{F}$ is in the representation $\bf{(\overline{N_c},
N_c', 1)}$ while 
the $G$ is in the representation $\bf{(1, N_c',\overline{N_c''})}$ and 
the $\widetilde{G}$ is in the representation $\bf{(1, \overline{N_c'},
N_c'')}$,
under the gauge group. 
The quarks $Q$ and $\widetilde{Q}$ are in the representation 
$\bf{(N_c, 1, 1)}$ and $\bf{(\overline{N_c}, 1, 1)}$ 
respectively, 
the quarks $Q'$ and $\widetilde{Q}'$ are in the representation 
$\bf{(1, N_c', 1)}$ and $\bf{(1, \overline{N_c'}, 1)}$ 
respectively, and 
the quarks $Q''$ and $\widetilde{Q}''$ are in the representation 
$\bf{(1, 1, N_c'')}$ and $\bf{(1, 1, \overline{N_c''})}$ 
respectively,
 under the gauge group.

The mass terms for each fundamental quarks 
can be added by displacing each D6-branes along 
$
v$
direction leading to their coordinates 
$v = + 
v_{D6_{-\theta}}(+v_{D6_{-\theta'}})[+v_{D6_{-\theta''}}]$ respectively  
while the quartic terms for each fundamental quarks 
can be added also by rotating each D6-branes
by an angle 
$-\theta(-\theta')[-\theta'']$ in $(w,v)$-plane respectively.
Then, the general 
superpotential by adding the above deformations is
given by
\bea
W_{elec} & = &
\left[ \beta_1   \tr (F \widetilde{F})^2
+\beta_2   \tr (F \widetilde{F} G \widetilde{G}) 
+\beta_3   \tr (G \widetilde{G})^2 \right]
 +   \frac{\alpha}{2} \tr (Q \widetilde{Q})^2 - m \tr Q
\widetilde{Q}  \nonu \\
 & + & \frac{\alpha'}{2} \tr (Q' \widetilde{Q}')^2 - m' \tr Q'
\widetilde{Q}' 
 +  \frac{\alpha''}{2} \tr (Q'' \widetilde{Q}'')^2 - m'' \tr Q''
\widetilde{Q}'' 
\label{elecsuper2}
\eea 
where the extra parameters are similarly described as 
the following geometric quantities
\bea
 \alpha'' \equiv \frac{\tan \theta''}{\Lambda''}, \qquad
m'' \equiv \frac{v_{D6_{-\theta''}}}{2\pi \ell_s^2}.
\nonu
\eea
The first three terms of (\ref{elecsuper2}), in general, 
are due to the rotation angles $\omega_1$ and
$\omega_3$  
of the first and third NS-brane 
with respect to the second NS-brane and 
the rotation angles $\omega_2$ and
$\omega_4$  
of the second and fourth NS-brane 
with respect to the third NS-brane. 
We consider the limit where $\omega_1=\omega_3=
\frac{\pi}{2}=\omega_2=\omega_4$ and $\beta_i(i=1, 2, 3)$ will vanish.
Although 
the relative displacement of each two color D4-branes can 
be added in the superpotential,
we focus on the particular limit $m_F =0 = m_G$.

Then the ${\cal N}=1$ supersymmetric electric brane
configuration for the superpotential 
(\ref{elecsuper2}) 
in type IIA string theory is given as follows and let us draw this
brane structure in
Figure 4 explicitly:

$\bullet$
Two NS5-branes in $(012345)$ directions 

$\bullet$ 
Two NS5'-branes in  $(012389)$ directions 

$\bullet$
$N_f$ $D6_{-\theta}$-branes in (01237)
directions and
two other directions in $(v,w)$-plane

$\bullet$
$N_f'$ 
$D6_{-\theta'}$-branes in (01237)
directions and
two other directions in $(v,w)$-plane

$\bullet$
$N_f''$ 
$D6_{-\theta''}$-branes in (01237)
directions and
two other directions in $(v,w)$-plane

$\bullet$
$N_c$-, $N_c'$- and $N_c''$-color D4-branes in $(01236)$ directions  

%%%%%%%%%%%%%
%Figure 4
%%%%%%%%%%%%%
%%%%%%%%%%%%%%%%%%%%%%%%%%%%%%%%%%%%%%%%%%%%%%%%%%%%%%%%%%%%%%%%%%%%%%
%%%%%%%%%%%%%%%%%%%%%%%%%%%%%%%%%%%%%%%%%%%%%%%%%%%%%%%%%%%%%%%%%%%%%%
\begin{figure}[ht]
   \epsfxsize=4.5in 
\centerline{\epsffile{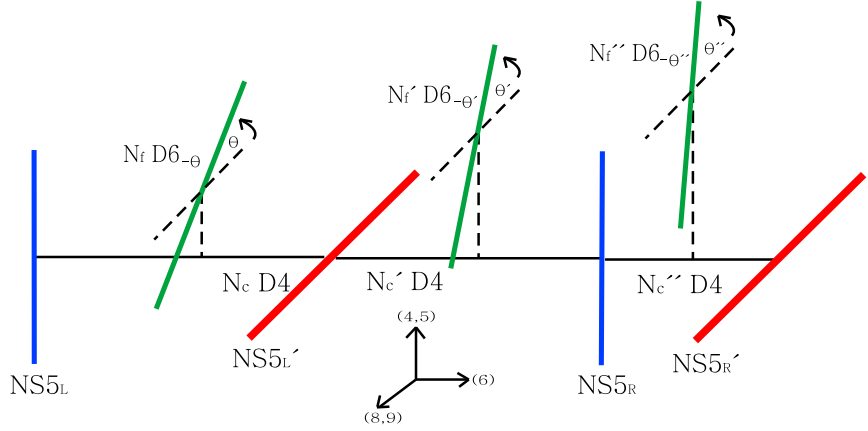}}
   \caption[FIG. \arabic{figure}.]{ 
The  ${\cal N}=1$ supersymmetric 
electric brane configuration for the gauge group $SU(N_c) \times
SU(N_c') \times SU(N_c'')$ 
with bifundamentals $F, \widetilde{F}, G, \widetilde{G}$, and 
fundamentals $Q, \widetilde{Q}, Q', \widetilde{Q}', Q'',
\widetilde{Q}''$. 
A 
rotation of $N_f(N_f')[N_f'']$ D6-branes in $(w,v)$-plane
corresponds to 
a quartic term for the fundamentals $Q, \widetilde{Q}(Q',
\widetilde{Q}')[Q'', \widetilde{Q}'']$ while 
a displacement of $N_f(N_f')[N_f'']$ D6-branes in $+v$ direction corresponds to a
mass term for the fundamentals $Q, \widetilde{Q}(Q',
\widetilde{Q}')[Q'', \widetilde{Q}'']$. Let us take $v_{D6_{-\theta}}
< v_{D6_{-\theta'}} < v_{D6_{-\theta''}}$ and 
$\theta < \theta' < \theta''$.}
\end{figure}
%%%%%%%%%%%%%%%%%%%%%%%%%%%%%%%%%%%%%%%%%%%%%%%%%%%%%%%%%%%%%%%%%%%%%
%%%%%%%%%%%%%%%%%%%%%%%%%%%%%%%%%%%%%%%%%%%%%%%%%%%%%%%%%%%%%%%%%%%%%

%%%%%%%%%%%%%%%%%%%%%%%%%%%%%%%%%%%%%%%%%%%%%%%%%%%%%%%%%%%%%%%%%%%%%%%%%%%
\subsection{Magnetic theory }
%%%%%%%%%%%%%%%%%%%%%%%%%%%%%%%%%%%%%%%%%%%%%%%%%%%%%%%%%%%%%%%%%%%%%%%%

The $NS5_R'$-brane 
starts out with linking number $l_e=\frac{N_f}{2} - N_c''$
from Figure 4 and after duality 
this $NS5_R'$-brane ends up with linking number 
$l_m = -\frac{N_f}{2} + \widetilde{N}_c-N_f'-N_f''$ from Figure 5.
We consider only the particular brane motion where
$N_f$ $D6_{-\theta}$-branes meet 
the $NS5_L'$-brane and the $NS5_R$-brane with {\it no angles}. That is, 
the  $D6_{-\theta}$-branes become D6-branes 
when they meet with the $NS5_L'$-brane instantaneously and then
after that
they come back to the original $D6_{-\theta}$-branes.
Moreover, these  $D6_{-\theta}$-branes become $D6_{-\frac{\pi}{2}}$-branes 
when they meet with the $NS5_R$-brane instantaneously and then
after that
they come back to the original $D6_{-\theta}$-branes.
Therefore, in this dual process, there is {\it no} creation of D4-branes.
Of course, the $D6_{-\theta}$-branes meet the $NS5_R'$-brane with an angle.
Similarly, 
the $N_f'$ $D6_{-\theta'}$-branes meet 
the $NS5_L'$-brane with no angles and 
the $N_f''$ $D6_{-\theta''}$-branes meet 
the $NS5_L'$-brane and the $NS5_R$-brane with no angles.
Then the dual color number $\widetilde{N}_c$
is given by $\widetilde{N}_c = N_f+N_f'+N_f''-N_c''$. 

The $NS5_R$-brane 
starts out with linking number $l_e=N_c'' - N_c'$
and after duality 
this $NS5_R$-brane ends up with linking number 
$l_m = \widetilde{N}_c'-\widetilde{N}_c$.
As we observed above, 
we consider only the particular brane motion where
all the  $D6_{-\theta, -\theta', -\theta''}$-branes 
become $D6_{-\frac{\pi}{2}}$-branes 
when they meet with the $NS5_R$-brane instantaneously and after that 
they come back to the original  $D6_{-\theta, -\theta', -\theta''}$-branes.
Therefore, in this dual process, there is no creation of D4-branes.
Then it turns out that the dual color number $\widetilde{N}_c'$
is given by $\widetilde{N}_c' = N_f+N_f'+N_f''-N_c'$. 

The $NS5_L$-brane 
starts out with linking number $l_e=N_c - \frac{(N_f'+N_f'')}{2}$
and after duality 
this $NS5_L$-brane ends up with linking number 
$l_m = -\widetilde{N}_c''+ N_f + \frac{(N_f'+N_f'')}{2}$.
As we observed above, 
we consider only the particular brane motion where
all the  $D6_{-\theta', -\theta''}$-branes meet the $NS5_L$-brane 
{\it with angles}. 
Then it turns out that the dual color number $\widetilde{N}_c''$
is given by $\widetilde{N}_c'' = N_f+N_f'+N_f''-N_c$. 

Then one obtains the folllowing dual color numbers
\bea
\widetilde{N}_c   = N_f+N_f'+N_f''-N_c'', \qquad
\widetilde{N}_c'  = N_f+N_f'+N_f''-N_c', \qquad
\widetilde{N}_c'' = N_f +N_f'+N_f''-N_c.
\nonu
\eea

The low energy theory \cite{Brodie,BH} on the three color D4-branes 
has $SU(\widetilde{N}_c) \times SU(\widetilde{N}_c') \times
SU(\widetilde{N}_c'')$ 
gauge group and  
$N_f$-fundamental dual quarks $q'', \widetilde{q}''$,
$N_f'$-fundamental dual quarks $q', \widetilde{q}'$, 
$N_f''$-fundamental dual quarks $q, \widetilde{q}$,
bifundamentals $f, \widetilde{f}, g, \widetilde{g}$
and various gauge singlets.
The $f$ is in the representation 
$\bf{(\widetilde{N}_c, \overline{\widetilde{N}_c'}, 1)}$ while 
the $\widetilde{f}$ is in the representation $\bf{(\overline{\widetilde{N}_c},
\widetilde{N}_c', 1)}$ and 
the $g$ is in the representation $\bf{(1, \widetilde{N}_c', 
\overline{\widetilde{N}_c''})}$ while 
the $\widetilde{g}$ is in the representation $\bf{(1, \overline{\widetilde{N}_c'},
\widetilde{N}_c'')}$, under the dual gauge group. 
The $N_f''$ flavors $q$ and $\widetilde{q}$ are in the representation 
$\bf{(\widetilde{N}_c, 1, 1)}$ and $\bf{(\overline{\widetilde{N}_c},
  1, 1)}$ 
respectively under the gauge group and 
in the representation 
$\bf{(\overline{N_f''}, 1)}$ and $\bf{(1, N_f'')}$ 
respectively under the flavor group $SU(N_f'')_L \times SU(N_f'')_R$.
Similarly, 
the $N_f'$ flavors $q'$ and $\widetilde{q}'$ are in the representation 
$\bf{(1, \widetilde{N}_c', 1)}$ and $\bf{(1,
  \overline{\widetilde{N}_c'}, 1)}$ 
respectively under the gauge group
and in the representation 
$\bf{(\overline{N_f'}, 1)}$ and $\bf{(1, N_f')}$ 
respectively under the flavor group $SU(N_f')_L \times SU(N_f')_R$.
The $N_f$ flavors $q''$ and $\widetilde{q}''$ are in the representation 
$\bf{(1, 1, \widetilde{N}_c'')}$ and $\bf{(1, 1, \overline{\widetilde{N}_c''})}$ 
respectively under the gauge group
and in the representation 
$\bf{(\overline{N_f}, 1)}$ and $\bf{(1, N_f)}$ 
respectively under the flavor group $SU(N_f)_L \times SU(N_f)_R$.

In particular, a magnetic meson field 
$
M_0 \equiv Q \widetilde{Q}
$
is $N_f \times N_f$ matrix and comes from 
4-4 strings of $N_f$ flavor D4-branes created when $N_f$ $D6_{-\theta}$-branes
meet the $NS5_R'$-brane, 
a magnetic meson field 
$
M'_0 \equiv Q' \widetilde{Q}'
$
is $N_f' \times N_f'$ matrix and comes from 
4-4 strings of $N_f'$ flavor D4-branes created when 
$N_f'$ $D6_{-\theta'}$-branes
meet the $NS5_L$-brane
 and 
a magnetic meson field 
$
M''_0 \equiv Q'' \widetilde{Q}''
$
is $N_f'' \times N_f''$ matrix and comes from 
4-4 strings of $N_f''$ flavor D4-branes created
when $N_f''$ $D6_{-\theta''}$-branes
meet the $NS5_L$-brane.

Then the most general magnetic superpotential, for
the case where $N_f(N_f')[N_f'']$ $D6_{-\theta}$-branes(
$D6_{-\theta'}$-branes)[$D6_{-\theta''}$-branes] meet 
the NS-branes {\it with angles},
is given by  
\bea
W_{dual} & = & \left[ (f \widetilde{f})^2 +
f\widetilde{f} g \widetilde{g} + (g \widetilde{g})^2 
  \right]
 \nonu \\
& + & \left( \frac{\alpha}{2} \tr M_0^2 - m M_0  \right) + \left(
\frac{\alpha'}{2} 
\tr {M_0'}^2 - m' M_0' \right)
+ \left( \frac{\alpha''}{2} 
\tr {M_0''}^2 - m'' M_0'' \right) \nonu \\
& + & \left[ M_0 \widetilde{q}'' (f \widetilde{f})^2  q''  
+ 
M_0' \widetilde{q}' (f \widetilde{f})^2   q'
+ 
M_0'' \widetilde{q} (f \widetilde{f})^2   q \right]  
\nonu \\
&+ & \left[
M_2 \widetilde{q}'' f \widetilde{f} q'' + M_4 \widetilde{q}'' q'' +
M_{2,F}' \widetilde{q}' g \widetilde{g} q' +
M_{2,G}' \widetilde{q}' f \widetilde{f} q' 
+ M_4' \widetilde{q}' q' + 
M_2'' \widetilde{q} f \widetilde{f} q \right. \nonu \\
& + & M_4'' \widetilde{q} q  +   
P_1 \widetilde{q}' f \widetilde{f} f \widetilde{q}'' +
\widetilde{P}_1 q' \widetilde{f} f \widetilde{f} q'' +
P_2 \widetilde{q} f  \widetilde{g} q'' +
\widetilde{P}_2 q \widetilde{f} g  \widetilde{q}'' +
P_3 \widetilde{q}' f  \widetilde{q}'' \nonu \\
& + &
\left. \widetilde{P}_3 q' \widetilde{f} q''
 +  
R_1 \widetilde{q}' f \widetilde{f} f \widetilde{q} +
\widetilde{R}_1 q' \widetilde{f} f \widetilde{f} q +
R_3 \widetilde{q}' f  \widetilde{q} +
\widetilde{R}_3 q' \widetilde{f} q \right]
\label{mag3}
\eea
where the mesons are given by \cite{Brodie,BH}
\bea
M_0 & \equiv &  Q  \widetilde{Q},\quad  
M_0'  \equiv Q' \widetilde{Q}', \quad
M_0''  \equiv Q'' \widetilde{Q}'', \quad 
M_2  \equiv Q  \widetilde{F} F \widetilde{Q}, \quad 
M_4   \equiv Q  (\widetilde{F} F)^2  \widetilde{Q},
\nonu \\
M_{2,F}'  & \equiv & Q'  \widetilde{F} F \widetilde{Q}',  \quad
M_{2,G}'  \equiv Q'  \widetilde{G} G \widetilde{Q}', \quad
M_4'   \equiv Q'  (\widetilde{F} F)^2  \widetilde{Q}', \quad
M_2''  \equiv Q''  \widetilde{G} G \widetilde{Q}'', \nonu \\ 
M_4''   & \equiv & Q''  (\widetilde{G} G)^2 
\widetilde{Q}'',
\quad
P_1  \equiv  Q \widetilde{F} Q', \quad
\widetilde{P}_1  \equiv  \widetilde{Q} F \widetilde{Q}', \quad
P_2 \equiv Q \widetilde{F} G \widetilde{Q}'', 
\nonu \\
\widetilde{P}_2 & \equiv & \widetilde{Q} F \widetilde{G} Q'', \quad
P_3  \equiv  Q \widetilde{F} F \widetilde{F} Q', \quad
\widetilde{P}_3   \equiv  \widetilde{Q} F \widetilde{F} F \widetilde{Q}', 
\quad
R_1  \equiv  Q' \widetilde{G} Q'', \nonu \\
\widetilde{R}_1  & \equiv &  \widetilde{Q}' G \widetilde{Q}'', \quad
R_3  \equiv  Q' \widetilde{G} G \widetilde{G} Q'', \quad
\widetilde{R}_3   \equiv  \widetilde{Q}' G \widetilde{G} G \widetilde{Q}''.
\nonu
\eea
The first two lines of (\ref{mag3}) are dual expressions for the electric
superpotential (\ref{elecsuper2}) and the corresponding meson fields
$M_0, M_0', M_0''$ are replaced and the remaining
lines of (\ref{mag3}) are the analogs of the cubic term
superpotential between
the meson and dual quarks in Seiberg duality. 

When the $N_f$ $D6_{-\theta}$-branes meet the $NS5_L'$-brane and 
$NS5_R$-brane,
no creation of D4-branes  
implies that there is no $M_2$- or $M_4$-term 
in the above superpotential
(\ref{mag3}).
The mesons  
$M_2$ and $M_4$ originate from  $SU(N_c)$ chiral mesons
$Q\widetilde{Q}$  when one
dualizes the $SU(N_c)$ gauge group first by moving the $NS5_L'$-brane
to the left of the $NS5_L$-brane. That is, the fluctuations of
strings stretching between the $N_f$ ``flavor'' D4-branes provide
these meson fields. 
After the additional dual procedures, 
the cubic terms in the
superpotential arise as 
$M_2$-dependent  and $M_4$-dependent terms 
where $M_2$ has 
extra $\widetilde{F} F$ fields and 
$M_4$ has an
extra $(\widetilde{F} F)^2$ fields, 
besides $Q \widetilde{Q}$, 
due to the further dualizations. The $M_2$-term in the superpotential 
has an extra 
$f \widetilde{f}$ factor 
besides $\widetilde{q}'' q''$.

When the $N_f''$ $D6_{-\theta''}$-branes meet the $NS5_L'$-brane and 
$NS5_R$-brane,
no creation of D4-branes  
implies that there is no $M_2''$- or $M_4''$-term 
in the above superpotential
(\ref{mag3}).
The mesons  
$M_2''$ and $M_4''$ originate from  $SU(N_c'')$ chiral mesons
$Q''\widetilde{Q}''$  when one
dualizes the $SU(N_c'')$ gauge group first by moving the $NS5_R$-brane
to the right of the $NS5_R'$-brane. That is, the fluctuations of
strings stretching between the $N_f''$ ``flavor'' D4-branes provide
these meson fields. 
After the additional dual procedures, 
the cubic terms in the
superpotential arise as 
$M_2''$-dependent  and $M_4''$-dependent terms 
where $M_2''$ has 
extra $\widetilde{G} G$ fields  and 
$M_4''$ has 
extra $(\widetilde{G} G)^2 $ fields, 
besides $Q'' \widetilde{Q}''$, 
due to the further dualizations. The $M_2''$-term in the
superpotential has an extra 
$f \widetilde{f}$ factor 
besides $\widetilde{q} q$.

Similarly, 
when 
the $N_f'$ $D6_{-\theta'}$-branes 
meet the $NS5_L'$-brane, the $NS5_R$-brane, or $NS5_R'$-brane with no angles, 
there is no $M_4'$ term in the above superpotential
(\ref{mag3}) \footnote{In \cite{BH}, they introduced the
$2N_f'$ full D4-branes without changing the linking number in order to
satisfy the correct dual color numbers in the gauge theory side
analysis. This has led to 
the fact that there are  $2N_f'$  D4-branes 
connecting the  $NS5_{R}$-brane and $N_f'$ $D6_{-\theta'}$-branes,
after duality. See Figure 5.
In other words, these extra $2N_f'$ D4-branes were needed
for the existence of meson fields $M_{2,F}'$ and $M_{2,G}'$. 
In our construction, we do not
need these extra 
$2N_f'$ full D4-branes because we do not want to have these unwanted
meson fields $M_{2,F}'$ and $M_{2,G}'$.}.
These meson fields 
originate from  $SU(N_c')$ chiral mesons
$Q'\widetilde{Q}'$  when one
dualizes the $SU(N_c')$ gauge group first by interchanging the $NS5_L'$-brane
and the $NS5_R$-brane each other. The
strings stretching between the $N_f'$ ``flavor'' D4-branes provide
this meson. 
After the additional dual procedures, 
the cubic terms in the superpotential arise as 
$M_{2,F}'$-term, $M_{2,G}'$-term and $M_4'$-term 
in (\ref{mag3}) where $M_{2,F}'$ has 
extra $\widetilde{F} F $ fields,
$M_{2,G}'$ has 
extra $\widetilde{G} G$ fields, 
and 
$M_4'$ has extra $(\widetilde{F} F)^2 $ fields,
besides $Q' \widetilde{Q}'$, due to the further dualizations.
The $M_{2,F}'$-term in the superpotential has an extra 
$g \widetilde{g}$  factor 
while 
$M_{2,G}'$-term has an extra 
$f \widetilde{f}$ factor, besides
$\widetilde{q}' q'$.

Furthermore, 
when the $N_f$ $D6_{-\theta}$-branes, the 
$N_f'$ $D6_{-\theta'}$-branes, the $NS5_{L}'$-brane and 
the $NS5_R$-brane
meet each other with no angles,
no $P_1$- and $P_3$- or 
$\widetilde{P}_1$- and $\widetilde{P}_3$-dependent terms occur in the
superpotential (\ref{mag3}).
These mesons 
originate from  $SU(N_c')$ chiral mesons
$\widetilde{F} Q'$ and $ F \widetilde{Q}'$ when one
dualizes the $SU(N_c')$  first by interchanging the $NS5_L'$-brane
and the $NS5_R$-brane. 
The
strings stretching between the $N_f'$ flavor D4-branes and $N_c$
color D4-branes give rise to these 
$N_f'$ $SU(N_c)$ fundamentals and 
$N_f'$ $SU(N_c)$ antifundamentals. 
After the additional dual procedures,  these cubic terms arise as 
these meson terms where there exist  extra 
$ \widetilde{f} f \widetilde{q}'', f \widetilde{f} q'',
\widetilde{q}''$ 
and $q''$ in the interactions of $P_1, \widetilde{P}_1, P_3$ and 
$\widetilde{P}_3$ in the superpotential and the mesons have 
extra $Q, \widetilde{Q}, Q \widetilde{F} F, \widetilde{Q} 
F \widetilde{F}$ fields respectively, 
due to the further dualizations. 

When the $N_f'$ $D6_{-\theta'}$-branes, the 
$N_f''$ $D6_{-\theta''}$-branes, the $NS5_{L}'$-brane and 
the $NS5_R$-brane
meet each other with no angles,
no $R_1$- and $R_3$- or 
$\widetilde{R}_1$- and $\widetilde{R}_3$-dependent terms arise in the
superpotential (\ref{mag3}).
These mesons 
originate from  $SU(N_c'')$ chiral mesons
$\widetilde{G} Q''$ and $ G \widetilde{Q}''$ when one
dualizes the $SU(N_c'')$  first by moving the $NS5_R$-brane
to the right of the $NS5_R'$-brane. 
The
strings stretching between the $N_f''$ flavor D4-branes and $N_c'$
color D4-branes give rise to these 
$N_f''$ $SU(N_c')$ fundamentals and 
$N_f''$ $SU(N_c')$ antifundamentals. 
After the additional dual procedures,  these cubic terms arise as 
these meson terms  where 
$R_1, \widetilde{R}_1, R_3$ and 
$\widetilde{R}_3$ have 
extra $Q', \widetilde{Q}', Q' \widetilde{G} G, \widetilde{Q}' 
G \widetilde{G}$ fields respectively, 
due to the further dualizations. 

Finally, 
when the $N_f$ $D6_{-\theta}$-branes, the 
$N_f''$ $D6_{-\theta''}$-branes, the $NS5_{L}'$-brane and 
the $NS5_R$-brane
meet each other with no angles,
no $P_2$- or 
$\widetilde{P}_2$-dependent terms arise in the
superpotential (\ref{mag3}).
These mesons 
originate from  $SU(N_c)$ chiral mesons
$Q \widetilde{F} $ and $ \widetilde{Q} F$ when one
dualizes the $SU(N_c)$  first by moving the $NS5_L'$-brane
to the left of the $NS5_L$-brane. 
The
strings stretching between the $N_f$ flavor D4-branes and $N_c'$
color D4-branes give rise to these 
$N_f$ $SU(N_c')$ fundamentals and 
$N_f$ $SU(N_c')$ antifundamentals. 
After the additional dual procedures,  these cubic terms arise as 
these meson terms  where there exist  extra 
$ \widetilde{g} q'', g \widetilde{q}''$ 
in the interactions of $P_2$ and 
$\widetilde{P}_2$ in the superpotential and the mesons have 
extra $G \widetilde{Q}'', \widetilde{G} Q''$ fields respectively, 
due to the further dualizations. 

Then the reduced magnetic superpotential in our case  
by taking the first three lines of (\ref{mag3}) 
is given by 
\bea
W_{dual} & = &  \left[ M_0 \widetilde{q}'' (f \widetilde{f})^2  q''  
 + \frac{\alpha}{2} \tr M_0^2 - m M_0  
\right]+ \left[
M_0' \widetilde{q}' (f \widetilde{f})^2   q'
+\frac{\alpha'}{2} 
\tr {M_0'}^2 - m' M_0' 
\right] \nonu \\
&+ & \left[ M_0'' \widetilde{q} (f \widetilde{f})^2   q +
 \frac{\alpha''}{2} 
\tr {M_0''}^2 - m'' M_0'' \right].
\label{dualW1}
\eea

%%%%%%%%%%%%%
%Figure 5
%%%%%%%%%%%%%
%%%%%%%%%%%%%%%%%%%%%%%%%%%%%%%%%%%%%%%%%%%%%%%%%%%%%%%%%%%%%%%%%%%%%%
%%%%%%%%%%%%%%%%%%%%%%%%%%%%%%%%%%%%%%%%%%%%%%%%%%%%%%%%%%%%%%%%%%%%%%
\begin{figure}[ht]
   \epsfxsize=4.5in 
\centerline{\epsffile{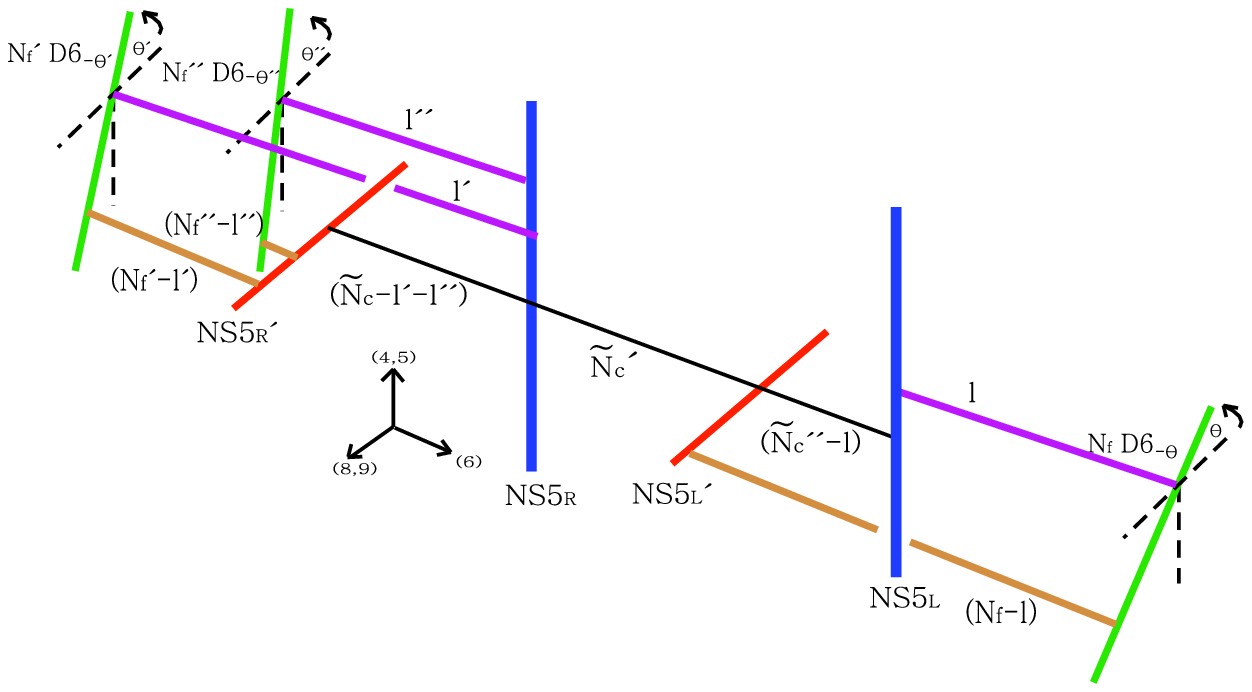}}
   \caption[FIG. \arabic{figure}.]{ 
The  ${\cal N}=1$ supersymmetric
magnetic brane configuration corresponding to Figure 4 with 
a splitting and a reconnection 
between D4-branes when the gravitational potential of
the NS5-brane is ignored. 
The $N_f$ flavor D4-branes connecting between
$D6_{-\theta}$-branes and $NS5_L'$-brane are splitting into $(N_f-l)$- and
$l$- D4-branes, 
the $N_f'$ flavor D4-branes connecting between
$D6_{-\theta'}$-branes and $NS5_R'$-brane are 
splitting into $(N_f'-l')$- and
$l'$- D4-branes, and 
the $N_f''$ flavor D4-branes connecting between
$D6_{-\theta''}$-branes and $NS5_R'$-brane are 
splitting into $(N_f''-l'')$- and
$l''$- D4-branes. 
}
\end{figure}
%%%%%%%%%%%%%%%%%%%%%%%%%%%%%%%%%%%%%%%%%%%%%%%%%%%%%%%%%%%%%%%%%%%%%
%%%%%%%%%%%%%%%%%%%%%%%%%%%%%%%%%%%%%%%%%%%%%%%%%%%%%%%%%%%%%%%%%%%%%

For the supersymmetric vacua, one can compute the F-term equations for
this superpotential (\ref{dualW1}) 
and the F-terms for $M_0, q'', \widetilde{q}'', M_0', q',
\widetilde{q}', M_0'', q, \widetilde{q}, f$ and $\widetilde{f}$ 
are given by
\bea
&& \widetilde{q}'' (f \widetilde{f})^2  q''  
-m + \alpha M_0   =  0, \qquad
(M_0  \widetilde{q}'' f \widetilde{f}) f \widetilde{f} =0, \qquad 
f \widetilde{f} (f \widetilde{f} q'' M_0) =0, \nonu \\
&&\widetilde{q}' (f \widetilde{f})^2   q'  -m' + 
\alpha' M_0'  =  0, \qquad
(M_0' \widetilde{q}' f \widetilde{f}) f \widetilde{f} =0, \qquad 
f \widetilde{f} (f \widetilde{f} q' M_0') =0, \nonu \\
&& \widetilde{q} (f \widetilde{f})^2   q  -m'' + 
\alpha'' M_0''  =  0, \qquad
(M_0'' \widetilde{q} f \widetilde{f}) f \widetilde{f}  =0, \qquad 
 f \widetilde{f} (f \widetilde{f}  q M_0'') =0, \nonu \\
&&  \widetilde{f} (f \widetilde{f} q''  M_0) \widetilde{q}'' +
\widetilde{f} q'' (M_0 \widetilde{q}'' f \widetilde{f}) + 
\widetilde{f} (f \widetilde{f}  q' M_0') \widetilde{q}' \nonu \\
&& +\widetilde{f}  q' 
(M_0' \widetilde{q}' f \widetilde{f}) + 
\widetilde{f} (f \widetilde{f}  q
M_0'') \widetilde{q} + 
\widetilde{f}  q (M_0''  \widetilde{q} f \widetilde{f}) =  0, \nonu \\
&&  (f \widetilde{f} q'' M_0)  \widetilde{q}'' f +    q''
(M_0  \widetilde{q}'' f \widetilde{f}) f +  
(f \widetilde{f}  q' M_0') \widetilde{q}' f \nonu \\
&& +
q' (M_0' \widetilde{q}' f \widetilde{f}) f +  (f \widetilde{f}  q M_0'')
\widetilde{q} f+ q' (M_0''
  \widetilde{q} f \widetilde{f}) f =  0.
\nonu
\eea
From this, it is easy to see that the last two equations are satisfied
if the second, third, fifth, sixth, eighth, and ninth are 
satisfied:$M_0  \widetilde{q}'' f \widetilde{f}=0 =
\cdots =f \widetilde{f}  q M_0''$.

The theory has many nonsupersymmetric meta-stable ground states and 
when we rescale the meson fields as
$
M_0 = h \Lambda \Phi_0, M_0' = h' \Lambda' \Phi_0'$ 
and 
$M''_0 = h'' \Lambda'' \Phi''_0$,
then the Kahler potential for $\Phi_0, \Phi_0'$ and $\Phi''_0$ 
is canonical and the magnetic
quarks are canonical near the origin of field space.
Then the magnetic superpotential (\ref{dualW1}) 
can be rewritten as
\bea
W_{mag} & = & \left[ h \Phi_0  \widetilde{q}''   (f 
\widetilde{f})^2  q'' 
 +  
\frac{h^2 \mu_{\phi}}{2} \tr \Phi_0^2- h \mu^2 \tr \Phi_0 \right]
 +  
\left[ h' \Phi_0'     \widetilde{q}'  (f \widetilde{f})^2  q' 
 +  
\frac{{h'}^2 \mu_{\phi}'}{2} \tr {\Phi_0'}^2- 
h' {\mu'}^2 \tr \Phi_0'\right]
\nonu \\
 & + & 
\left[ h'' \Phi_0''     \widetilde{q}  (f \widetilde{f})^2  q 
 +  
\frac{{h''}^2 \mu_{\phi}''}{2} \tr {\Phi_0''}^2- 
h'' {\mu''}^2 \tr \Phi_0''\right]
\nonu
\label{finalW1}
\eea
where
$
\mu^2 = m \Lambda, {\mu'}^2 =m' \Lambda',  {\mu''}^2 =m'' \Lambda''$ and  
$\mu_{\phi} = \alpha \Lambda^2, 
\mu_{\phi}' = \alpha' {\Lambda'}^2, \mu_{\phi}'' = \alpha'' {\Lambda''}^2$.

Now one splits 
the $(N_f-l) \times (N_f-l)$
block  at the lower right corner of $h\Phi_0$ and $\widetilde{q}'' 
(f \widetilde{f})^2  q''$ 
into blocks of 
size $n$ and $(N_f-l-n)$,
one decomposes 
the $(N_f'-l') \times (N_f'-l')$
block  at the lower right corner of $h' \Phi'_0$ and $\widetilde{q}'   
(f \widetilde{f})^2  q'$ 
into blocks of 
size $n'$ and $(N_f'-l'-n')$
and 
one splits 
the $(N_f''-l'') \times (N_f''-l'')$
block  at the lower right corner of $h'' \Phi_0''$ and $\widetilde{q} 
(f \widetilde{f})^2  q$ 
into blocks of 
size $n''$ and $(N_f''-l''-n'')$
as follows \cite{GK0710}:
\bea
h\Phi_0 & = &  \left(
\begin{array}{ccc}
0_l & 0 & 0  \\
0 & h \Phi_n & 0 \\
0 & 0 & \frac{\mu^2}{\mu_{\phi}} {\bf 1}_{N_f-l-n}
\end{array}
\right), 
h' \Phi_0'  =   \left(
\begin{array}{ccc}
0_{l'} & 0 & 0  \\
0 & h' \Phi_{n'}' & 0 \\
0 & 0 & \frac{{\mu'}^2}{\mu_{\phi}'} {\bf 1}_{N_f'-l'-n'}
\end{array}
\right), \label{vac2} 
\\
h'' \Phi_0''  & = &   \left(
\begin{array}{ccc}
0_{l''} & 0 & 0  \\
0 & h'' \Phi_{n''}'' & 0 \\
0 & 0 & \frac{{\mu''}^2}{\mu_{\phi}''} {\bf 1}_{N_f''-l''-n''}
\end{array}
\right), 
\widetilde{q}''  (f \widetilde{f})^2  q''   =  \left(
\begin{array}{ccc}
\mu^2 {\bf 1}_l & 0 & 0  \\
0 & \widetilde{ \varphi}'' (y \widetilde{y})^2  \varphi''  &  0 \\
0 & 0 & 0_{N_f-l-n}
\end{array}
\right), \nonu \\
\widetilde{q}' (f \widetilde{f})^2  q'  & = & \left(
\begin{array}{ccc}
{\mu'}^2 {\bf 1}_{l'} & 0 & 0  \\
0 & \widetilde{\varphi}' (y \widetilde{y})^2   \varphi'  &  0 \\
0 & 0 & 0_{N_f'-l'-n'}
\end{array}
\right),
\widetilde{q} (f \widetilde{f})^2  q   =  \left(
\begin{array}{ccc}
{\mu''}^2 {\bf 1}_{l''} & 0 & 0  \\
0 & \widetilde{\varphi} (y \widetilde{y})^2   \varphi  &  0 \\
0 & 0 & 0_{N_f''-l''-n''}
\end{array}
\right).
\nonu
\eea
Here $\varphi''$ and $\widetilde{\varphi}''$ 
are $n \times (\widetilde{N}_c''-l)$
dimensional matrices,   
$\varphi'$ and $\widetilde{\varphi}'$ are $n' 
\times (\widetilde{N}_c'-l')$
dimensional matrices and
 $\varphi$ and $\widetilde{\varphi}$ 
are $n'' \times (\widetilde{N}_c-l'')$
dimensional matrices.
In the brane configuration shown in Figure 6, 
$\varphi''$ and $\widetilde{\varphi}''$ correspond to 
fundamental strings connecting the $n$ flavor D4-branes and
$(\widetilde{N}_c''-l)$
color D4-branes,  
$\varphi'$ and $\widetilde{\varphi}'$ correspond to 
fundamental strings connecting the $n'$ flavor D4-branes and
$(\widetilde{N}_c'-l')$
color D4-branes and 
$\varphi$ and $\widetilde{\varphi}$ correspond to 
fundamental strings connecting the $n''$ flavor D4-branes and
$(\widetilde{N}_c-l'')$
color D4-branes.
The $\Phi_n$ and $\widetilde{\varphi}'' (y \widetilde{y})^2  
 \varphi''$
are $n \times n$ matrices,  
$\Phi'_{n'}$ and $\widetilde{\varphi}' 
(y \widetilde{y})^2  \varphi'$
are $n' \times n'$ matrices and
$\Phi''_{n''}$ and $\widetilde{\varphi} 
(y \widetilde{y})^2  \varphi$
are $n'' \times n''$ matrices.

The supersymmetric ground state corresponds to
$
h\Phi_n= \frac{\mu^2}{\mu_{\phi}} {\bf 1}_{n}, 
\widetilde{\varphi}''   y \widetilde{y}  
=0=y  \widetilde{y} \varphi''$,  
$h'\Phi'_{n'}= \frac{{\mu'}^2}{\mu_{\phi}'} {\bf 1}_{n'}, 
\widetilde{\varphi}' y \widetilde{y} =0=y \widetilde{y} 
\varphi'$ and 
$h''\Phi''_{n''}= \frac{{\mu''}^2}{\mu_{\phi}''} {\bf 1}_{n''}, 
\widetilde{\varphi} y \widetilde{y}  =0=y \widetilde{y} 
\varphi$.
The $l$ of the
$N_f$-flavor D4-branes are reconnected with $l$-color
D4-branes
and the resulting $l$ D4-branes 
stretch from the $D6_{-\theta}$-branes to
the $NS5_L$-brane directly 
and the intersection point between the 
$l$ D4-branes and the $D6_{-\theta}$-branes is given by 
$(v, w)=(+v_{D6_{-\theta}}, 0)$.
This
corresponds to  exactly the $l$'s eigenvalues from 
zeros of 
$h\Phi_0$ in (\ref{vac2}).
Now the remaining $(N_f-l)$-flavor D4-branes between 
the $D6_{-\theta}$-branes and 
the $NS5_L'$-brane correspond to the eigenvalues 
of $h\Phi_0$ in (\ref{vac2}), i.e.,   
$\frac{\mu^2}{\mu_{\phi}} {\bf 1}_{N_f-l}$.
The intersection point between the 
$(N_f-l)$ D4-branes and the $NS5_L'$-branes is given 
by $(v, w)=(0, +v_{D6_{-\theta}} \cot \theta)$ 
from trigonometric 
geometry.

Similarly, 
the $l'$ of the
$N_f'$-flavor D4-branes are reconnected with $l'$-color
D4-branes
and the resulting $l'$ D4-branes 
stretch from the $D6_{-\theta'}$-branes to
the $NS5_R$-brane directly 
and the intersection point between the 
$l'$ D4-branes and the $D6_{-\theta'}$-branes is given by 
$(v, w)=(+v_{D6_{-\theta'}}, 0)$.
This
corresponds to  exactly the $l'$'s eigenvalues from 
zeros of 
$h'\Phi'_0$ in (\ref{vac2}).
Now the remaining $(N_f'-l')$-flavor D4-branes between 
the $D6_{-\theta'}$-branes and 
the $NS5_R'$-brane correspond to the eigenvalues 
of $h'\Phi'_0$ in (\ref{vac2}), i.e.,   
$\frac{{\mu'}^2}{\mu_{\phi}'} {\bf 1}_{N_f'-l'}$.
The intersection point between the 
$(N_f'-l')$ D4-branes and the $NS5_R'$-branes is given 
by $(v, w)=(0, +v_{D6_{-\theta'}} \cot \theta')$ from trigonometric 
geometry.

The $l''$ of the
$N_f''$-flavor D4-branes are reconnected with $l''$-color
D4-branes
and the resulting $l''$ D4-branes 
stretch from the $D6_{-\theta''}$-branes to
the $NS5_R$-brane directly 
and the intersection point between the 
$l''$ D4-branes and the $D6_{-\theta''}$-branes is given by 
$(v, w)=(+v_{D6_{-\theta''}}, 0)$.
This
corresponds to  exactly the $l''$'s eigenvalues from 
zeros of 
$h''\Phi''_0$ in (\ref{vac2}).
Now the remaining $(N_f''-l'')$-flavor D4-branes between 
the $D6_{-\theta''}$-branes and 
the $NS5_R'$-brane correspond to the eigenvalues 
of $h'' \Phi''_0$ in (\ref{vac2}), i.e.,   
$\frac{{\mu''}^2}{\mu_{\phi}''} {\bf 1}_{N_f''-l''}$.
The intersection point between the 
$(N_f''-l'')$ D4-branes and the $NS5_R'$-branes is given 
by $(v, w)=(0, +v_{D6_{-\theta''}} \cot \theta'')$ from trigonometric 
geometry.

Now the full one loop potential containing $\Phi_n, \Phi'_{n'}, \Phi''_{n''},
\Phi_n^{\dagger}, {\Phi'}^{\dagger}_{n'}$ and
${\Phi''}^{\dagger}_{n''}$ with $\mu_{\phi} << \mu << \Lambda_m$,
$\mu_{\phi}' << \mu' << \Lambda_m'$ and $\mu_{\phi}'' << \mu'' << \Lambda_m''$, 
by combining the superpotential 
and the vacuum expectation values for the fields, 
takes the form
\bea
V  & = &  
|h \Phi_n \widetilde{\varphi}'' y \widetilde{y}|^2   
+  |h  y \widetilde{y} \varphi'' \Phi_n|^2
  +  
| h \widetilde{\varphi}''  (y \widetilde{y})^2 
\varphi''-h \mu^2 {\bf 1}_{n} + 
h^2 \mu_{\phi} \Phi_n|^2 
\nonu \\ 
& + & |h' \Phi_{n'}'  \widetilde{\varphi}' y  \widetilde{y}|^2   
+  |h'  y \widetilde{y} \varphi' \Phi_{n'}' |^2
  +  
|   h' \widetilde{\varphi}' (y \widetilde{y})^2  
\varphi' -h' {\mu'}^2 {\bf 1}_{n'} + 
{h'}^2 \mu_{\phi}' \Phi_{n'}'|^2
\nonu \\
& + & |h'' \Phi_{n''}''  \widetilde{\varphi} y  \widetilde{y}|^2   
+  |h''  y \widetilde{y} \varphi \Phi_{n''}'' |^2
  +  
|   h'' \widetilde{\varphi} (y \widetilde{y})^2  
\varphi -h'' {\mu''}^2 {\bf 1}_{n''} + 
{h''}^2 \mu_{\phi}'' \Phi_{n''}''|^2 
\nonu \\
& + &  b |h^2 \mu|^2 \tr \Phi_n^{\dagger} \Phi_n
 + b' |{h'}^2 \mu'|^2 
\tr {\Phi'}^{\dagger}_{n'} 
\Phi_{n'}' + b'' |{h''}^2 \mu''|^2 
\tr {\Phi''}^{\dagger}_{n''} 
\Phi_{n''}'' 
\nonu
\eea
where $b = \frac{(\ln 4-1)}{8\pi^2} \widetilde{N}_c$,  
$b' = \frac{(\ln 4-1)}{8\pi^2} \widetilde{N}_c'$ and 
$b'' = \frac{(\ln 4-1)}{8\pi^2} \widetilde{N}_c''$.
Differentiating this potential with respect to 
$\Phi_n^{\dagger}$, ${\Phi'}^{\dagger}_{n'}$ and 
${\Phi''}^{\dagger}_{n''}$ and putting 
$\widetilde{\varphi}'' y \widetilde{y} =
0 =  y \widetilde{y} \varphi''$, 
$\widetilde{\varphi}' y \widetilde{y} =0 =  y \widetilde{y}
\varphi'$ and 
$\widetilde{\varphi} y \widetilde{y}  =0 =  y \widetilde{y}
\varphi$, 
one obtains
\bea
h \Phi_n 
& \simeq &  \frac{ \mu_{\phi}}{b }
{\bf 1}_n   \qquad  \mbox{or} \qquad
M_n \simeq \frac{\alpha \Lambda^3}{\widetilde{N}_c} {\bf 1}_{n}, 
\nonu \\
h' \Phi_{n'}' 
& \simeq &  \frac{ \mu_{\phi}'}{b' }
{\bf 1}_{n'}  \qquad  \mbox{or} \qquad
M'_{n'} \simeq \frac{\alpha' {\Lambda'}^3}{\widetilde{N}_c'} {\bf
  1}_{n'}, 
\nonu \\
h'' \Phi_{n''}'' 
& \simeq &  \frac{ \mu_{\phi}''}{b'' }
{\bf 1}_{n''}  \qquad  \mbox{or} \qquad
M''_{n''} \simeq \frac{\alpha'' {\Lambda''}^3}{\widetilde{N}_c''} {\bf
  1}_{n''},
\nonu
\eea
corresponding to the $w$ coordinates of $n$ curved flavor D4-branes between 
the $D6_{-\theta}$-branes and the $NS5_L'$-brane,  
the $w$ coordinates of $n'$ curved flavor D4-branes between 
the $D6_{-\theta'}$-branes and the $NS5_R'$-brane,
the $w$ coordinates of $n''$ curved flavor D4-branes between 
the $D6_{-\theta''}$-branes and the $NS5_R'$-brane
respectively.
Since $ \frac{ \mu_{\phi}}{b } << \frac{\mu^2}{\mu_{\phi}}$,  
$ \frac{ \mu_{\phi}'}{b' } << \frac{{\mu'}^2}{\mu_{\phi}'}$ and 
$ \frac{ \mu_{\phi}''}{b'' } << \frac{{\mu''}^2}{\mu_{\phi}''}$, the
$n$-, $n'$- and $n''$-
curved D4-branes are nearer to $w=0$ at which the $NS5_L$-brane or the
$NS5_R$-brane is located.

%%%%%%%%%%%%%
%Figure 6
%%%%%%%%%%%%%
%%%%%%%%%%%%%%%%%%%%%%%%%%%%%%%%%%%%%%%%%%%%%%%%%%%%%%%%%%%%%%%%%%%%%%
%%%%%%%%%%%%%%%%%%%%%%%%%%%%%%%%%%%%%%%%%%%%%%%%%%%%%%%%%%%%%%%%%%%%%%
\begin{figure}[ht]
   \epsfxsize=4.5in 
\centerline{\epsffile{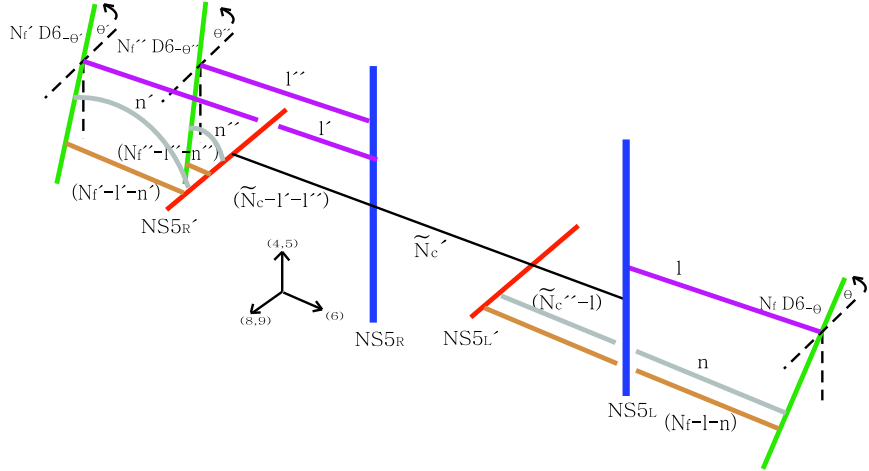}}
   \caption[FIG. \arabic{figure}.]{ 
The  nonsupersymmetric
meta-stable 
magnetic brane configuration corresponding to Figure 4 with 
a misalignment between D4-branes when the gravitational potential of
the NS5-brane is considered. 
The $(N_f-l)$ flavor D4-branes in Figure 5 connecting between
$D6_{-\theta}$-branes and $NS5_L'$-brane are further 
splitting into $(N_f-l-n)$- and
$n$- D4-branes,
the $(N_f'-l')$ flavor D4-branes  connecting between
$D6_{-\theta'}$-branes and $NS5_R'$-brane are further 
splitting into $(N_f'-l'-n')$- and
$n'$- D4-branes, and 
the $(N_f''-l'')$ flavor D4-branes  connecting between
$D6_{-\theta''}$-branes and $NS5_R'$-brane are further 
splitting into $(N_f''-l''-n'')$- and
$n''$- D4-branes. The shape of $n$ ``curved'' D4-branes is more clear when
the $D6_{-\theta}$-branes are moved into the left hand side of 
$NS5_L$-brane, as in 
\cite{Ahn08-1}.}
\end{figure}
%%%%%%%%%%%%%%%%%%%%%%%%%%%%%%%%%%%%%%%%%%%%%%%%%%%%%%%%%%%%%%%%%%%%%
%%%%%%%%%%%%%%%%%%%%%%%%%%%%%%%%%%%%%%%%%%%%%%%%%%%%%%%%%%%%%%%%%%%%%

%%%%%%%%%%%%%%%%%%%%%%%%%%%%%%%%%%%%%%%%%%%%%%%%%%%%%%%%%%%%%%%%%%%%%%%%%%%
\subsection{Addition of adjoint fields }
%%%%%%%%%%%%%%%%%%%%%%%%%%%%%%%%%%%%%%%%%%%%%%%%%%%%%%%%%%%%%%%%%%%%%%%%%%%

When we consider higher order superpotential for the adjoint fields 
whose degree is greater than 
four(i.e., the number of NS5-branes or NS5'-branes $k > 3$), 
there exist more meson fields.
For odd $k$, the chiral ring truncates \cite{BH}.
In general, it is not known how  
the deformations for the meson fields $Q X_1^j \widetilde{Q}, 
Q X_2^j \widetilde{Q} $ or $Q''
X_3^j \widetilde{Q}''$ arise geometrically 
in the type IIA brane 
configuration.
Therefore, it is not clear how to add these deformations in an
electric theory or a magnetic theory.

%%%%%%%%%%%%%%%%%%%%%%%%%%%%%%%%%%%%%%%%%%%%%%%%%%%%%%%%%%%%%%%%
%%%%%%%%%%%%%%%%%%%%%%%%%%%%%%%%%%%%%%%%%%%%%%%%%%%%%%%%%%%%%%%%
\section{
$Sp(N_c) \times SO(2N_c') \times Sp(N_c'')$ with $2N_f$-fund.,
$2N_f'$-vectors,  $2N_f''$-fund.,  and bifund.}
%%%%%%%%%%%%%%%%%%%%%%%%%%%%%%%%%%%%%%%%%%%%%%%%%%%%%%%%%%%%%%%%
%%%%%%%%%%%%%%%%%%%%%%%%%%%%%%%%%%%%%%%%%%%%%%%%%%%%%%%%%%%%%%%%

Let us add the O4-plane to the brane configuration of previous section. 

%%%%%%%%%%%%%%%%%%%%%%%%%%%%%%%%%%%%%%%%%%%%%%%%%%%%%%%%%%%%%%%%%%%%%%%%%%%
\subsection{Electric theory}
%%%%%%%%%%%%%%%%%%%%%%%%%%%%%%%%%%%%%%%%%%%%%%%%%%%%%%%%%%%%%%%%%%%%%%%%%%%

The type IIA supersymmetric electric
brane configuration \cite{Ahn97,Ahn07-8} corresponding to 
${\cal N}=1$ $Sp(N_c) \times SO(2N_c') \times Sp(N_c'')$ 
gauge theory  with  
$2N_f$-fundamental fields $Q$,
$2N_f'$-vectors $Q'$,
$2N_f''$-fundamental fields $Q''$,
bifundamentals $F$ and $G$  
can be described as two NS5-branes, two
NS5'-branes, 
$2N_c$-, $2N_c'$- and $2N_c''$-D4-branes, $2N_f$-, $2N_f'$- and 
$2N_f''$-D6-branes and an orientifold 4-plane. 
The $F$ is in the representation $\bf{(2N_c, 2N_c', 1)}$ while 
the $G$ is in the representation $\bf{(1, 2N_c', 2N_c'')}$, 
under the gauge group. 
The quarks $Q$ are in the representation 
$\bf{(2N_c, 1, 1)}$, 
the quarks $Q'$ are in the representation 
$\bf{(1, 2N_c', 1)}$,
the quarks $Q''$ are in the representation 
$\bf{(1, 1, 2N_c'')}$,
 under the gauge group.

The mass terms for each quarks 
can be added by displacing each D6-branes along 
$
v$
direction leading to their coordinates 
$v = \pm 
v_{D6_{-\theta}}( \pm v_{D6_{-\theta'}})[ \pm v_{D6_{-\theta''}}]$ respectively  
while the quartic terms for each quarks 
can be added also by rotating each D6-branes
by an angle 
$-\theta(-\theta')[-\theta'']$ in $(w,v)$-plane respectively.
Then, the general 
superpotential by adding the above deformations is
given by
\bea
W_{elec} & = &
\left[ \beta_1   \tr F^4
+\beta_2   \tr (F^2 G^2) 
+\beta_3   \tr G^4  \right] 
 +  \frac{\alpha}{2} \tr (Q Q)^2 - m \tr Q Q 
\nonu \\
 &+& \frac{\alpha'}{2} \tr (Q' Q')^2 - m' \tr Q'
Q' 
+  \frac{\alpha''}{2} \tr (Q'' Q'')^2 - m'' \tr Q'' Q''. 
\label{elecsuper3}
\eea
We consider the limit where $\omega_1=\omega_3=
\frac{\pi}{2}=\omega_2=\omega_4$ and $\beta_i(i=1, 2, 3)$ will vanish.
Although 
the relative displacement of each two color D4-branes can 
be added in the superpotential,
we focus on the particular limit $m_F =0 = m_G$.
The mass matrix $m$ is antisymmetric, the mass matrix $m'$ is 
symmetric and the mass matrix $m''$ is antisymmetric.

Then the ${\cal N}=1$ supersymmetric electric brane
configuration for the superpotential 
(\ref{elecsuper3}) 
in type IIA string theory is given as follows:

$\bullet$
Two NS5-branes in $(012345)$ directions 

$\bullet$ 
Two NS5'-branes in  $(012389)$ directions 

$\bullet$
$2N_f$ $D6_{-\theta}$-branes in (01237)
directions and
two other directions in $(v,w)$-plane

$\bullet$
$2N_f'$ 
$D6_{-\theta'}$-branes in (01237)
directions and
two other directions in $(v,w)$-plane

$\bullet$
$2N_f''$ 
$D6_{-\theta''}$-branes in (01237)
directions and
two other directions in $(v,w)$-plane

$\bullet$
$2N_c$-, $2N_c'$- and $2N_c''$-color D4-branes in $(01236)$ directions  

$\bullet$ $O4^{\pm}$-planes in (01236) directions

The corresponding brane configuration can be obtained from the
previous section by considering the correct mirrors based on the
O4-plane action.

%%%%%%%%%%%%%%%%%%%%%%%%%%%%%%%%%%%%%%%%%%%%%%%%%%%%%%%%%%%%%%%%%%%%%%%%%%%
\subsection{Magnetic theory }
%%%%%%%%%%%%%%%%%%%%%%%%%%%%%%%%%%%%%%%%%%%%%%%%%%%%%%%%%%%%%%%%%%%%%%%%%%

It is straightforward to compute the 
dual color numbers by considering the D4-brane charge of an
orientifold 4-plane.

The $NS5_R'$-brane 
starts out with linking number $l_e=\frac{2N_f}{2} - 2N_c''-2$
and after duality 
this $NS5_R'$-brane ends up with linking number 
$l_m = -\frac{2N_f}{2} + 2\widetilde{N}_c-2N_f'-2N_f''+2$.
We consider only the particular brane motion where
$N_f$ $D6_{-\theta}$-branes meet 
the $NS5_L'$-brane and the $NS5_R$-brane with {\it no} angles(and their
mirrors). 
That is, 
the  $D6_{-\theta}$-branes become D6-branes 
when they meet with the $NS5_L'$-brane instantaneously and then
after that
they come back to the original $D6_{-\theta}$-branes.
Moreover, these  $D6_{-\theta}$-branes become $D6_{-\frac{\pi}{2}}$-branes 
when they meet with the $NS5_R$-brane instantaneously and then
after that
they come back to the original $D6_{-\theta}$-branes.
Therefore, in this dual process, there is no creation of D4-branes.
Of course, the $D6_{-\theta}$-branes meet the $NS5_R'$-brane with an angle.
Similarly, 
the $N_f'$ $D6_{-\theta'}$-branes meet 
the $NS5_L'$-brane with {\it no} angles and 
the $N_f''$ $D6_{-\theta''}$-branes meet 
the $NS5_L'$-brane and the $NS5_R$-brane with {\it no} angles.
Then the dual color number $2\widetilde{N}_c$
is given by $2\widetilde{N}_c = 2N_f+2N_f'+2N_f''-2N_c''-4$. 

The $NS5_R$-brane 
starts out with linking number $l_e=2N_c'' - 2N_c'+2$
and after duality 
this $NS5_R$-brane ends up with linking number 
$l_m = 2\widetilde{N}_c'-2\widetilde{N}_c-2$.
As we observed above, 
we consider only the particular brane motion where
all the  $D6_{-\theta, -\theta', -\theta''}$-branes 
become $D6_{-\frac{\pi}{2}}$-branes 
when they meet with the $NS5_R$-brane instantaneously and after that 
they come back to the original  $D6_{-\theta, -\theta', -\theta''}$-branes.
Therefore, in this dual process, there is {\it no} creation of D4-branes.
Then it turns out that the dual color number $2\widetilde{N}_c'$
is given by $2\widetilde{N}_c' = 2N_f+2N_f'+2N_f''-2N_c'+4$. 

The $NS5_L$-brane 
starts out with linking number $l_e=2N_c - \frac{(2N_f'+2N_f'')}{2}+2$
and after duality 
this $NS5_L$-brane ends up with linking number 
$l_m = -2\widetilde{N}_c''+ 2N_f + \frac{(2N_f'+2N_f'')}{2}-2$.
As we observed above, 
we consider only the particular brane motion where
all the  $D6_{-\theta', -\theta''}$-branes meet the $NS5_L$-brane 
{\it with angles}. 
Then it turns out that the dual color number $2\widetilde{N}_c''$
is given by $2\widetilde{N}_c'' = 2N_f+2N_f'+2N_f''-2N_c-4$. 
Then one obtains the folllowing dual color numbers
\bea
2\widetilde{N}_c   & = & 2N_f+2N_f'+2N_f''-2N_c''-4, \nonu \\
2\widetilde{N}_c'  & = & 2N_f+2N_f'+2N_f''-2N_c'+4, \nonu \\
2\widetilde{N}_c'' & = & 2N_f +2N_f'+2N_f''-2N_c-4.
\nonu
\eea

The low energy theory  on the three color D4-branes 
has $Sp(\widetilde{N}_c) \times SO(2\widetilde{N}_c') \times
Sp(\widetilde{N}_c'')$ 
gauge group and  
$2N_f$-fundamental dual quarks $q''$,
$2N_f'$-vectors dual quarks $q'$, 
$2N_f''$-fundamental dual quarks $q$,
bifundamentals $f, g$
and various gauge singlets.
The $f$ is in the representation 
$\bf{(2\widetilde{N}_c, 2\widetilde{N}_c', 1)}$ while 
the $g$ is in the representation $\bf{(1, 2\widetilde{N}_c', 
2\widetilde{N}_c'')}$,
under the dual gauge group. 
The $2N_f''$ flavors $q$ are in the representation 
$\bf{(2\widetilde{N}_c, 1, 1)}$, 
the $2N_f'$ flavors $q'$ are in the representation 
$\bf{(1, 2\widetilde{N}_c', 1)}$, and
the $2N_f$ flavors $q''$ are in the representation 
$\bf{(1, 1, 2\widetilde{N}_c'')}$, 
under the gauge group.
In particular, a magnetic meson field 
$
M_0 \equiv Q Q
$
is $2N_f \times 2N_f$ matrix and comes from 
4-4 strings of $2N_f$ flavor D4-branes(when $2N_f$
$D6_{-\theta}$-branes meet the $NS5_R'$-brane), 
a magnetic meson field 
$
M'_0 \equiv Q' Q'
$
is $2N_f' \times 2N_f'$ matrix and comes from 
4-4 strings of $2N_f'$ flavor D4-branes(when $2N_f'$
$D6_{-\theta'}$-branes meet the $NS5_L$-brane) and 
a magnetic meson field 
$
M''_0 \equiv Q'' Q''
$
is $2N_f'' \times 2N_f''$ matrix and comes from 
4-4 strings of $2N_f''$ flavor D4-branes(when $2N_f''$
$D6_{-\theta''}$-branes meet the $NS5_L$-brane).

Then the most general magnetic superpotential, for   
the case where $2N_f(2N_f')[2N_f'']$ 
$D6_{-\theta}$-branes($D6_{-\theta'}$-branes)[$D6_{-\theta}$-branes] meet 
the NS-branes {\it with angles},
is given by  
\bea
W_{dual} & = & \left[ f^4 +
f^2 g^2 + g^4 \right]
 \nonu \\
& + & \left( \frac{\alpha}{2} \tr M_0^2 - m M_0  \right) + \left(
\frac{\alpha'}{2} 
\tr {M_0'}^2 - m' M_0' \right)
+ \left( \frac{\alpha''}{2} 
\tr {M_0''}^2 - m'' M_0'' \right) \nonu \\
& + & \left[ M_0 q'' f^4 q''  
+ 
M_0' q' f^4  q'
+ 
M_0'' q f^4  q \right]  
\nonu \\
&+ & \left[
M_2 q'' f^2 q'' + M_4 q'' q'' +
M_{2,F}' q' g^2 q' +
M_{2,G}' q' f^2 q' 
+ M_4' q' q' + 
M_2'' q f^2 q \right. \nonu \\
& + & \left. M_4'' q q  +   
P_1 q' f^3 q''  +
P_2 q f g q'' +
P_3 q' f  q'' 
+
R_1 q' f^4 q +
R_3 q' f q  \right]
\label{mag4}
\eea
where the mesons are given by 
\bea
M_0 & \equiv &  Q  Q,\quad  
M_0'  \equiv Q' Q', \quad
M_0''  \equiv Q'' Q'', \quad 
M_2  \equiv Q  F^2 Q, \quad 
M_4   \equiv Q F^4 Q,
\nonu \\
M_{2,F}'  & \equiv & Q'  F^2 Q',  \quad
M_{2,G}'  \equiv Q' G^2 Q', \quad
M_4'   \equiv Q'  F^4 Q', \quad
M_2''  \equiv Q'' G^2 Q'', \nonu \\ 
M_4''   & \equiv & Q''  G^4 Q'',
\quad
P_1  \equiv  Q F Q', \quad
P_2 \equiv Q F G Q'', \quad
P_3  \equiv  Q F^3 Q', \nonu \\
R_1  & \equiv &  Q' G Q'', \quad
R_3  \equiv  Q' G^3 Q''.
\nonu
\eea
The first two lines of (\ref{mag4}) are dual expressions for the electric
superpotential (\ref{elecsuper3}) and the corresponding meson fields
$M_0, M_0', M_0''$ are replaced and the remaining
lines of (\ref{mag4}) are the analogs of the cubic term
superpotential between
the meson and dual quarks in Seiberg duality. 

When the $N_f$ $D6_{-\theta}$-branes meet the $NS5_L'$-brane and 
$NS5_R$-brane,
{\it no} creation of D4-branes  
implies that there is no $M_2$- or $M_4$-term 
in the above superpotential
(\ref{mag4})(and their mirrors).
The mesons  
$M_2$ and $M_4$ originate from  $Sp(N_c)$ chiral mesons
$Q Q$  when one
dualizes the $Sp(N_c)$ gauge group first by moving the $NS5_L'$-brane
to the left of the $NS5_L$-brane. That is, the fluctuations of
strings stretching between the $2N_f$ ``flavor'' D4-branes provide
these meson fields. 
After the additional dual procedures, 
the cubic terms in the
superpotential arise as 
$M_2$-dependent  and $M_4$-dependent terms 
where $M_2$ has 
extra $F^2 $ fields and 
$M_4$ has 
extra $F^4$ fields, 
besides $Q Q$, 
due to the further dualizations. The $M_2$-term in the superpotential has an extra 
$f^2$ factor 
besides $q'' q''$.

When the $N_f''$ $D6_{-\theta''}$-branes meet the $NS5_L'$-brane and 
$NS5_R$-brane,
no creation of D4-branes  
implies that there is no $M_2''$- or $M_4''$-term 
in the above superpotential
(\ref{mag4})(and their mirrors).
The mesons  
$M_2''$ and $M_4''$ originate from  $Sp(N_c'')$ chiral mesons
$Q'' Q''$  when one
dualizes the $Sp(N_c'')$ gauge group first by moving the $NS5_R$-brane
to the right of the $NS5_R'$-brane. That is, the fluctuations of
strings stretching between the $2N_f''$ ``flavor'' D4-branes provide
these meson fields. 
After the additional dual procedures, 
the cubic terms in the
superpotential arise as 
$M_2''$-dependent  and $M_4''$-dependent terms 
where $M_2''$ has 
extra $G^2$ fields and 
$M_4''$ has 
extra $G^4$ fields, 
besides $Q'' Q''$, 
due to the further dualizations. The $M_2''$-term in the
superpotential has an extra 
$f^2$ factor 
besides $q q$.

Similarly, 
when 
the $N_f'$ $D6_{-\theta'}$-branes 
meet the $NS5_L'$-brane, the $NS5_R$-brane, or $NS5_R'$-brane with no angles, 
there is no $M_4'$ term in the above superpotential
(\ref{mag4})(and their mirrors) \footnote{In general, 
$4N_f'$ full D4-branes without changing the linking number should be
added in order to
satisfy the correct dual color numbers in the gauge theory side
analysis. This has led to 
the fact that there are  $4N_f'$  D4-branes 
connecting the  $NS5_{R}$-brane and $N_f'$ $D6_{-\theta'}$-branes,
after duality. 
In other words, these extra $4N_f'$ D4-branes were needed
for the existence of meson fields $M_{2,F}'$ and $M_{2,G}'$. 
In our construction, we do not
need these extra 
$4N_f'$ full D4-branes because we do not want to have these unwanted
meson fields $M_{2,F}'$ and $M_{2,G}'$.}.
These meson fields 
originate from  $SO(2N_c')$ chiral mesons
$Q' Q'$  when one
dualizes the $SO(N_c')$ gauge group first by interchanging the $NS5_L'$-brane
and the $NS5_R$-brane each other. The
strings stretching between the $2N_f'$ ``flavor'' D4-branes provide
this meson. 
After the additional dual procedures, 
the cubic terms in the superpotential arise as 
$M_{2,F}'$-term, $M_{2,G}'$-term and $M_4'$-term 
in (\ref{mag4}) where $M_{2,F}'$ has 
extra $F^2 $ fields,
$M_{2,G}'$ has 
extra $G^2$ fields, 
and 
$M_4'$ has extra $F^4 $ fields,
besides $Q' Q'$, due to the further dualizations.
The $M_{2,F}'$-term in the superpotential has  extra 
$g^2$ factor 
while 
$M_{2,G}'$-term has  extra 
$f^2$ dependent factor, besides
$q' q'$.

Furthermore, 
when the $N_f$ $D6_{-\theta}$-branes, the 
$N_f'$ $D6_{-\theta'}$-branes, the $NS5_{L}'$-brane and 
the $NS5_R$-brane
meet each other with no angles,
no $P_1$- and $P_3$-dependent terms arise in the
superpotential (\ref{mag4}).
These mesons 
originate from  $SO(2N_c')$ chiral mesons
$F Q'$ when one
dualizes the $SO(2N_c')$  first by interchanging the $NS5_L'$-brane
the $NS5_R$-brane. 
The
strings stretching between the $2N_f'$ flavor D4-branes and $2N_c$
color D4-branes give rise to these 
$2N_f'$ $Sp(N_c)$ fundamentals. 
After the additional dual procedures,  these cubic terms arise as 
these meson terms  where there exist  extra 
$f^2 q''$
and $q''$ of interactions in $P_1$ and 
$P_3$ in the superpotential and these mesons have 
extra $Q, Q F^2 $ fields respectively, 
due to the further dualizations. 

When the $N_f'$ $D6_{-\theta'}$-branes, the 
$N_f''$ $D6_{-\theta''}$-branes, the $NS5_{L}'$-brane and 
the $NS5_R$-brane
meet each other with no angles,
no $R_1$- and $R_3$-dependent terms arise in the
superpotential (\ref{mag4}).
These mesons 
originate from  $Sp(N_c'')$ chiral mesons
$ G Q''$ when one
dualizes the $Sp(N_c'')$  first by moving the $NS5_R$-brane
to the right of the $NS5_R'$-brane. 
The
strings stretching between the $2N_f''$ flavor D4-branes and $2N_c'$
color D4-branes give rise to these 
$2N_f''$ $SO(2N_c')$ vectors. 
After the additional dual procedures,  these cubic terms arise as 
these meson terms  where 
$R_1$ and 
$R_3$ have 
extra $Q', Q' G^2$ fields respectively, 
due to the further dualizations. 

Finally, 
when the $N_f$ $D6_{-\theta}$-branes, the 
$N_f''$ $D6_{-\theta''}$-branes, the $NS5_{L}'$-brane and 
the $NS5_R$-brane
meet each other with no angles,
no $P_2$-dependent term occurs in the
superpotential (\ref{mag4}).
These mesons 
originate from  $Sp(N_c)$ chiral mesons
$Q F $ when one
dualizes the $Sp(N_c)$  first by moving the $NS5_L'$-brane
to the left of the $NS5_L$-brane. 
The
strings stretching between the $2N_f$ flavor D4-branes and $2N_c'$
color D4-branes give rise to these 
$2N_f$ $SO(2N_c')$ vectors. 
After the additional dual procedures,  these cubic terms arise as 
these meson terms  where there exist  extra 
$ g q''$ in the $P_2$ interaction term and this meson has 
extra $G Q''$ fields respectively, 
due to the further dualizations. 

Then the reduced magnetic superpotential in our case  
by taking the first three lines of (\ref{mag4}) 
is given by 
\bea
W_{dual} & = &  \left[ M_0 q'' f^4 q''  
 + \frac{\alpha}{2} \tr M_0^2 - m M_0  
\right]+ \left[
M_0' q' f^4  q'
+\frac{\alpha'}{2} 
\tr {M_0'}^2 - m' M_0' 
\right] \nonu \\
&+ & \left[ M_0'' q f^4  q +
 \frac{\alpha''}{2} 
\tr {M_0''}^2 - m'' M_0'' \right].
\label{dualWdual}
\eea

For the supersymmetric vacua, one can compute the F-term equations for
this superpotential (\ref{dualWdual}) 
and the F-terms for $M_0, q'', M_0', q', M_0'', q$ and $f$ 
are given by
\bea
&& q'' f^4 q''  -m + \alpha M_0   =  0, \qquad
(M_0  q'' f^2) f^2 +
f^2 (f^2 q'' M_0) =0, \nonu \\
&& q' f^4  q'  -m' + \alpha' M_0'  =  0, \qquad
(M_0' q' f^2) f^2 +
f^2 (f^2 q' M_0') =0, \nonu \\
&& q f^4  q  -m'' + \alpha'' M_0''  =  0, \qquad
(M_0'' q f^2) f^2 +
 f^2 (f^2  q M_0'') =0, \nonu \\
&&  f (f^2 q''  M_0) q'' +
f q'' (M_0 q'' f^2) + 
f (f^2  q' M_0') q' 
+f  q' 
(M_0' q' f^2) + 
f (f^2  q
M_0'') q  \nonu \\
&& + 
f  q (M_0''  q f^2) 
+  (f^2 q'' M_0)  q'' f +    q''
(M_0  q'' f^2) f +  
(f^2  q' M_0') q' f  +
q' (M_0' q' f^2) f \nonu \\
&& +  (f^2  q M_0'') q f+ q' (M_0''
  q f^2) f =  0.
\nonu
\eea
From this, it is easy to see that the last equation is satisfied
if the second, fourth and sixth are 
satisfied:$M_0  q'' f^2=0=\cdots=f^2  q M_0''$.

%%%%%%%%%%%%%
%Figure 5
%%%%%%%%%%%%%

The theory has many nonsupersymmetric meta-stable ground states and 
when we rescale the meson fields as
$
M_0 = h \Lambda \Phi_0, M_0' = h' \Lambda' \Phi_0'$ 
and 
$M''_0 = h'' \Lambda'' \Phi''_0$,
then the Kahler potential for $\Phi_0, \Phi_0'$ and $\Phi''_0$ 
is canonical and the magnetic
quarks are canonical near the origin of field space.
Then the magnetic superpotential (\ref{dualWdual}) 
can be rewritten as
\bea
W_{mag} & = & \left[ h \Phi_0  q''   f^4 q'' 
 +  
\frac{h^2 \mu_{\phi}}{2} \tr \Phi_0^2- h \mu^2 \tr \Phi_0 \right]
 +  
\left[ h' \Phi_0'    q'  f^4 q' 
 +  
\frac{{h'}^2 \mu_{\phi}'}{2} \tr {\Phi_0'}^2- 
h' {\mu'}^2 \tr \Phi_0'\right]
\nonu \\
 & + & 
\left[ h'' \Phi_0''  q  f^4 q 
 +  
\frac{{h''}^2 \mu_{\phi}''}{2} \tr {\Phi_0''}^2- 
h'' {\mu''}^2 \tr \Phi_0''\right]
\nonu
\eea
where
$
\mu^2 = m \Lambda, {\mu'}^2 =m' \Lambda',  {\mu''}^2 =m'' \Lambda''$ and  
$\mu_{\phi} = \alpha \Lambda^2, 
\mu_{\phi}' = \alpha' {\Lambda'}^2, \mu_{\phi}'' = \alpha'' {\Lambda''}^2$.

Now one splits 
the $2(N_f-l) \times 2(N_f-l)$
block  at the lower right corner of $h\Phi_0$ and $q'' 
f^4  q''$ 
into blocks of 
size $2n$ and $2(N_f-l-n)$,
one decomposes 
the $2(N_f'-l') \times 2(N_f'-l')$
block  at the lower right corner of $h' \Phi'_0$ and $q'   
f^4 q'$ 
into blocks of 
size $2n'$ and $2(N_f'-l'-n')$
and 
one splits 
the $2(N_f''-l'') \times 2(N_f''-l'')$
block  at the lower right corner of $h''\Phi_0''$ and $q 
f^4 q$ 
into blocks of 
size $2n''$ and $2(N_f''-l''-n'')$
as follows:
\bea
h\Phi_0 & = &  \left(
\begin{array}{ccc}
0_{2l} & 0 & 0  \\
0 & h \Phi_{2n} & 0 \\
0 & 0 & \frac{\mu^2}{\mu_{\phi}} {\bf 1}_{N_f-l-n} \otimes i \sigma_2
\end{array}
\right), h' \Phi_0'  =   \left(
\begin{array}{ccc}
0_{2l'} & 0 & 0  \\
0 & h' \Phi_{2n'}' & 0 \\
0 & 0 & \frac{{\mu'}^2}{\mu_{\phi}'} {\bf 1}_{N_f'-l'-n'} \otimes \sigma_3
\end{array}
\right), \nonu \\
h'' \Phi_0''  & = &   \left(
\begin{array}{ccc}
0_{2l''} & 0 & 0  \\
0 & h'' \Phi_{2n''}'' & 0 \\
0 & 0 & \frac{{\mu''}^2}{\mu_{\phi}''} {\bf 1}_{N_f''-l''-n''} \otimes
i \sigma_2
\end{array}
\right), 
q''  f^4 q''   =  \left(
\begin{array}{ccc}
\mu^2 {\bf 1}_{2l} & 0 & 0  \\
0 & \varphi'' y^4 \varphi''  &  0 \\
0 & 0 & 0_{2(N_f-l-n)}
\end{array}
\right), \nonu \\
q' f^4 q'  & = & \left(
\begin{array}{ccc}
{\mu'}^2 {\bf 1}_{2l'} & 0 & 0  \\
0 & \varphi' y^4  \varphi'  &  0 \\
0 & 0 & 0_{2(N_f'-l'-n')}
\end{array}
\right),
q f^4 q   =  \left(
\begin{array}{ccc}
{\mu''}^2 {\bf 1}_{2l''} & 0 & 0  \\
0 & \varphi y^4  \varphi  &  0 \\
0 & 0 & 0_{2(N_f''-l''-n'')}
\end{array}
\right).
\label{vac3}
\eea
Here $\varphi''$ 
is $2n \times 2(\widetilde{N}_c''-l)$
dimensional matrices,   
$\varphi'$ is $2n' 
\times 2(\widetilde{N}_c'-l')$
dimensional matrices and
$\varphi$ is 
$2n'' \times 2(\widetilde{N}_c-l'')$
dimensional matrices.
The 
$\varphi''$ corresponds to 
fundamental strings connecting the $2n$ flavor D4-branes and
$2(\widetilde{N}_c''-l)$
color D4-branes,  
$\varphi'$ corresponds to 
fundamental strings connecting the $2n'$ flavor D4-branes and
$2(\widetilde{N}_c'-l')$
color D4-branes and 
$\varphi$ corresponds to 
fundamental strings connecting the $2n''$ flavor D4-branes and
$2(\widetilde{N}_c-l'')$
color D4-branes.
The $\Phi_{2n}$ and $\varphi'' y^4 \varphi''$
are $2n \times 2n$ matrices,  
$\Phi'_{2n'}$ and $\varphi' 
y^4 \varphi'$
are $2n' \times 2n'$ matrices and
$\Phi''_{2n''}$ and $\varphi 
y^4 \varphi$
are $2n'' \times 2n''$ matrices.

The supersymmetric ground state corresponds to
$
h\Phi_{2n}= \frac{\mu^2}{\mu_{\phi}} {\bf 1}_{n} \otimes i \sigma_2, 
\varphi''   y^2  
=0=y^2 \varphi''$,  
$h'\Phi'_{2n'}= \frac{{\mu'}^2}{\mu_{\phi}'} {\bf 1}_{n'} \otimes \sigma_3, 
\varphi' y^2 =0=y^2 
\varphi'$ and 
$h''\Phi''_{2n''}= \frac{{\mu''}^2}{\mu_{\phi}''} {\bf 1}_{n''}
\otimes i \sigma_2, 
\varphi y^2  =0=y^2 
\varphi$.
The $l$ of the upper
$N_f$-flavor D4-branes are reconnected with $l$-color
D4-branes
and the resulting $l$ D4-branes 
stretch from the upper $D6_{-\theta}$-branes to
the $NS5_L$-brane directly 
and the intersection point between the 
$l$ D4-branes and the $D6_{-\theta}$-branes is given by 
$(v, w)=(+v_{D6_{-\theta}}, 0)$.
The mirrors are located at $(v, w)=(-v_{D6_{-\theta}}, 0)$.
This
corresponds to  exactly the $2l$'s eigenvalues from 
zeros of 
$h\Phi_0$ in (\ref{vac3}).
Now the remaining upper $(N_f-l)$-flavor D4-branes between 
the upper $D6_{-\theta}$-branes and 
the $NS5_L'$-brane correspond to the eigenvalues 
of $h\Phi_0$ in (\ref{vac3}), i.e.,   
$\frac{\mu^2}{\mu_{\phi}} {\bf 1}_{N_f-l}$.
The intersection point between the 
upper $(N_f-l)$ D4-branes and the $NS5_L'$-branes is given 
by $(v, w)=(0, +v_{D6_{-\theta}} \cot \theta)$ 
from trigonometric 
geometry.
The mirrors are located at 
$(v, w)=(0, -v_{D6_{-\theta}} \cot \theta)$.

Similarly, 
the $l'$ of the upper 
$N_f'$-flavor D4-branes are reconnected with $l'$-color
D4-branes
and the resulting $l'$ D4-branes 
stretch from the upper $D6_{-\theta'}$-branes to
the $NS5_R$-brane directly 
and the intersection point between the 
$l'$ D4-branes and the upper $D6_{-\theta'}$-branes is given by 
$(v, w)=(+v_{D6_{-\theta'}}, 0)$.
The mirrors are located at 
$(v, w)=(-v_{D6_{-\theta'}}, 0)$.
This
corresponds to  exactly the $2l'$'s eigenvalues from 
zeros of 
$h'\Phi'_0$ in (\ref{vac3}).
Now the remaining upper $(N_f'-l')$-flavor D4-branes between 
the upper $D6_{-\theta'}$-branes and 
the $NS5_R'$-brane correspond to the eigenvalues 
of $h'\Phi'_0$ in (\ref{vac3}), i.e.,   
$\frac{{\mu'}^2}{\mu_{\phi}'} {\bf 1}_{N_f'-l'}$.
The intersection point between the 
upper $(N_f'-l')$ D4-branes and the $NS5_R'$-branes is given 
by $(v, w)=(0, +v_{D6_{-\theta'}} \cot \theta')$ from trigonometric 
geometry.
The mirrors are located at 
$(v, w)=(0, -v_{D6_{-\theta'}} \cot \theta')$.

The $l''$ of the upper 
$N_f''$-flavor D4-branes are reconnected with $l''$-color
D4-branes
and the resulting $l''$ D4-branes 
stretch from the upper $D6_{-\theta''}$-branes to
the $NS5_R$-brane directly 
and the intersection point between the 
$l''$ D4-branes and the upper $D6_{-\theta''}$-branes is given by 
$(v, w)=(+v_{D6_{-\theta''}}, 0)$.
The mirrors are located at $(v, w)=(-v_{D6_{-\theta''}}, 0)$.
This
corresponds to  exactly the $2l''$'s eigenvalues from 
zeros of 
$h''\Phi''_0$ in (\ref{vac3}).
Now the remaining upper 
$(N_f''-l'')$-flavor D4-branes between 
the upper $D6_{-\theta''}$-branes and 
the $NS5_R'$-brane correspond to the eigenvalues 
of $h''\Phi''_0$ in (\ref{vac3}), i.e.,   
$\frac{{\mu''}^2}{\mu_{\phi}''} {\bf 1}_{N_f''-l''}$.
The intersection point between the 
upper 
$(N_f''-l'')$ D4-branes and the $NS5_R'$-branes is given 
by $(v, w)=(0, +v_{D6_{-\theta''}} \cot \theta'')$ from trigonometric 
geometry.
The mirrors are located at 
$(v, w)=(0, -v_{D6_{-\theta''}} \cot \theta'')$.

Now the full one loop potential containing $\Phi_{2n}, 
\Phi'_{2n'}, \Phi''_{2n''}$, 
by combining the superpotential 
and the vacuum expectation values for the fields, 
takes the form
\bea
V  & = &  
|h \Phi_{2n} \varphi'' y^2 |^2   
+  |h  y^2 \varphi'' \Phi_{2n}|^2
  +  
| h \varphi''  y^4 
\varphi''-h \mu^2 {\bf 1}_{2n} + 
h^2 \mu_{\phi} \Phi_{2n}|^2 
\nonu \\ 
& + & |h' \Phi_{2n'}'  \varphi' y^2 |^2   
+  |h'  y^2 \varphi' \Phi_{2n'}' |^2
  +  
|   h' \varphi' y^4 
\varphi' -h' {\mu'}^2 {\bf 1}_{2n'} + 
{h'}^2 \mu_{\phi}' \Phi_{2n'}'|^2
\nonu \\
& + & |h'' \Phi_{2n''}''  \varphi y^2 |^2   
+  |h''  y^2 \varphi \Phi_{2n''}'' |^2
  +  
|   h'' \varphi y^4 
\varphi -h'' {\mu''}^2 {\bf 1}_{2n''} + 
{h''}^2 \mu_{\phi}'' \Phi_{2n''}''|^2 
\nonu \\
& + &  b |h^2 \mu|^2 \tr \Phi_{2n} \Phi_{2n}
 + b' |{h'}^2 \mu'|^2 
\tr {\Phi'}_{2n'} 
\Phi_{2n'}' + b'' |{h''}^2 \mu''|^2 
\tr {\Phi''}_{2n''} 
\Phi_{2n''}'' 
\nonu
\eea
where $b = \frac{(\ln 4-1)}{8\pi^2} \widetilde{N}_c$,  
$b' = \frac{(\ln 4-1)}{8\pi^2} \widetilde{N}_c'$ and 
$b'' = \frac{(\ln 4-1)}{8\pi^2} \widetilde{N}_c''$.
Differentiating this potential with respect to 
$\Phi_{2n}$, ${\Phi'}_{2n'}$ and 
${\Phi''}_{2n''}$ and putting 
$\varphi'' y^2 =
0 =  y^2 \varphi''$, 
$\varphi' y^2 =0 =  y^2
\varphi'$ and 
$\varphi y^2  =0 =  y^2
\varphi$, 
one obtains
\bea
h \Phi_{2n} 
& \simeq &  \frac{ \mu_{\phi}}{b }
{\bf 1}_n  \otimes i \sigma_2 \qquad  \mbox{or} \qquad
M_{2n} \simeq \frac{\alpha \Lambda^3}{\widetilde{N}_c} {\bf 1}_{n}
\otimes i \sigma_2, 
\nonu \\
h' \Phi_{2n'}' 
& \simeq &  \frac{ \mu_{\phi}'}{b' }
{\bf 1}_{n'}  \otimes \sigma_3 \qquad  \mbox{or} \qquad
M'_{2n'} \simeq \frac{\alpha' {\Lambda'}^3}{\widetilde{N}_c'} {\bf
  1}_{n'} \otimes \sigma_3, 
\nonu \\
h'' \Phi_{2n''}'' 
& \simeq &  \frac{ \mu_{\phi}''}{b'' }
{\bf 1}_{n''} \otimes i \sigma_2 \qquad  \mbox{or} \qquad
M''_{2n''} \simeq \frac{\alpha'' {\Lambda''}^3}{\widetilde{N}_c''} {\bf
  1}_{n''} \otimes i \sigma_2,
\nonu
\eea
corresponding to the $w$ coordinates of $2n$ curved flavor D4-branes between 
the $D6_{-\theta}$-branes and the $NS5_L'$-brane,  
the $w$ coordinates of $2n'$ curved flavor D4-branes between 
the $D6_{-\theta'}$-branes and the $NS5_R'$-brane,
the $w$ coordinates of $2n''$ curved flavor D4-branes between 
the $D6_{-\theta''}$-branes and the $NS5_R'$-brane
respectively.

%%%%%%%%%%%%%%%%%%%%%%%%%%%%%%%%%%%%%%%%%%%%%%%%%%%%%%%%%%%%%%%%%%%%%%%%%%%
\subsection{Addition of adjoint fields }
%%%%%%%%%%%%%%%%%%%%%%%%%%%%%%%%%%%%%%%%%%%%%%%%%%%%%%%%%%%%%%%%%%%%%%%%%%%

When we consider higher order superpotential for the adjoint fields 
whose degree is greater than 
four(i.e., the number of NS5-branes or NS5'-branes $k > 3$), 
there exist more meson fields.
Since it is not known how  
the deformations for the meson fields $Q X_1^j Q, 
Q X_2^j Q $ or $Q''
X_3^j Q''$ arise geometrically 
in the type IIA brane 
configuration,
it is not clear how to add these deformations in an
electric theory or a magnetic theory.

%%%%%%%%%%%%%%%%%%%%%%%%%%%%%%%%%%%%%%%%%%%%%%%%%%%%%%%%%%%%%%%%
\subsection{
$SO(2N_c) \times Sp(N_c') \times SO(2N_c'')$ with $2N_f$-vectors,
$2N_f'$-fund.,  $2N_f''$-vectors,  and bifund.}
%%%%%%%%%%%%%%%%%%%%%%%%%%%%%%%%%%%%%%%%%%%%%%%%%%%%%%%%%%%%%%%%

In this case, compared to the previous case, the charge of 
orientifold 4-plane is reversed. Correspondingly, 
the orthogonal(symplectic) 
gauge group is changed into symplectic(orthogonal) gauge group.
The only difference appears in 
the symmetric and antisymmetric property for the meson field and 
adjoint field. We do not present this here but it can be done by
following the prescription of previous section.  

%%%%%%%%%%%%%%%%%%%%%%%%%%%%%%%%%%%%%%%%%%%%%%%%%%%%%%%%%%%%%%%%%%%%%%%%%%%%%%%
%%%%%%%%%%%%%%%%%%%%%%%%%%%%%%%%%%%%%%%%%%%%%%%%%%%%%%%%%%%%%%%%%%%%%%%%%%%%%%%%
\section{
Conclusions and outlook }
%%%%%%%%%%%%%%%%%%%%%%%%%%%%%%%%%%%%%%%%%%%%%%%%%%%%%%%%%%%%%%%%%%%%%%%%%%%%%%%%
%%%%%%%%%%%%%%%%%%%%%%%%%%%%%%%%%%%%%%%%%%%%%%%%%%%%%%%%%%%%%%%%%%%%%%%%%%%%%%%%
 
We have found the type IIA
nonsupersymmetric meta-stable brane configurations
for 
two gauge group theory 
with fundamentals, bifundamentals and adjoints, 
three gauge group theory with fundamentals and bifundamentals,
and their orientifold 4-plane generalizations.
It would be interesting to discover whether
the meta-stable brane configurations exist for 
the theory with no D6-branes 
\cite{Ahn07-5,Ahn07-6,Ahn07-7,Ahn07-10} when we take the dual for the
whole gauge groups. For
the three gauge group theory with fundamentals and bifundamentals, it
is also possible to take the dual only for two gauge groups among three.
Then it would be interesting to see how the meta-stable brane
configurations occur.

\vspace{.7cm}

%%%%%%%%%%%%%%%%%%%%%%%%%%%%%%%%%%%%%%%%%%%%%%%%%%%%%%%%%%%%%%%
\centerline{\bf Acknowledgments}
%%%%%%%%%%%%%%%%%%%%%%%%%%%%%%%%%%%%%%%%%%%%%%%%%%%%%%%%%%%%%%%

%I would like to thank 
%D. Kutasov 
%for discussions. 
This work was supported by the 
National Research Foundation of Korea(NRF) grant 
funded by the Korea government(MEST)(No. 2009-0084601).
%This work was supported by grant No.
%R01-2006-000-10965-0 from the Basic Research Program of the Korea
%Science \& Engineering Foundation.  
%I would like to thank KIAS(Korea Institute for 
%Advanced Study) for hospitality  where
%this work was undertaken. 

\end{document}